\pdfoutput=1
\documentclass[11pt,twoside,a4paper,cmspaper,final,collab]{cms-tdr}
\def\svnVersion{4d408fd5}\def\svnDate{2025/12/15}\def\cmsCernNoTag{CERN-EP-2025-130}\def\cmsCernDate{\today}\def\cmsMessage{Published in Physical Review D as \href{http://dx.doi.org/10.1103/8lw4-6lyk}{\doi{10.1103/8lw4-6lyk}.}}
\begin{document}\cmsNoteHeader{SMP-23-003}

\newlength\cmsTabSkip\setlength{\cmsTabSkip}{1ex}
\providecommand{\cmsTable}[1]{\resizebox{\textwidth}{!}{#1}}
\ifthenelse{\boolean{cms@external}}{\providecommand{\cmsLeft}{upper\xspace}}{\providecommand{\cmsLeft}{left\xspace}}
\ifthenelse{\boolean{cms@external}}{\providecommand{\cmsRight}{lower\xspace}}{\providecommand{\cmsRight}{right\xspace}}

\providecommand{\Pellpr}{{\HepParticle{\Pell}{}{\prime}}\xspace}

\newcommand {\br}           {\ensuremath{\mathcal{B}}\xspace}
\newcommand {\zprime}       {\PZpr}
\newcommand {\pp}           {\ensuremath{\Pp\Pp}\xspace}
\newcommand {\zzprime}      {\ensuremath{\PZ^{(\prime)}}\xspace}
\newcommand {\taue}         {\ensuremath{\PGt_{\Pe}}\xspace}
\newcommand {\taumu}        {\ensuremath{\PGt_{\PGm}}\xspace}
\newcommand {\emu}          {\ensuremath{\Pe\PGm}\xspace}
\newcommand {\etau}         {\ensuremath{\Pe\PGt}\xspace}
\newcommand {\mutau}        {\ensuremath{\PGm\PGt}\xspace}
\newcommand {\tautau}       {\ensuremath{\PGt\PGt}\xspace}
\newcommand {\ztautau}      {\ensuremath{\PZ\to\tautau}\xspace}
\newcommand {\zmumu}        {\ensuremath{\PZ\to\PGm\PGm}\xspace}
\newcommand {\zee}          {\ensuremath{\PZ\to\Pe\Pe}\xspace}
\newcommand {\zll}          {\ensuremath{\PZ\to\Pell\Pell}\xspace}
\newcommand {\zmumugamma}  {\ensuremath{\PZ\to\PGm\PGm^*\to\PGm\PGm\PGg}\xspace}
\newcommand {\zmutau}       {\ensuremath{\PZ\to\PGm\PGt}\xspace}
\newcommand {\zmutauh}      {\ensuremath{\PZ\to\PGm\tauh}\xspace}
\newcommand {\zmutaue}      {\ensuremath{\PZ\to\PGm\PGt_{\Pe}}\xspace}
\newcommand {\zetau}        {\ensuremath{\PZ\to\Pe\PGt}\xspace}
\newcommand {\zetauh}       {\ensuremath{\PZ\to\Pe\tauh}\xspace}
\newcommand {\zetaumu}      {\ensuremath{\PZ\to\Pe\PGt_{\PGm}}\xspace}
\newcommand {\zemu}         {\ensuremath{\PZ\to\Pe\PGm}\xspace}
\newcommand {\zltauh}        {\ensuremath{\PZ\to\Pell\tauh}\xspace}
\newcommand {\zltaul}        {\ensuremath{\PZ\to\Pell\PGt_{\Pell}}\xspace}
\newcommand {\zltau}        {\ensuremath{\PZ\to\Pell\PGt}\xspace}
\newcommand {\zpemu}        {\ensuremath{\zprime\to\Pe\PGm}\xspace}
\newcommand {\zzpemu}       {\ensuremath{\zzprime\to\Pe\PGm}\xspace}
\newcommand {\htautau}      {\ensuremath{\PH\to\tautau}\xspace}
\newcommand {\hww}          {\ensuremath{\PH\to\PW\PW}\xspace}
\newcommand {\jtotau}       {\ensuremath{j\to\tauh}\xspace}
\newcommand {\mcoll}        {\ensuremath{m^{{\text{col}}}}\xspace}
\newcommand {\metau}        {\ensuremath{m_{\Pe\PGt}}\xspace}
\newcommand {\mmutau}       {\ensuremath{m_{\PGm\PGt}}\xspace}
\newcommand {\mltau}        {\ensuremath{m_{\Pell\PGt}}\xspace}
\newcommand {\mvis}         {\mltau}
\newcommand {\memu}         {\ensuremath{m_{\Pe\PGm}}\xspace}
\newcommand {\ptmu}         {\ensuremath{\pt^{\PGm}}\xspace}
\newcommand {\pte}          {\ensuremath{\pt^{\Pe}}\xspace}
\newcommand {\met}          {\ptmiss}
\newcommand {\metvec}       {\ptvecmiss}
\newcommand {\mt}           {\mT}
\newcommand {\mtlep}        {\ensuremath{\mT(\Pell\PGt,\met)}\xspace}
\newcommand {\hatpt}        {\ensuremath{\hat{p}_{\mathrm{T}}}\xspace}

\newcommand{\zjets}{\ensuremath{\PZ\!+\!\text{jets}}\xspace}
\newcommand{\wjets}{\ensuremath{\PW\!+\!\text{jets}}\xspace}
\newcommand{\gjets}{\ensuremath{\PGg\!+\!\text{jets}}\xspace}
\newcommand{\ww}{\ensuremath{\PW\PW}\xspace}

\cmsNoteHeader{SMP-23-003}

\title{Search for charged lepton flavor violating \texorpdfstring{\PZ}{Z} and \texorpdfstring{\PZpr}{Z'} boson decays in proton-proton collisions at \texorpdfstring{$\sqrt{s}=13\TeV$}{sqrt(s) = 13 TeV}}

\date{\today}

\abstract{
A search for flavor violating decays of the \PZ boson to charged leptons is performed using data from proton-proton collisions at $\sqrt{s}=13\TeV$ collected with the CMS detector at the LHC, corresponding to an integrated luminosity of 138\fbinv.  Each of the decays \zemu, \zetau, and \zmutau is considered. The data are consistent with the backgrounds expected from standard model processes.  For the \zemu channel the observed (expected) 95\% confidence level upper limit on the branching fraction is 1.9 (2.0)$\times 10^{-7}$, which is the most stringent direct limit to date on this process; the corresponding limits for the \zetau and \zmutau channels are 13.8 (11.4)$\times 10^{-6}$ and 12.0 (5.3)$\times 10^{-6}$, respectively.  Additionally, the $\Pe\PGm$ final state is used to search for lepton flavor violating decays of \PZpr resonances in the mass range from 110 to 500\GeV.  No significant excess is observed above the predicted background levels.
}

\hypersetup{%
pdfauthor={CMS Collaboration},%
pdftitle={Search for charged lepton flavor violating Z and Z' boson decays in proton-proton collisions at sqrt(s) = 13 TeV},%
pdfsubject={CMS},%
pdfkeywords={CMS, charged lepton flavor violation}}

\maketitle

\section{Introduction}

The standard model (SM) has been highly successful in describing nature on the small distance scale.
An important exception is the neutrino sector, where in the SM neutrinos are massless, while the observation of neutrino flavor oscillations~\cite{Super-Kamiokande:1998uiq,Super-Kamiokande:2001ljr,SNO:2002tuh,KamLAND:2002uet} indicates that
neutrinos have small but nonzero masses, and that their individual flavor is not a conserved quantity.
Generation number violation has also been observed in the quark sector, but not to date among charged leptons.
In an extended version of the SM that includes neutrino masses, the anticipated rates of charged lepton flavor violation (CLFV) induced by neutrino mixing
are vanishingly small; \eg, the branching fraction
$\br(\PZ\to \Pe\PGm) \approx \mathcal{O}(10^{-60})$~\cite{Illana:2000ic}.
Therefore any observation of a CLFV process would be direct evidence of physics beyond the SM (BSM).

Potentially measurable rates of CLFV in various processes, including \PZ boson decay, are predicted in BSM scenarios such as the seesaw mechanism, and supersymmetry and leptoquark models~\cite{Marciano:2008zz, Calibbi:2017uvl}.
In addition, new particles predicted by these and other BSM models may exhibit CLFV through their decays.  The new particles include additional gauge
bosons such as a \zprime boson, which would generally be
expected to be heavier than the \PZ boson~\cite{Langacker:2008yv}.
Axion-like particles have also been suggested as
potential BSM particles that may decay into flavor-violating states~\cite{Cornella:2019uxs,Bauer:2021mvw}.

Searches for CLFV have been conducted in low-energy processes including $\PGm\to \Pe\PGg$~\cite{MEGII:2023ltw}, $\PGm \to \Pe\Pe\Pe$~\cite{SINDRUM:1987nra}, $\PGmm\mathrm{N}(A,Z) \to \Pem \mathrm{N}(A,Z)$~\cite{SINDRUMII:2006dvw}, $\Pep \Pem \to \PGmp (\Pep) \PGtm$~\cite{CLEO:2008lxu,BaBar:2006xxh}, $\PGt\to \Pell\PGg$ ($\Pell = \Pe, \PGm$)~\cite{BaBar:2009hkt},
and $\PGt\to \Pellpr\Pellp\Pellm$~\cite{Hayasaka:2010np}.
Limits from these reactions lead to indirect constraints on the \PZ boson branching fractions $\br(\zemu) < 10^{-12}$, $\br(\zetau) < 10^{-7}$, and $\br(\zmutau) < 2\times 10^{-7}$~\cite{Marciano:2008zz}.
However, direct measurements of CLFV \PZ decays are needed to rule out anomalous couplings or cancellations that could evade these limits~\cite{Nussinov:2000nm}.
Searches for such decays have been performed by experiments at the CERN
LEP~\cite{OPAL:1995grn,DELPHI:1996iox} and LHC~\cite{ATLAS:2020zlz,ATLAS:2021bdj,ATLAS:2022uhq} colliders.  Current limits on the branching fractions for \zemu, \zetau, and \zmutau are $2.6\times10^{-7}$~\cite{ATLAS:2022uhq}, $5.0\times10^{-6}$~\cite{ATLAS:2021bdj}, and $6.5\times10^{-6}$~\cite{ATLAS:2021bdj}, respectively.

This paper presents a search for decays of the \PZ boson in the CLFV channels \zemu, \zetau, and \zmutau, together with a scan of the $\Pe\PGm$ invariant mass over the range 110--500\GeV to search for potential heavy neutral resonances.
The proton-proton (\pp) collision data at $\sqrt{s} = 13\TeV$ were collected with the CMS detector in 2016--2018, and correspond to an integrated luminosity of 138\fbinv.  New multivariate techniques extend the reach beyond previous results.

The signal extraction makes use of two main variables:  the invariant mass of the (visible) daughters of the \zzprime boson candidate, and a multivariate discriminator, implemented as a boosted decision tree (BDT), that combines kinematic observables chosen to discriminate signal from background. 
For the \emu channel, where the \zzprime boson is fully reconstructed, we extract the signal yield with a fit to the \emu mass distributions, relying mainly on the sidebands to determine the background under the signal peak.  These distributions are taken from separate subsamples that vary in purity, as measured by the BDT discriminator.  To distinguish signal from background in the \etau and \mutau channels, where the peak in the invariant mass distribution of the visible \PZ boson daughters is less sharply defined, we instead rely primarily on distributions in the BDT discriminant.  These distributions are taken from subsamples defined by ranges in the visible invariant mass chosen according to the dominance of the signal or of various backgrounds.  

In Section~\ref{sec:detector} we describe the CMS detector and particle reconstruction, and in Section~\ref{sec:simulation} the simulation of signal and background processes.  The event selection is given in Section~\ref{sec:selection}, followed by the signal extraction for the \zemu and \zltau channels in Sections~\ref{sec:zemu_analysis} and \ref{sec:zxtau_analysis}, respectively.  Systematic uncertainties are discussed in Section~\ref{sec:systematics}, and the results in Section~\ref{sec:results}.  We conclude with a summary in Section~\ref{sec:summary}.  Tabulated results are provided in the HEPData record for this analysis~\cite{hepdata}.

\section{The CMS detector and event reconstruction}
\label{sec:detector}

The central feature of the CMS apparatus is a superconducting solenoid of 6\unit{m} internal diameter, providing a magnetic field of 3.8\unit{T}. Within the solenoid volume are a silicon pixel and strip tracker, a lead tungstate crystal electromagnetic calorimeter (ECAL), and a brass and scintillator hadron calorimeter (HCAL), each composed of a barrel and two endcap sections. Forward calorimeters extend the coverage in pseudorapidity ($\eta$) provided by the barrel and endcap detectors.
The ECAL consists of 75\,848 lead tungstate crystals, which provide pseudorapidity coverage of $\abs{\eta} < 1.48 $ in its barrel region and $1.48 < \abs{\eta} < 3.00$ in the two endcap regions. Preshower detectors consisting of two planes of silicon sensors interleaved with a total of three radiation lengths of lead are located in front of each endcap detector.
Muons are measured in gas-ionization detectors embedded in the steel flux-return yoke outside the solenoid.
A more detailed description of the CMS detector, together with a definition of the coordinate system used and the relevant kinematic variables, is reported in Refs.~\cite{CMS:2008xjf,CMS:2023gfb}.

The silicon tracker used in 2016 measured charged particles within the range $\abs{\eta} < 2.5$.
For nonisolated particles of transverse momentum \pt between 1 and 10\GeV and $\abs{\eta} < 1.4$, the track resolutions were typically 1.5\% in \pt and 25--90
(45--150)\mum in the transverse (longitudinal) impact parameter~\cite{CMS:2014pgm}.
At the start of 2017, a new pixel detector was installed~\cite{Phase1Pixel}, which provided expanded acceptance, to
$\abs{\eta} < 3.0$, and improved transverse impact parameter resolution of 20--75\mum~\cite{DP-2020-049}.
The primary vertex is taken to be the vertex corresponding to the hardest scattering in the event, evaluated
using tracking information alone, as described in Section 9.4.1 of Ref.~\cite{CMS-TDR-15-02}.

The global event reconstruction (also called particle-flow event reconstruction~\cite{CMS:2017yfk}) aims to reconstruct and identify each individual particle in an event, with an optimized combination of all subdetector information. In this process, the identification (ID) of the particle type (photon, electron, muon, charged hadron, neutral hadron) plays an important role in the determination of the particle direction and energy. Photons are identified as ECAL energy clusters not linked to the extrapolation of any charged-particle trajectory to the ECAL. Electrons are identified as a primary charged-particle track and potentially many ECAL energy clusters corresponding to this track extrapolation to the ECAL and to possible bremsstrahlung photons emitted along the way through the tracker material. Muons are identified as tracks in the central tracker consistent with either a track or several hits in the muon system, and associated with calorimeter deposits compatible with the muon hypothesis. Charged hadrons are identified as charged-particle tracks neither identified as electrons, nor as muons. Finally, neutral hadrons are identified as HCAL energy clusters not linked to any charged hadron trajectory, or as a combined ECAL and HCAL energy excess with respect to the expected charged-hadron energy deposit.

For each event, hadronic jets are clustered from these reconstructed particles using the infrared- and collinear-safe anti-\kt algorithm~\cite{Cacciari:2008gp, Cacciari:2011ma} with a distance parameter of 0.4. Jet momentum is determined as the vectorial sum of all particle momenta in the jet, and is found from simulation to be, on average, within 5--10\% of the true momentum over the entire spectrum and detector acceptance.

Additional \pp interactions within the same or nearby bunch crossings (pileup) can contribute additional tracks and calorimetric energy depositions to the jet momentum.  To mitigate this effect, charged particles identified to be originating from pileup vertices are discarded, and an offset correction is applied to correct for remaining contributions~\cite{CMS:2020ebo}.

Jet energy corrections are derived from simulation to bring the measured response of jets equal, on average, to that of particle-level jets. In situ measurements of the momentum balance in dijet, \gjets, \zjets, and strong production of jets (QCD multijet) events are used to account for any residual differences in the jet energy scale between data obtained with the detector (``data'') and simulation~\cite{CMS:2016lmd}. The jet energy resolution amounts typically to 15--20\% at 30\GeV, 10\% at 100\GeV, and 5\% at 1\TeV~\cite{CMS:2016lmd}. Additional selection criteria are applied to each jet to remove jets potentially dominated by anomalous contributions from various subdetector component or reconstruction failures.

Jets associated with \PQb hadron decays are identified within the tracker acceptance of $\abs{\eta}<2.5$ using the \textsc{DeepJet} deep neural
network ID algorithms~\cite{Chatrchyan:2012jua,CMS-DP-2018-058,Bols:2020bkb}.
We use both tight and loose working points of this algorithm.
Their efficiencies for \PQb jets (light-flavor quark and gluon jets) have been measured in \ttbar events to be 84 (10)\% for the loose working point and 52 (0.1)\% for the tight working point.

Hadronic tau lepton decays (\tauh) are reconstructed from jets using the hadrons-plus-strips algorithm~\cite{CMS:2018jrd}, which combines 1 or 3 tracks with energy deposits in the calorimeters to identify the various hadronic \PGt decay modes. Neutral pions are reconstructed as clusters with dynamic size in $(\eta,\phi)$, from reconstructed electrons and photons; the size of the dimension in the azimuthal angle $\phi$ (`` strip'') varies as a function of the \pt of the electron or photon candidate, because of the dispersion in the magnetic field.
To distinguish between genuine \tauh decays and jets originating from the hadronization of quarks or gluons, and from electrons or muons, we use the \textsc{DeepTau} algorithm~\cite{CMS:2022prd}. Information from all individual reconstructed particles near the \tauh axis is combined with properties of the \tauh candidate and the event.  The rate of a jet to be misidentified as a \tauh candidate depends on the \pt and quark flavor of the jet.
In a data sample enriched in \wjets events this misidentification (misID) rate is measured to be about 0.25\%, with an ID efficiency for genuine \tauh of 60\%, at the working point and kinematic regime of our signal regions (SRs)~\cite{CMS:2022prd}.
The rate for electrons (muons) to be misidentified as \tauh is 2.60 (0.03)\% for a genuine \tauh ID efficiency of 80 (${>}99$)\%.

The electron momentum is estimated by combining the energy measurement in the ECAL with the corresponding momentum measurement in
the tracker~\cite{CMS-DP-2020-021}. The momentum resolution for electrons with $\pt \approx 45\GeV$ from $\PZ \to \Pe \Pe$ decays ranges
from 1.6 to 5\%. The resolution is generally better in the barrel region than in the endcaps, and it also depends on the bremsstrahlung energy emitted by the electron as it traverses the material upstream of the ECAL~\cite{CMS:2020uim,CMS-DP-2020-021}.

Muons are measured in the range $\abs{\eta} < 2.4$, with detection planes made using three technologies: drift tubes, cathode strip chambers, and resistive-plate chambers.
The efficiency for the reconstruction and ID of muons is greater than 96\%.
Once these muon candidates are matched to tracks measured in the silicon tracker, the relative \pt resolution for muons with \pt up to 100\GeV is 1\% in the barrel and 3\% in the endcaps.
The \pt resolution in the barrel is below 7\% for muons with \pt up to 1\TeV~\cite{CMS:2018rym}.

The missing transverse momentum vector \ptvecmiss is computed as the negative vector sum of the transverse momenta of all the particle-flow candidates in an event; its magnitude is denoted as \ptmiss.
The \ptvecmiss is modified to account for corrections to the energy scale of the reconstructed jets in the event.
To reduce the dependence of \ptvecmiss on pileup, we apply the pileup-per-particle-identification algorithm~\cite{Bertolini:2014bba}, which weights the particle-flow candidates entering the \ptvecmiss sum by their probability of originating from the primary vertex~\cite{CMS:2019ctu}.

Events of interest are selected with a two-tiered trigger system. The first level, composed of custom hardware processors, uses information from the calorimeters and muon detectors to select events at a rate of around 100\unit{kHz} within a fixed latency of about 4\mus~\cite{CMS:2020cmk}. The second level, known as the high-level trigger, comprises a farm of processors running a version of the full event reconstruction software optimized for fast processing, and reduces the event rate to around 1\unit{kHz} before data storage~\cite{CMS:2016ngn,CMS:2024aqx}.

\section{Simulated event samples}
\label{sec:simulation}
Monte Carlo simulated samples are used to optimize the analysis and to model various background sources.
Events arising from \wjets are simulated, for the 2016 (2017--2018) data taking periods, with the {\MGvATNLO}\,2.2.2 (2.4.2)~\cite{Alwall:2014hca,Alwall:2007fs}
event generator at leading order (LO) accuracy.
Drell--Yan (DY), $\PW\PGg$, and $\PW\PW\PW$ events are simulated with the same generators at next-to-LO (NLO) accuracy.
The
{\POWHEG}\,v2.0~\cite{Nason:2004rx,Frixione:2007vw,Alioli:2010xd,Alioli:2009je,Re:2010bp,Alioli:2008tz,Nason:2009ai,Melia:2011tj,Nason:2013ydw,Luisoni:2013cuh} program at NLO is used to generate events from \ttbar, $\PQt\PW$,
single top quark, and \PH boson production
(where \PH denotes the 125-\GeV Higgs boson), as well as
diboson events originating from
$\PW\PW$, $\PW\PH$ or $\PZ\PH$ production.
The $\PW\PZ$ and $\PZ\PZ$ events are generated at LO using the \PYTHIA 8.205~\cite{Sjostrand:2014zea} package.
The simulated background samples are normalized using cross section calculations
performed at NLO or next-to-NLO (NNLO) precision~\cite{Melnikov:2006kv,Czakon:2011xx,Kidonakis:2013zqa,Kidonakis:2013zqa,Campbell:2011bn,Gehrmann:2014fva}.

Signal processes are generated at LO using \PYTHIA.
For each CLFV decay channel (\zemu, \zetau, or \zmutau), a tree-level coupling is added, with no coupling to the other decay channels.
Events for the off-\PZ mass scan are generated with \PYTHIA, including only the contribution of a vector boson \PZpr, and not considering any mixing with the \PZ boson or photon.
The decay amplitude is calculated using tree-level couplings to the $\Pe\PGm$ final state with no other decay channels.
The resonances are generated in the narrow-width approximation, \ie, with a width of $0.1\GeV$, far below the experimental resolution.
Events are simulated with masses of 100, 125, 150, 175, 200, 300, 400, and 500\GeV.

In all samples the hadronization, as well as the underlying event, are modeled with \PYTHIA using the
CUETP8M1~\cite{Khachatryan:2015pea} (CP5~\cite{CMS:2019csb}) tune for the 2016 (2017-18) data-taking years.  The LO (NLO) simulations use the MLM (FxFx) scheme for matching the products of the matrix-element calculations to the parton shower simulation~\cite{Frederix:2012ps}.
The parton distribution functions used with the matrix-element generators are taken from the NNPDF3.0~\cite{NNPDF:2014otw} set at LO or NLO, except for the \ww samples for the 2017--2018 data-taking years where NNPDF3.1\cite{NNPDF:2017mvq} is used; for \PYTHIA the NNPDF2.3LO~\cite{Ball:2013hta} set is used.

The CMS detector response is simulated using \GEANTfour~\cite{Agostinelli:2002hh}.
Simulated samples are reconstructed with the same software packages as used for collision
data. To account for effects of pileup, minimum-bias events simulated with \PYTHIA are added to the hard scattering process, with a distribution in the number of vertices matching that in the data.

\section{Event selection}
\label{sec:selection}

An event is required to pass either a single-electron or single-muon
trigger.  To ensure that the trigger is fully efficient (${>}95\%$) for the events selected for analysis, an electron (muon) matched to the single-electron (muon) trigger is required to have \pt exceeding a threshold of 29 (25)\GeV for data collected in 2016,
35 (28)\GeV for data collected in 2017, and 34 (25)\GeV for data collected in 2018.

For all three search channels we make a selection of events
with two lepton candidates (\Pe, \PGm, or \tauh) of different flavor and opposite charge.
These events are sorted into three separate data channels: $\Pe\PGm$ (from which the \zzpemu, \zetaumu, and \zmutaue candidates are selected), $\Pe\tauh$, and $\PGm\tauh$. To suppress SM backgrounds that produce pairs of leptons, events are rejected if they contain additional electrons or muons, and for the $\Pe\tauh$ and $\PGm\tauh$ channels, also if they contain additional \tauh candidates.
In the search for the \zetau or \zmutau
signal we include both a hadronic \PGt decay category ($\Pe\tauh$ or $\PGm\tauh$, respectively) and a different-flavor leptonic \PGt decay category ($\Pe\taumu$ or $\PGm\taue$).
As each CLFV \PZ boson decay
signal is the target of a separate search, there is no requirement that the \zetaumu, \zmutaue, and \zemu samples be nonoverlapping.

To select a light lepton \Pell (\Pe or \PGm) that is produced in the decay of a \PZ boson or tau lepton and not from a hadron within a quark jet, we construct the relative isolation variable 
$I^{\Pell}_{\text{rel}}$, defined as a function of the lepton's transverse momentum $\pt^{\Pell}$ as
\begin{equation}
  I^{\Pell}_{\text{rel}} = \frac{\sum{\!\pt^{\text{charged}}} + \max{\!\left[0,\, \sum{\!\pt^{\text{neutral}}} + \sum{\!\pt^{\PGg}} - \pt^{\mathrm{PU}}\right]}}{\pt^{\Pell}},
  \label{eq:lepton_iso}
\end{equation}
where the sums include the particles, excluding the lepton itself, within a cone of size $\Delta R \equiv \sqrt{\smash[b]{(\Delta\eta)^2 + (\Delta \phi)^2}} = 0.4\, (0.3)$ about the electron (muon).
Here $\pt^{\text{charged}}$ refers to charged hadrons, $\pt^{\text{neutral}}$ to neutral hadrons, $\pt^{\PGg}$ to photons, and $\pt^{\mathrm{PU}}$ to the estimated neutral contribution from pileup.
Electrons and muons are required to pass a tight isolation requirement, $I_{\text{rel}}^{\Pell} < 0.15$.

To select prompt electrons and muons we require their transverse and longitudinal impact parameters, $d_{xy}$ and $d_{z}$, to be less than 0.2 and 0.5\cm, respectively.
For further suppression of the background from nonprompt electrons (primarily from QCD multijet and \wjets events), we require electrons to have $\pt > 15\GeV$.  Electron candidates with ECAL clusters in the gap region between the barrel and endcap ($1.444 < \abs{\eta} < 1.566$) are rejected.

The \tauh candidates are required to have $\pt > 20\GeV$ and $\abs{\eta} < 2.3$, with $\abs{d_{z}} < 0.2\cm$.  They are further required to pass the following working points of the \textsc{DeepTau} tagging algorithm~\cite{CMS:2022prd} for the suppression of:  muons (``tight''); jets (``tight'' for the SR, ``very loose'' for the control regions (CRs) used to estimate the misID background); and electrons (``loose'', except ``very tight'' for the $\Pe\tauh$ channel SR to suppress the background from \zee where one of the electrons is misidentified as a \tauh candidate).

Jets are required to have $\pt > 20\GeV$ and $\abs{\eta} < 3$.  Where jets are used in the event selection, jets within $\Delta R < 0.3$ of identified leptons are excluded from consideration.

Additional requirements specific to the individual channels are described in the corresponding sections below.

\section{Search for \texorpdfstring{$\PZ\to\Pe\mu$}{Z->e+mu}}
\label{sec:zemu_analysis}

The \zemu search uses the $\Pe\PGm$ data channel,
requiring exactly one electron and one muon candidate.  Both leptons are required to have $\abs{\eta} < 2.4$, and $\pt > 20\GeV$ in addition to the trigger threshold requirement noted in Section~\ref{sec:selection} above.
The invariant mass \memu of the dilepton system is required to satisfy $70 < \memu < 110\GeV$; this range includes sidebands around the \PZ boson mass ($91.1880\pm0.0020\GeV$~\cite{PDG2024}) for estimation of the background.
The sources of backgrounds near the \PZ boson mass are $\PW\PW$, \ttbar, DY production of \tautau, nonprompt $\Pe\PGm$, and \zmumugamma production.

Events from \ttbar (as well as subdominant single \PQt quark) production are suppressed by rejecting events with jets that satisfy the loose \PQb tagging~\cite{Bols:2020bkb}.
To suppress the \ztautau contribution we require, in addition to the requirements on the track impact parameters given in Section~\ref{sec:selection}, that their significances satisfy $\abs{d_{xy}}/\sigma_{d_{xy}} < 3$ and $\abs{d_{z}}/\sigma_{d_{z}} < 4.7$, with $\sigma_{d_{xy}}$ and $\sigma_{d_{z}}$ the respective uncertainties.
The efficiency reduction from these requirements on the track impact parameters is ${\approx} 5 \%$ on the signal, and ${\approx} 20 \%$ on those backgrounds that have tau leptons in the final state.

\begin{figure}[!htbp]
  \centering
\includegraphics[width=0.48\textwidth]{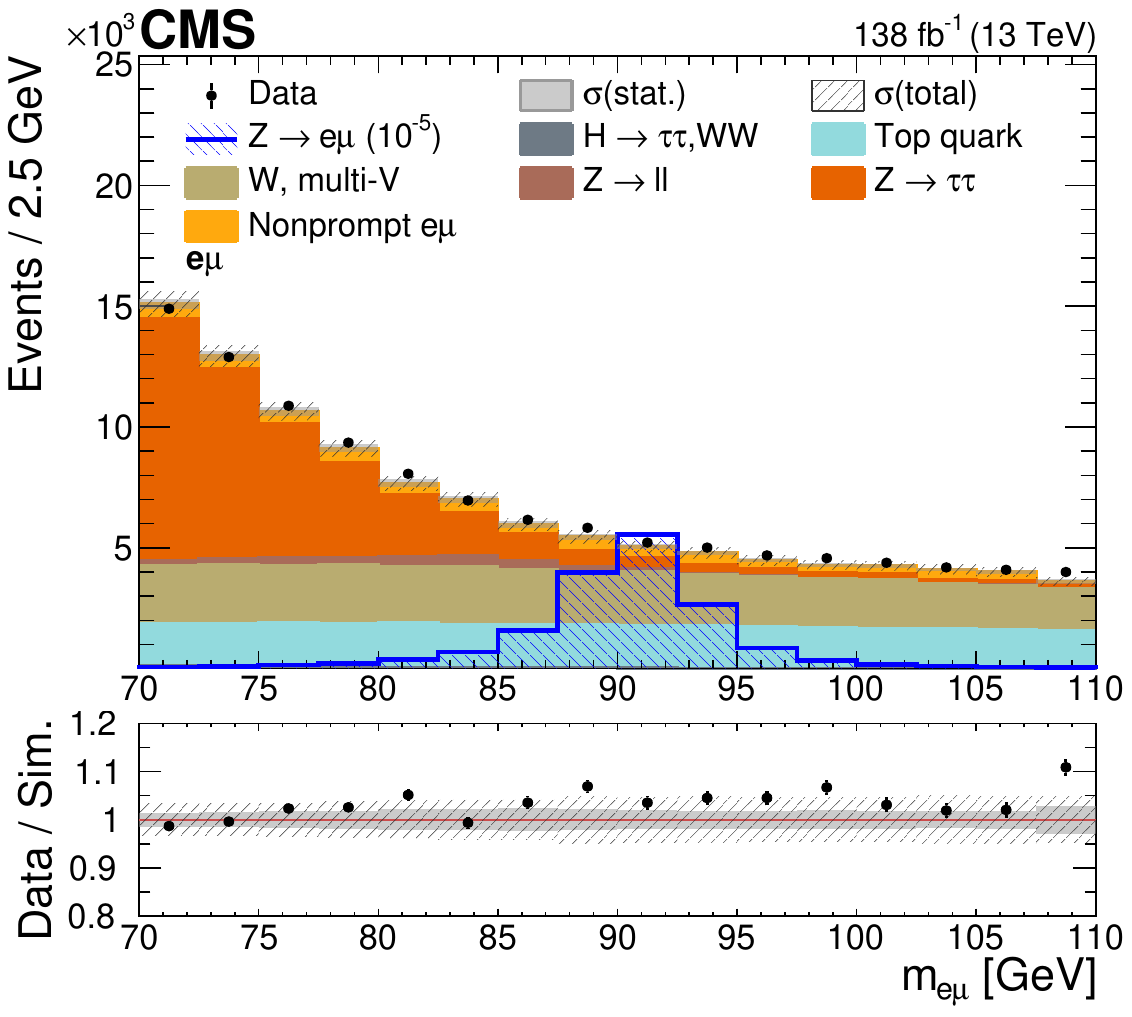}
  \caption{Invariant mass of the $\Pe\PGm$ system for data (points, with bars denoting statistical uncertainty) and simulated background (stacked filled histograms) for events passing the baseline selection.  In the legend, ``W, multi-V'' refers to \wjets events having a jet that is misidentified as a lepton, together with multiple vector boson production.  The hatched histogram shows a hypothetical \zemu signal normalized to a branching fraction of $10^{-5}$. The lower panel shows the ratio of the data to simulated background yields, with the statistical (combined systematic and statistical) uncertainty in the simulated yield indicated by the filled (hatched) gray band.}
  \label{fig:BEMu_Mll}
\end{figure}

To extract the signal yield we exploit the narrow resonance peak of the signal against a mainly smoothly falling background in the \memu spectrum.
The background shape arises from the sharply falling dominant contribution of $\PZ\to\taue\taumu$, combined with a nearly flat spectrum from $\PW\PW$, \ttbar, and nonprompt $\Pe\PGm$ events.
The distribution in \memu of the events surviving the baseline selection
described above is shown in Fig.~\ref{fig:BEMu_Mll}, for the data and for the simulated background components.  The shape of the simulated signal, normalized to a branching fraction of $10^{-5}$, is superimposed.

The final selection in the \zemu search makes use of a BDT trained to distinguish signal from background events.  The signal is then extracted from a fit of signal and background probability density functions (pdfs) to the \memu distribution, for three SRs defined as ranges of the BDT discriminant value.

\subsection{The BDT implementation for \texorpdfstring{$\PZ\to\Pe\mu$}{Z->e+mu}}
\label{sec:emuBDT}

While DY production of \PGt pairs is one of the main backgrounds, these events are concentrated in the region below the signal resonance.
In our SRs, both for the \PZ and \zprime cases, events from \ttbar and \ww production are dominant at the signal resonance mass.
Therefore, a single BDT is trained using simulated events
for the \zemu signal and for the background.  The background is represented by the \ww simulation, which is similar to \ttbar after our baseline selection.
We select disjoint training and testing samples, restricting \memu for the training sample to the \PZ-boson peak region, 86--96\GeV, to limit the influence of any correlations of the BDT features with \memu.
The BDT algorithm is implemented with the \textsc{XGBoost}~\cite{Chen:2016btl} package.
The hyperparameters of the BDT are optimized with a random grid search.
The input features aim to exploit the difference between signal and background in \ptmiss and other kinematic variables.  For the signal process, the \PZ boson is produced predominantly near threshold and decays to visible daughters, which are roughly collinear and have similar \pt, with little \ptmiss.  Backgrounds with leptonically decaying \PW bosons give rise to a characteristic distribution in the transverse mass \mT formed from the \PW daughter lepton and \ptmiss.
The transverse mass \mt(1, 2) is defined for a system of two particles (or of two systems of particles) with transverse momenta $\pt^{(1,2)}$ and azimuthal separation $\Delta\phi$ as
\begin{equation}
  \label{eq:mt}
  \mT(1, 2) = \sqrt{2 \pt^{(1)}\pt^{(2)} (1 - \cos{\Delta\phi})}.
\end{equation}

Since the signal extraction in the \zemu search treats the BDT and \memu as independent variables, we include only features having small correlations with \memu in the BDT.
From all the features tested, those that have the most discriminating power are selected:
\begin{enumerate}
  \item $\pt^{\text{trailing}}/\pt^{\text{leading}}$;
  \item $\pt^{\Pe\PGm}$;
  \item $\eta_{\Pe\PGm}$;
  \item $\mT(\met, \text{leading})$;
  \item $\mT(\met, \text{trailing})$;
  \item $\abs{\Delta\phi(\ptmiss,\Pe\PGm)}$; and
  \item \met.
\end{enumerate}
The terms ``leading'' and ``trailing'' denote the higher- and lower-\pt lepton, respectively.

\begin{figure}[!htbp]
        \centering
        \includegraphics[width=0.49\textwidth]{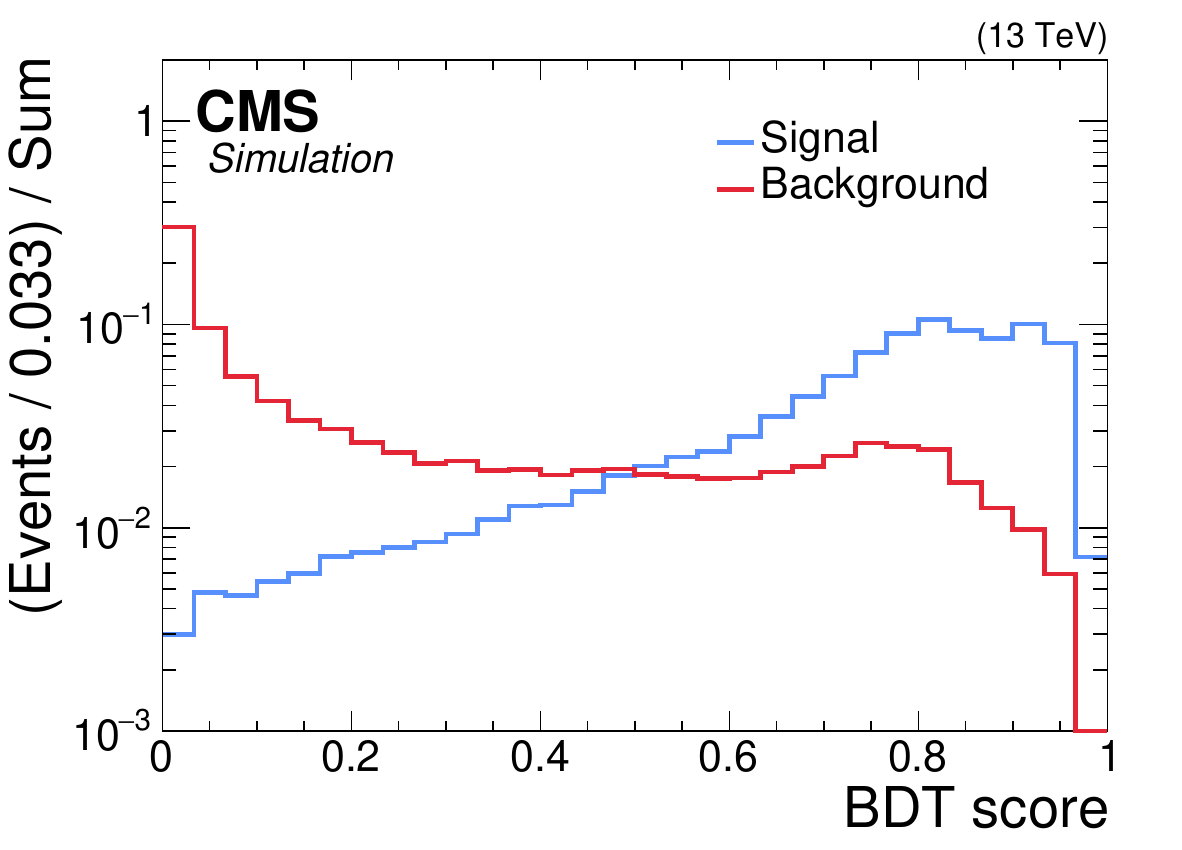}
        \includegraphics[width=0.49\textwidth]{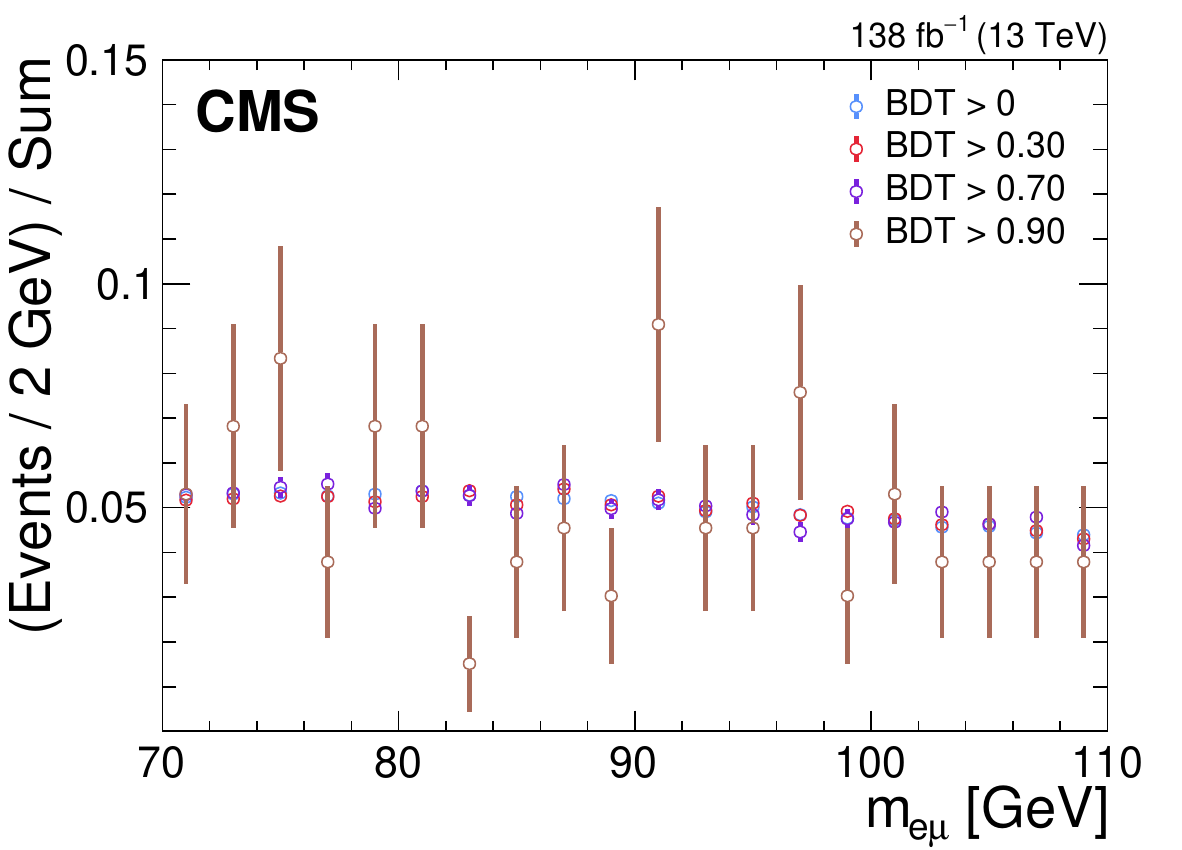}
        \caption{The \cmsLeft plot shows
        unity-normalized distributions of the \zemu BDT score for simulated events from the BDT test samples satisfying $70<\memu<110\GeV$.  The blue and red histograms represent the signal and \ww background, respectively.
        The \cmsRight plot displays the unity-normalized distribution of \memu for events in the \ttbar data CR used to check for BDT mass-spectrum bias, for several BDT thresholds; the vertical bars show the statistical uncertainties.
          }
        \label{fig:BEMu_MVA_Score}
\end{figure}

The resulting distributions in the discriminant value (BDT score) for the BDT
test events in the range 70--110\GeV in \memu are shown for signal and background in the \cmsLeft plot of Fig.~\ref{fig:BEMu_MVA_Score}.  We see from this plot that substantial discrimination between signal and background is achieved.

The \memu distribution for the \ww background test sample confirms that no structure is introduced by the training.  To check this conclusion in data,
we construct a CR enriched in \ttbar (and \ww) events by applying the SR selection but replacing the veto on any loose \PQb-tagged jet with a requirement of one tight \PQb-tagged jet.
The \cmsRight plot in Fig.~\ref{fig:BEMu_MVA_Score} shows the \memu shape for events in this CR for each of the BDT thresholds used to define the BDT binning.  The distribution is seen to be free of structure within uncertainties.
The BDT-score ranges chosen to define the SRs are 0.3--0.7, 0.7--0.9, and 0.9--1.0.  This binning optimized the sensitivity of the simultaneous fit to the \memu distributions in all of the SRs.

\subsection{Fit model for \texorpdfstring{$\PZ\to\Pe\mu$}{Z->e+mu}}
\label{sec:emuFit}

For the events within one of the BDT score categories, parametric functions of \memu are defined to describe the signal and backgrounds in that category.  The parameters of these functions are then determined from a fit to the binned data distribution, performed simultaneously in all categories.

The \zemu signal distribution is described by a double-sided Crystal Ball function~\cite{Oreglia:1980cs,Gaiser:1982yw}.
This function has a Gaussian core with independent power law tails defined such that the function is continuous and differentiable at all points.  Its parameters are determined from a fit to simulated signal events, and fixed in the fit to the data.

The nonresonant background spectrum is modeled by analytic functions obtained directly from fits to the data, which removes from the model any dependence
on simulation.  The forms of these functions are established with fits to the data in the \memu sidebands, defined as $70 < \memu < 86\GeV$ and $96 < \memu < 110\GeV$.
Rather than choosing a single function that best fits the data sidebands for each region, we employ the ``discrete profiling'' method~\cite{Dauncey:2014xga} in which an envelope of all well-fitting functions considered are included in the background model. The fit is then free to
choose the background function that minimizes the likelihood at each point of the fit, profiling over the discrete
nuisance parameter that represents the function choice.

Each of the individual functions included in the background model envelope is constructed from a broad, low-mass Gaussian component to represent the tail of the \ztautau spectrum,
and an additional function, chosen through statistical tests based on CRs in data, for the nonresonant processes.
This function has the form
\begin{equation}
  f(x) = f_g  g(x| \mu, \sigma) + (1 - f_g)  h_n(x | \vec{\theta}),
  \label{eq:emuBkg}
\end{equation}
where $g$ is the Gaussian pdf with mean $\mu$ and standard deviation $\sigma$,
$f_g$ its fraction in the model, and $h_n$ is the $n$th-order pdf of the function family $h$
with parameters $\vec{\theta}$.  The free parameters in the fit to the data are $f_g$, $\mu$, $\sigma$, and $\vec{\theta}$.
The $h_n$ function is chosen from among the Chebychev polynomial, exponential sum, or power law sum families.
The background functions that pass a $\chi^2$-probability test for fit consistency, $p(\chi^2) > 1\%$, are added to the model, and an F-test~\cite{Fisher:1922saa},
$p(\chi^2_{n} - \chi^2_{n+1}, \Delta n_\mathrm{d}) < 5\%$, is performed to determine the maximum function order for each family.  Here $\Delta n_\mathrm{d}$ is the number of degrees of freedom added by a step to the next higher-order function.  Tests performed with large numbers of simulated background events confirm that the (zero) signal strength extracted from the fit is free of bias.

The model in Eq.~(\ref{eq:emuBkg}) does not adequately describe the background from \zmumugamma events, which occur in cases where a muon loses a significant fraction of its energy to final-state radiation, thereby failing to reach the muon detectors, and
the resulting photon is misidentified as an electron.
This gives rise to a broad peak below the \PZ mass region that is not differentiated by the BDT since the background comes from true \PZ boson decays.
Therefore, this background needs to be accounted for separately in the final fit.
It is modeled with a double-sided Crystal Ball function, determined from fits to simulated \zmumu events that are reconstructed as \zemu.
We find satisfactory agreement between the simulation and data in a sample of pairs reconstructed as $\Pe\PGm$, but having the same sign (SS) of charge.  These events have a large fraction of electron candidates from misID background and exhibit a shape similar to that of the opposite-sign background.

\begin{figure*}[p!htb]
  \centering
  \includegraphics[width=0.49\textwidth]{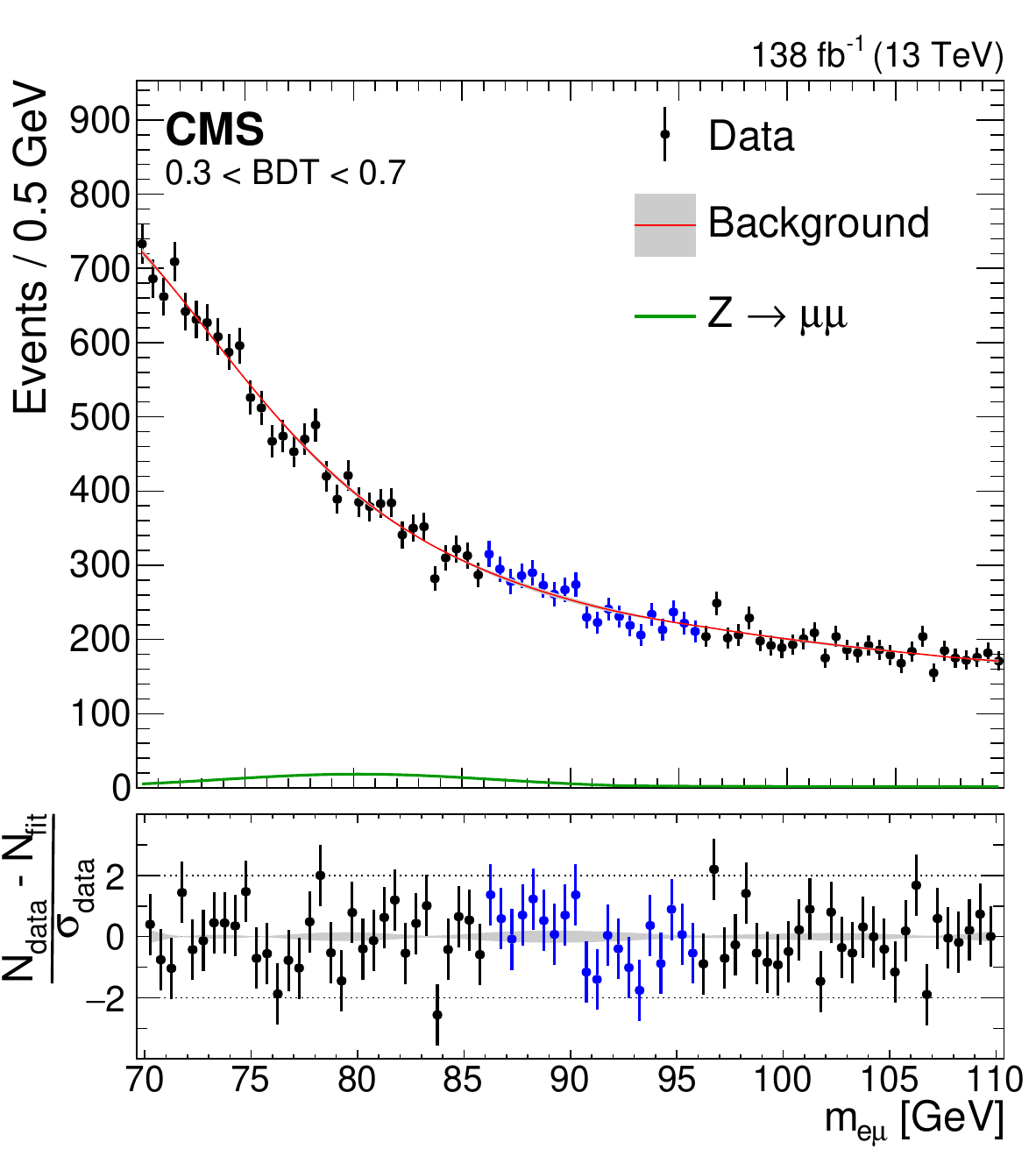}
  \includegraphics[width=0.49\textwidth]{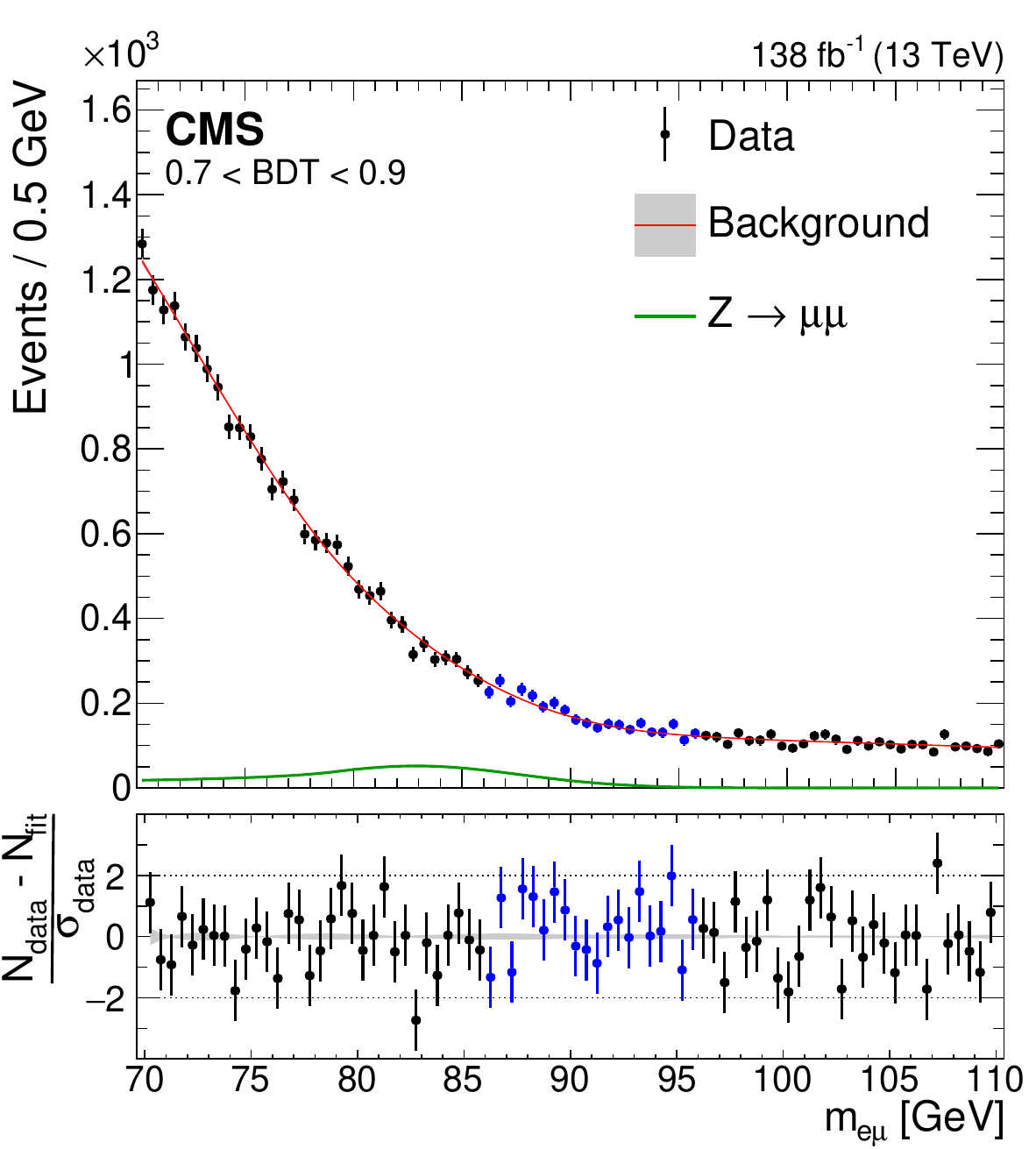}\\
  \includegraphics[width=0.49\textwidth]{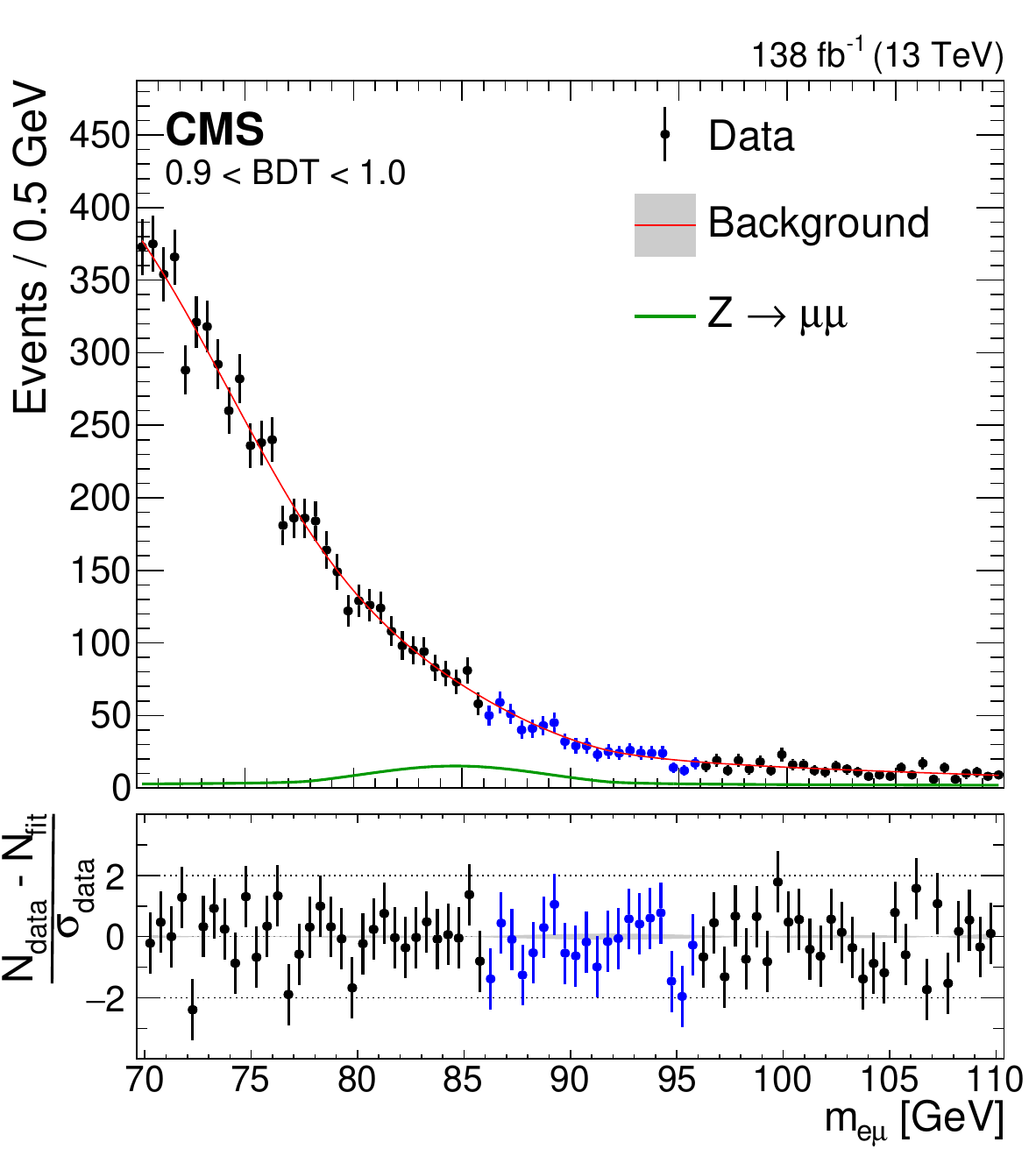}
  \caption{Fits of the data sidebands with the background functions for the BDT score ranges 0.3--0.7 (upper left), 0.7--0.9 (upper right), and 0.9--1.0 (lower).  In the upper panel of each plot, the black (blue) points with bars show the data with statistical uncertainties in the sideband (signal) region, the solid red line shows the average background prediction, and the green curve shows the \zmumugamma background component.  The lower panel in each plot shows the ratio of the difference between the data and the average background prediction to the uncertainty in the data, and the gray band shows the spread of the background estimates from the separate families of parametric functions.
  }
  \label{fig:zemu_env_fits}
\end{figure*}

The background model envelope from the fit to the \memu sidebands is shown in Fig.~\ref{fig:zemu_env_fits} for each BDT category.  The data points in blue in the figure are not revealed or used for the determination of the background shape.

\subsection{Application to the \texorpdfstring{\PZpr}{Z'} mass scan}

\begin{figure}[htbp]
  \centering
  \includegraphics[width=0.49\textwidth]{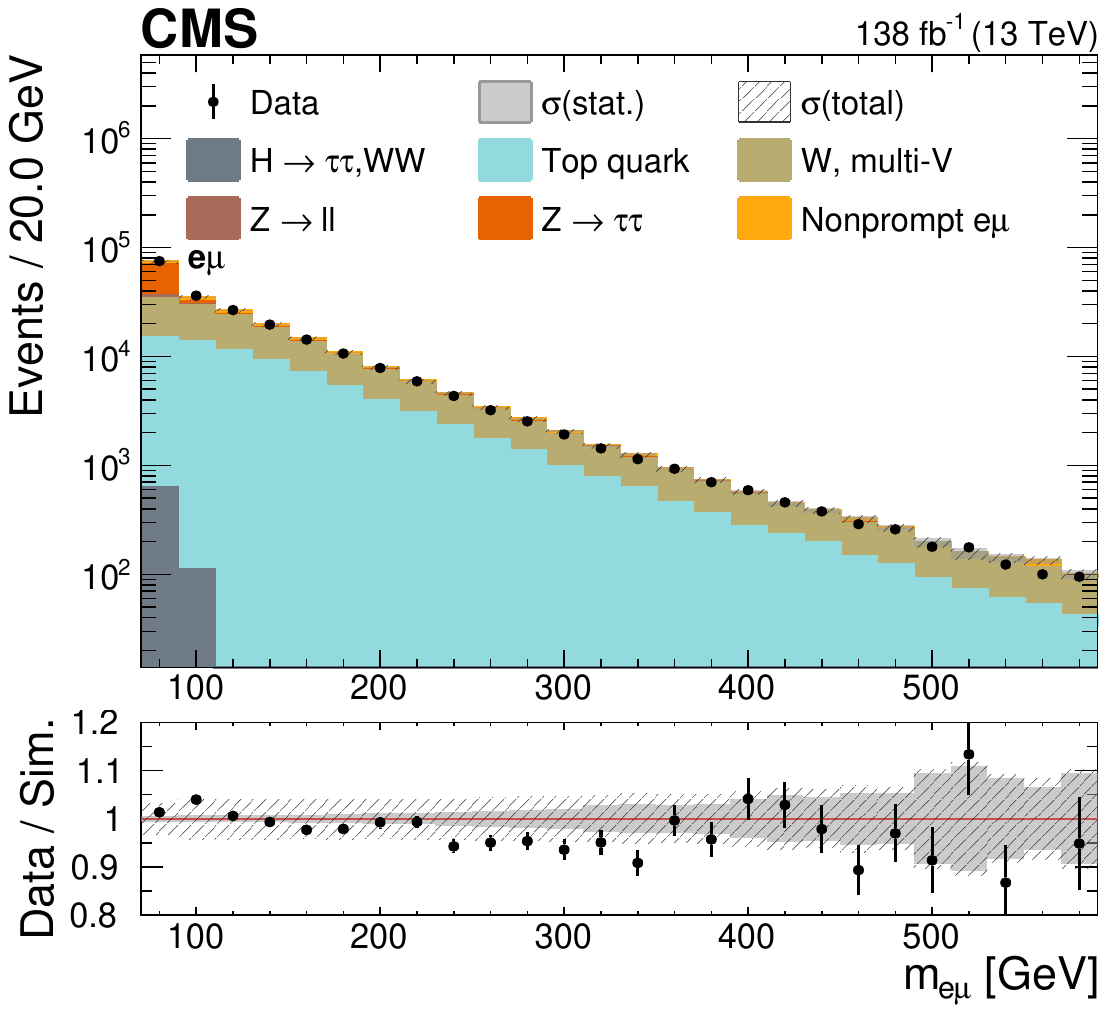}
  \caption{Invariant mass of the $\Pe\PGm$ system for data (points) and simulated background (stacked filled histograms), for events that pass the baseline selection except for that selection's upper limit on the invariant mass. The lower panel shows the ratio of the data to simulated yields, with the statistical (combined systematic and statistical) uncertainty in the simulated yield indicated by the filled (hatched) gray band.}
  \label{fig:zprime_mass_region}
\end{figure}

For the $\Pe\mu$ final state, a scan for new neutral resonances \zprime above the \PZ boson mass is also performed, using the same strategy and selection as in the \zemu search.
The extended-range \memu distribution in data and simulation is shown in Fig. \ref{fig:zprime_mass_region}.
The most important difference in procedure for this scan is that a sliding window in \memu is used for successive scan points.
With this strategy, potential masses of \zprime are explored from 110 to 500 \GeV, the upper limit being chosen because there are not enough events for a reliable fit above that value.  The size of each fit range is computed as twenty standard deviations of the mass resolution for the signal point considered.
The low end of the lowest fit range is 95\GeV, which evades the \zmumu and \ztautau background components, and thus allows the model to be simplified to just the function family $h$ of Eq.~(\ref{eq:emuBkg}).
For each mass point we define two BDT bins, 0.3--0.7 and 0.7--1.0, combining the upper two bins used for the \zemu search because of their smaller populations. 

\section{Search for \texorpdfstring{$\PZ\to\Pe\tau$}{Z->e+tau} and \texorpdfstring{$\PZ\to\mu\tau$}{Z->mu+tau}} {\label{sec:zxtau_analysis}

The \zetau and \zmutau searches each consider hadronic and leptonic \PGt-daughter categories.
In addition to the common selection described in Section~\ref{sec:selection},
in all four categories we require $40 < \mltau < 170 \GeV$, where \mltau is the invariant mass of the system comprising a prompt light-lepton candidate and the visible daughter(s) of a tau lepton.
To reduce backgrounds from \tautau, \wjets, \ww, and \ttbar production, we also require $\mT(\PGt,\met)/\mltau < 0.8$ and $\mT(\Pell\PGt,\met) < 70\, (90)\GeV$, in the hadronic (leptonic) final-state categories.

To further reduce the \tautau background, the prompt light lepton candidate must satisfy $\pt > 28\, (20)\GeV$ in the hadronic (leptonic) final-state categories.
Events are rejected if they contain \PQb-tagged jets, based on the tight (loose) working point of the \PQb jet ID for the hadronic (leptonic) \PGt-decay categories, to suppress backgrounds from \PQt quark processes.
The QCD multijet background becomes significant at low electron \pt in the \zetaumu category, and so we require $\pte > \ptmu$ in that category.
Additionally, the light leptons are required to have $\abs{\eta^{\Pell}} < 2.2\, (2.4)$ in the hadronic (leptonic) \PGt decay categories, for compatibility with the \tautau background estimation method described in Section~\ref{sec:embedding_method} below.

The search strategy for final states involving a tau lepton needs to take into account that the \PZ boson can be only partially reconstructed because of the neutrino(s) in the \PGt decay.  The signal and background spectra in \mltau are less distinct than those of \memu.  Here the background components are modeled separately, using data in CRs to the extent possible.  New variables derived from the visible daughters are developed to approximate the mass of the \PZ boson.  These and other discriminating variables are again combined into a BDT to optimize the signal extraction. 

\subsection{Background estimates for the \texorpdfstring{$\PZ\to\Pe\tau$}{Z->e+tau} and \texorpdfstring{$\PZ\to\mu\tau$}{Z->mu+tau} channels}
\label{sec:zxtau_backgrounds}

The \zetau and \zmutau searches rely on signal and background templates (binned pdfs). The dominant
background sources are \tautau production
and those arising from misidentified
leptons.  Other sources are leptonic \PW boson decays, such as those from \ww and \ttbar processes.

\subsubsection{Background from \texorpdfstring{$\tau\tau$}{Di-tau}}
\label{sec:embedding_method}

The production of tau lepton pairs, most importantly from DY \ztautau, is a background in all of the \zltau channels.
Neutrinos associated with the decays of both tau leptons contribute to \ptmiss, and the visible invariant
mass spectrum of the dilepton system peaks below the \PZ boson mass, with a high-side tail.
The \PGt candidate in a \zetau or \zmutau signal event would be kinematically identical to one from the \ztautau decay,
while the prompt lepton would have generally higher \pt than the daughter from a leptonically decaying \PGt lepton in the \ztautau event,
and there would be no contribution to \ptmiss from that leg of the \PZ boson decay.

The SM processes that produce tau lepton pairs are identical to those that produce muon pairs, and the latter are very cleanly reconstructed in the detector.  Therefore we use an ``embedding'' technique~\cite{CMS:2019pkt} to model \tautau backgrounds, except for \htautau production, which is treated separately.  In this method, we remove the recorded detector deposits associated with the muons from a data sample of muon pair production, and substitute those derived from simulated \tautau events, matching the kinematic properties of the simulated tau lepton with those of the muon it replaces.
The result is a hybrid data sample where the \PGt decay and reconstruction are simulated, but the \PGt production kinematics
and the rest of the event (pileup, \ptmiss, jets, etc.) are taken directly from data.
This approach significantly reduces simulation uncertainties, where the only remaining simulation-based uncertainties
are those related to the decay and reconstruction of the tau leptons.

\subsubsection{The \texorpdfstring{$\tau_{\mathrm{h}}$}{hadronic tau lepton} misID background}

The background from jets misidentified as \tauh (\jtotau) is estimated in the $\PGm\tauh$ and $\Pe\tauh$ channels using a
``misID-factor'' method, similar to those used in Refs.~\cite{CMS:2022kdi,CMS:2021rsq}.
The method uses an application region (AR) enriched in \jtotau events, which has the same kinematic selection as the SR but with an inverted jet-suppression ID (``antijet-ID'') requirement (very loose and not tight, using working points of the discriminator~\cite{CMS:2022prd}).
Shape-dependent transfer factors (TFs) are used to relate the rate of the \jtotau events in the AR to an estimate of the corresponding background rate in the SR.

Since the TFs depend on properties that might be experimentally inaccessible (\eg, the underlying parton flavor),
and depend on the background process (\eg, \wjets, QCD multijets), they are measured separately for each background source.
An overall TF is then evaluated by combining these factors weighted by the process composition in the CR:
\begin{equation*}
  f(\vec{\beta}) = \sum_{p\in {\text{processes}}}c_p(\vec{\beta}) f_p(\vec{\beta}).
\end{equation*}
The parameters $\vec{\beta}$ are derived from the kinematic properties of the \tauh and prompt \Pell candidates along with \ptmiss, and the functions $f$, $f_p$, and $c_p$ are, respectively, the overall TF to be evaluated for a given event in the AR, the TF for process $p$, and the expected composition fraction of process $p$ in the AR.

The process-dependent TFs $f_p$ are measured for \wjets, QCD multijet, and \ttbar background.  The \wjets TFs
are measured in a high-\mtlep
region, and the QCD multijet TFs are measured
in a region with SS $\Pell\tauh$ pairs and $0.05 < I^{\Pell}_{\text{rel}} < 0.15$.  The \ttbar TFs are determined using
simulation, as there is not a suitable measurement region in data. These measurement regions are
referred to as the determination regions.  The weights $c_p$ are estimated as functions of \mtlep, using simulated \jtotau events for \wjets and \ttbar; for QCD multijet background we use an SS data CR, correcting for non-QCD contributions in this CR by subtracting their simulated yields.

The $f_p$ TFs are measured as functions of the \tauh candidate \pt independently for each of four \tauh identified decay modes,
one- or three-prong with zero or one \PGpz.
There are also subleading kinematic dependencies in the TFs.  Corrections for these are derived by comparing the observation with the estimate in the tight antijet-ID subset of the determination region.
This is done for the light-lepton \pt, the \tauh $\abs{\eta}$, and $\abs{\Delta\phi(\Pell,\met)}$.
Additional corrections to account for the CR selection bias are measured in alternate CRs and applied as functions
of $I^{\Pell}_{\text{rel}}$, \mvis, and the BDT score defined in Section~\ref{ltauBDT}.

\subsubsection{Light-lepton misID background}

Electron and muon candidates in the $\Pe\PGt_\PGm$ and $\PGm\PGt_\Pe$ selections arising from misID of other particles are estimated using a method similar to the \jtotau estimation.
This background arises mainly from \wjets events, where one lepton comes from the \PW boson and one is a misidentified jet or a nonprompt
lepton originating from a jet, and from QCD multijet events, where each lepton is either a misidentified jet or a nonprompt
lepton originating from a jet.

The AR for the light-lepton misID background is the same as the SR, except with an SS selection.
The determination regions for the TFs are the same as for the SRs and ARs, except with a loose isolation requirement
($0.15 < I^\PGm_{\text{rel}} < 0.50$) on the muon candidate.
The TFs are measured as functions of $\Delta R(\Pe,\PGm)$ separately in categories of 0, 1, and ${>}1$ jets.
Subleading corrections to the $(\pte, \ptmu)$ distribution are measured in this determination region,
and CR bias corrections for the $(\pte, \ptmu)$ and BDT score (Section~\ref{ltauBDT}) distributions 
are measured in a loose electron isolation region.

\subsubsection{Backgrounds estimated from simulation}

The background from leptonic \PW decays in \ttbar and \ww production is relevant in all search channels,
though it is most impactful at high mass in the \emu data channel, and is estimated from simulation. The embedded \tautau samples already include the small
contribution from \ttbar and \ww decays to \tautau, and so these are omitted from the simulated background model.
The remaining \ttbar and \ww background processes are kept small by the event selection, which exploits several properties that distinguish them from signal.  Both \ttbar and \ww have large \ptmiss associated with each lepton leg, unlike the signal processes, and the \ttbar background is significantly suppressed by the veto of events with \PQb-tagged jets.

There are also small background contributions from other diboson and triboson processes, but these
have small cross sections and are suppressed by the rejection of events with additional light leptons.
The \htautau and \hww processes are also modeled with simulation, because these processes yield too few $\PGm\Pgm\ptmiss$ events in data for application of the \PGt embedding method.

Photons misidentified as electrons and light leptons misidentified as \tauh candidates are also modeled using simulated samples.
The background from DY \zee and \zmumu events where one light lepton is misidentified as a \tauh candidate
must be carefully separated from the signal as the prompt-lepton candidate leg is identical to that of the CLFV \PZ signal.

\subsection{Discriminating variables for \texorpdfstring{$\PZ\to\Pe\tau$ and $\PZ\to\mu\tau$}{Z->e+tau and Z->mu+tau}}

Because of the neutrinos in the \PGt decay, the invariant mass of the visible daughters in the \zltau channels is a less effective discriminating
variable than in the construction of \zemu against the backgrounds, of which \ztautau is the largest.
However, we can exploit the facts that since the decaying boson is much more massive than the \PGt, the \PGt is typically produced with a significant Lorentz boost, leading to a highly collimated decay, and that one of the \PZ boson daughters for the signal decay is a light lepton unaccompanied by neutrinos.
We utilize two different approaches to take advantage of these features.

\subsubsection{Collinear mass estimate}

The first approach is the collinear mass \mcoll, derived with the collimated \PGt decay approximation, in which {\metvec}
projected onto the visible tau lepton \ptvec equals the \ptvec of the neutrino(s). With the further assumption that the neutrino three-momentum lies along the three-momentum of the visible \PGt daughter(s), we then
obtain the neutrino four-vector
\begin{equation}
  \label{eq:neutrino_collinear}
  p^{\PGn} = \frac{\metvec\cdot\hatpt^{\text{vis}}}{\pt^{\text{vis}}} p^{\text{vis}},
\end{equation}
where $p^{\text{vis}}$ is the four-momentum of the visible \PGt daughter(s), and $\pt^{\text{vis}}$ and $\hatpt^{\text{vis}}$ are the magnitude and direction unit vector of its transverse momentum components, respectively.
The sign of $\metvec\cdot\hatpt^{\text{vis}}$ indicates whether the neutrino vector
is parallel or antiparallel to the visible \PGt daughter's momentum.
The collinear
mass is then the mass of the $p^{\Pell} + p^{\text{vis}} + p^\nu = p^{\Pell} + p^\PGt$ system. The \mcoll distribution
is shown in Fig.~\ref{fig:collinear_mass} for the \zmutau search. The true $\Delta\phi$ between the visible \PGt decay products and the associated neutrinos is typically small,
but because of uncertainties in the \met reconstruction this angle can be larger than $\pi/2$,
leading to antiparallel neutrino configurations.

\begin{figure}[!htbp]
  \centering
  \includegraphics[width=0.49\textwidth]{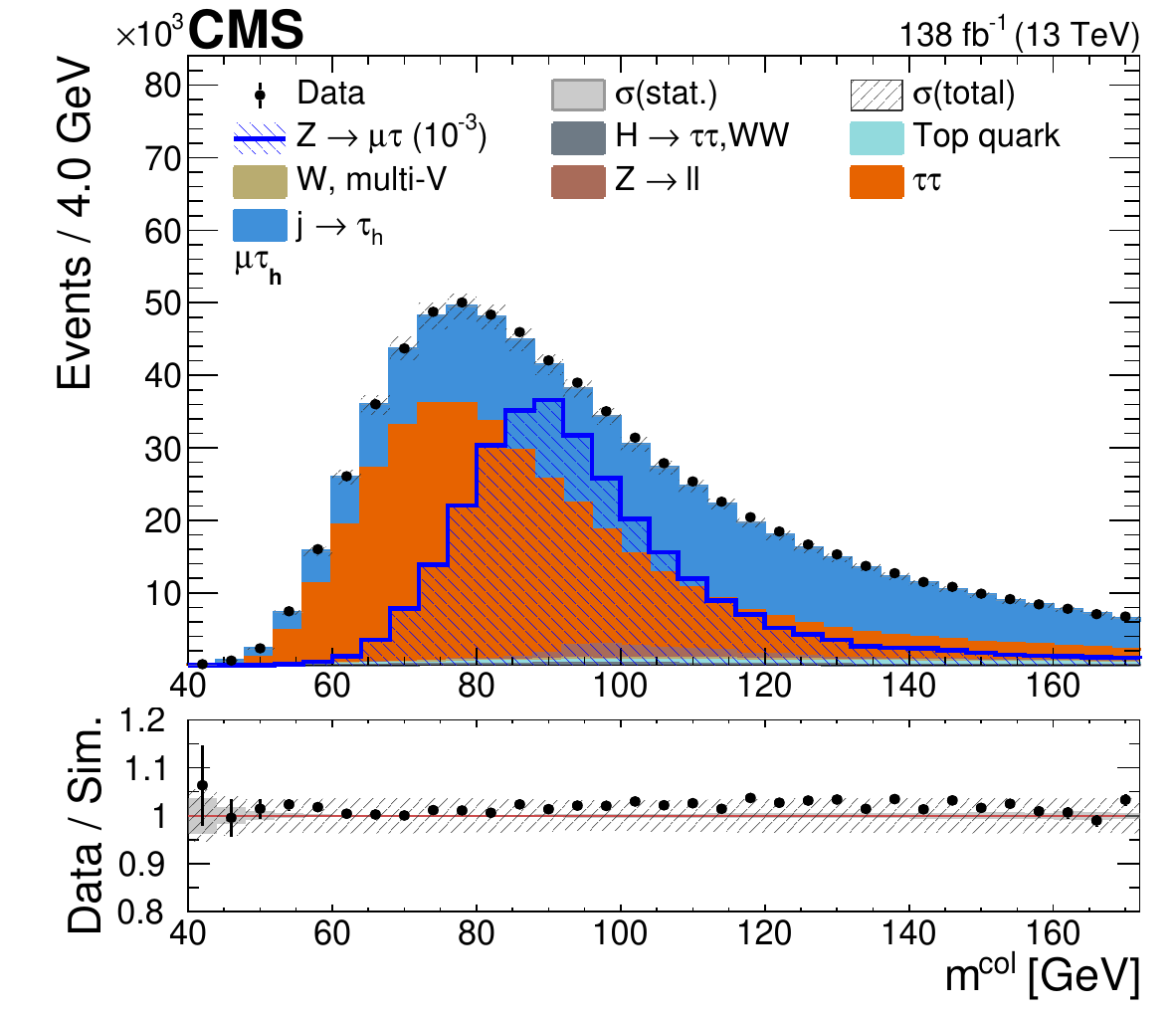}
  \includegraphics[width=0.49\textwidth]{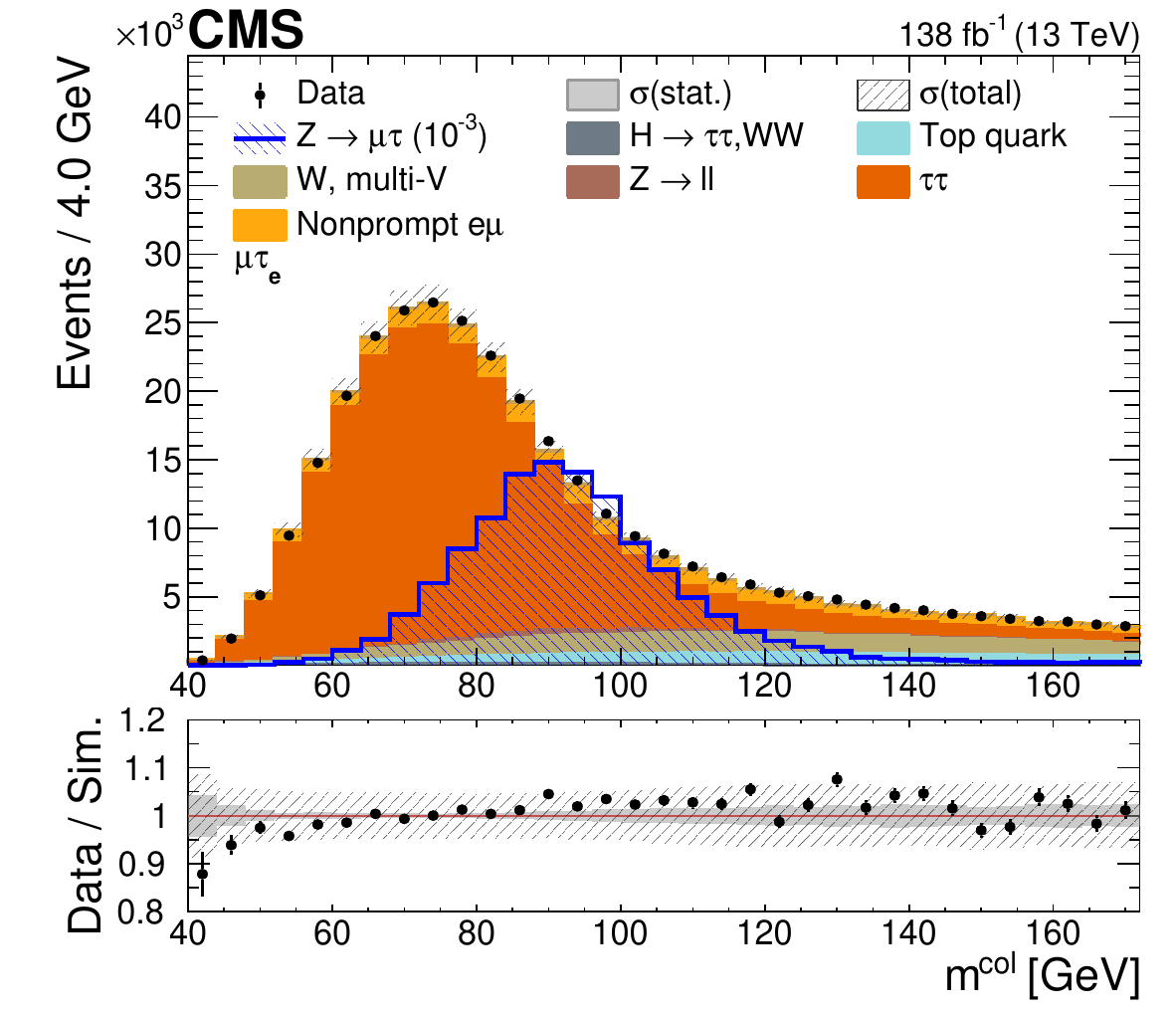}
  \caption{The $\PGm\tauh$ (\cmsLeft) and $\PGm\taue$ (\cmsRight) \mcoll distributions, for the data (black markers with bars showing the statistical uncertainties) and the simulated backgrounds (filled stacked histograms).  The hatched blue histogram shows the shape of the signal, normalized to a branching fraction of $10^{-3}$, for comparison.
  The lower panel shows the ratio of the data to simulated yields, with the statistical (combined systematic and statistical) uncertainty in the simulated yield indicated by the filled (hatched) gray band.}
  \label{fig:collinear_mass}
\end{figure}

\subsubsection{The tau lepton momentum scale variables \texorpdfstring{$\alpha_{1,2}$}{alpha(1,2)}}

Another approach is to use collinear approximations for the \PGt decays along with assumptions~\cite{Davidson:2012wn} that the \PZ boson is on shell and produced at threshold, \ie, that the \PZ boson daughters' \pt vanishes.  The latter constraint implies equality of the \PZ daughters' \pt, $\pt^{(1)} = \pt^{(2)}$.  This relation can be restated in terms of the four-momentum $p^{(i)}$ of the tau lepton and that of its daughters, defined as $p^{\text{vis},\,i} = p^{(i)}/\alpha_i$:    
\begin{equation}
  \label{eq:alpha_pt}
  \alpha_1\pt^{\text{vis},\,1} = \alpha_2 \pt^{\text{vis},\,2}.
\end{equation}

The on-shell \PZ boson assumption combined with four-momentum conservation in the decay provides a second relation between the $\alpha_i$:
\begin{equation}
  \label{eq:alpha_mass}
  m_\PZ^2 =  m^2_1 + m^2_2 + 2p^{(1)} \cdot p^{(2)} \approx 2p^{(1)} \cdot p^{(2)} = 2\alpha_1\alpha_2 p^{\text{vis},1}\cdot  p^{\text{vis},2}.
\end{equation}

From these equations we find
\begin{equation}
  \label{eq:alpha_equation}
  \alpha_{1,\,2} =  \sqrt{\left(\frac{\pt^{\text{vis},\,(2,\,1)}}{\pt^{\text{vis},\,(1,\,2)}}\right)\frac{m_\PZ^2}{2p^{\text{vis},\,1} \cdot p^{\text{vis},\,2}}}.
\end{equation}

For the signal decay \zltau we have $\alpha_1 = \alpha_{\Pell}=1$, whereas for the \ztautau background we have in general $\alpha_i>1$ for both \PZ boson daughters.
Figure~\ref{fig:alpha_variables} shows these estimates for \zmutauh, where for the signal events $\alpha_\PGm$ is centered at one, as expected, and provides considerable discrimination against both \ztautau and \tauh misID backgrounds.

\begin{figure}[!htbp]
  \centering
  \includegraphics[width=0.49\textwidth]{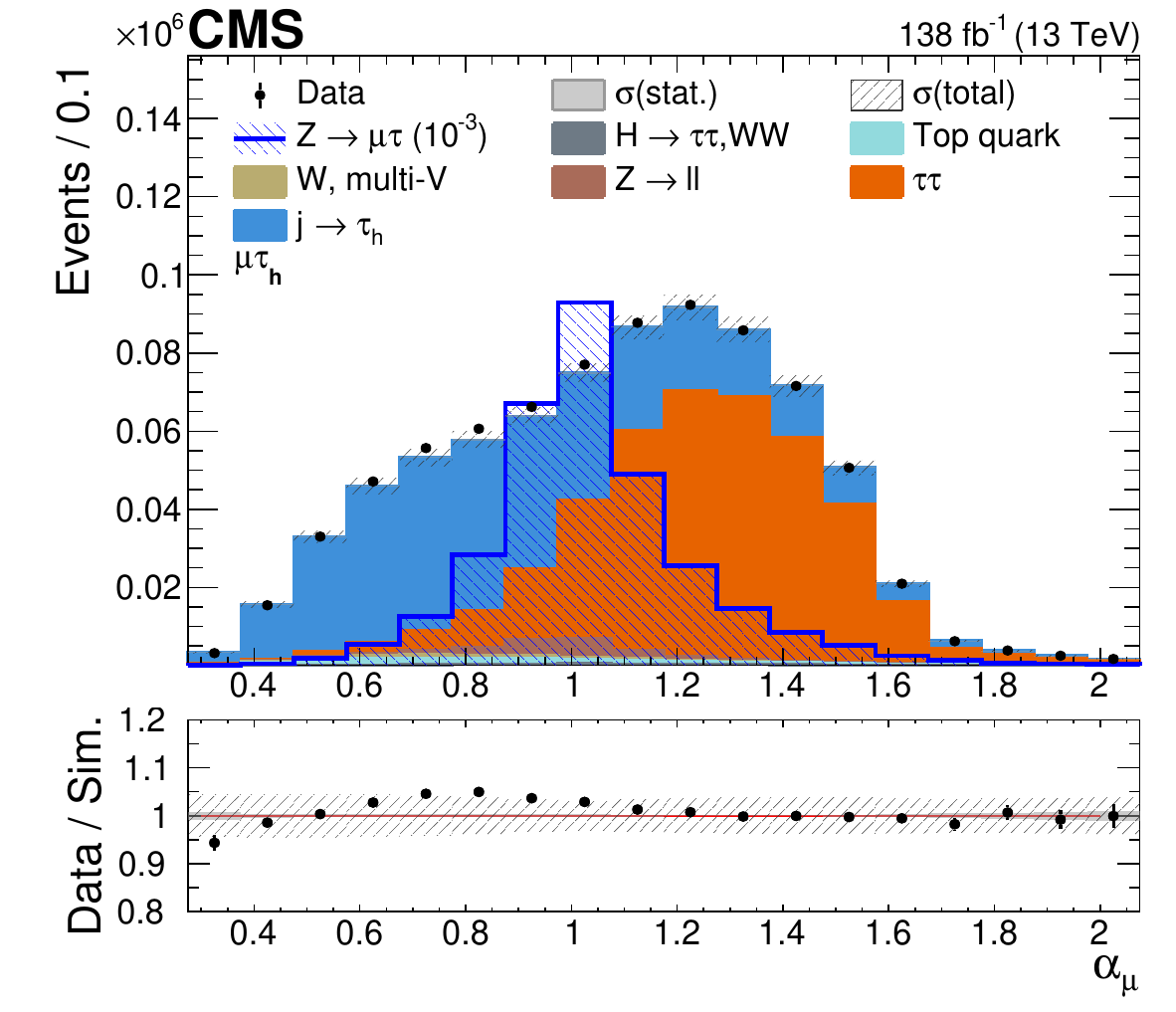}
  \includegraphics[width=0.49\textwidth]{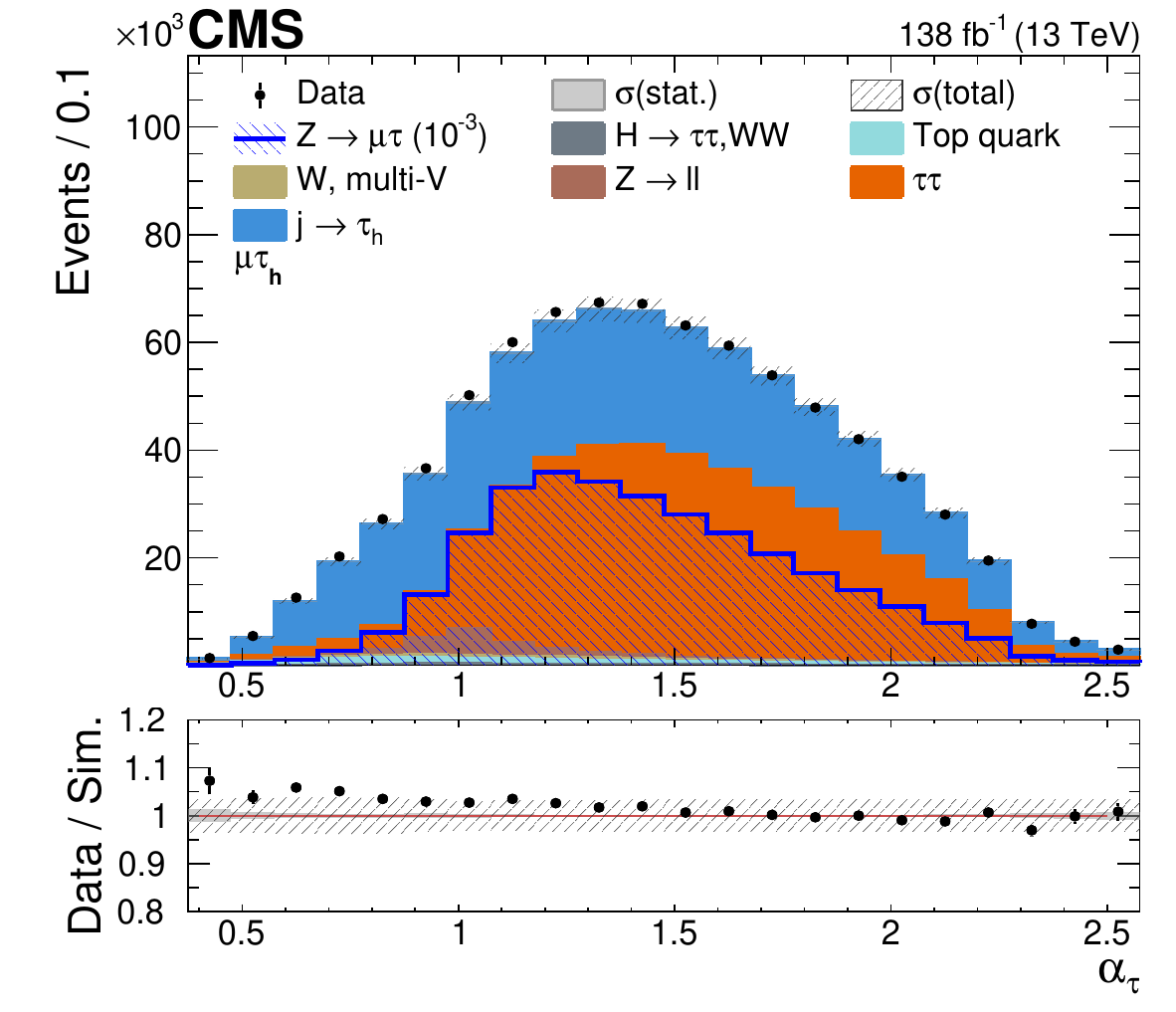}
  \caption{Distributions of $\PGm\tauh$ signal and estimated backgrounds in $\alpha_{\PGm}$ (\cmsLeft), $\alpha_{\PGt}$ (\cmsRight), for the data (black markers with bars showing the statistical uncertainties) and the simulated backgrounds (filled stacked histograms).  The hatched blue histogram shows the shape of the signal, normalized to a branching fraction of $10^{-3}$, for comparison.  The lower panel shows the ratio of the data to simulated yields, with the statistical (combined systematic and statistical) uncertainty in the simulated yield indicated by the filled (hatched) gray band.}
  \label{fig:alpha_variables}
\end{figure}

\subsection{The BDTs for the \texorpdfstring{$\PZ\to\Pe\tau$ and $\PZ\to\mu\tau$}{Z->e+tau and Z->mu+tau} decay channels}
\label{ltauBDT}

A BDT is trained for each final state of each signal, for a total of four BDTs (\zetauh, \zetaumu, \zmutauh, and \zmutaue).
The ROOT TMVA toolkit~\cite{TMVA:2007ngy} is used for this training.
The BDTs are constructed with input features that describe the kinematic properties of the \PZ boson candidate and of \ptmiss:
\begin{enumerate}
  \item \mltau;
  \item \mcoll;
  \item $\alpha_{\Pell}$ and $\alpha_{\PGt}$;
  \item $\pt^{\Pell} / \pt^{\PGt}$;
  \item $\pt^{\Pell\PGt}$;
  \item $\Delta\phi_{\Pell\PGt}$;
  \item $\mt(\Pell, \met)$;
  \item $\Delta\phi(\PGt, \met)$; and
  \item $\pt^{\text{jet}}$.
\end{enumerate}
The leading jet \pt is included here, as \PQb quark initiated jets in \ttbar decays that
either fail the \PQb tagging or are beyond the \PQb tagging region of $\abs{\eta} < 2.4$ typically
have higher \pt than jets associated with the signal process.

The BDT training uses \tautau simulations rather than the embedding samples.
Data events from the loose antijet ID (SS) region are used in the training
to model the \jtotau (nonprompt $\Pe\PGm$) background in the hadronic (leptonic) \PGt categories.
Events used for training, whether simulated events or CR data, are excluded from use in the background template estimates,
where the remaining events or data are scaled to compensate for the removed events.
To improve the discrimination against rare \zll background events in which one of the
leptons is misidentified as one of a different flavor, these events are given a larger weight in the training.
The BDT score distribution $O(x)$ is transformed using the signal BDT score cumulative distribution function $p(x) = \int_{-\infty}^{x} O(x^{\prime})\rd{x^{\prime}}/\int_{-\infty}^{+\infty} O(x^{\prime})\rd{x^{\prime}}$
to produce a useful shape that is less impacted by changes in the underlying machine-learning technique or hyperparameters.

\subsection{The \texorpdfstring{$\PZ\to\Pe\tau$ and $\PZ\to\mu\tau$}{Z->e+tau and Z->mu+tau} mass categories}
\label{sec:zxtau_mass_regions}

Several of the backgrounds in the \zetau and \zmutau search have large systematic uncertainties that impact
the search sensitivity. These are:
\begin{enumerate}
\item \tautau embedding uncertainties;
\item \jtotau and nonprompt $\Pe\PGm$ estimates from data; and
\item $\zll\to\Pell\tauh$ misID rate uncertainties.
\end{enumerate}

To improve background uncertainty constraints and isolate background- or signal-enriched regions, the \zetau and \zmutau
fits are split into the \mltau mass regions described in Table~\ref{tab:Tau_Mass_Regions}. The leptonic \PGt signal decays typically have lower mass
than the \tauh decays because of the extra neutrino produced, which leads to differences in the \tautau region selection.
Additionally, \zll is a significant background only in the hadronic channels, so there is no CR for these processes in the leptonic channels.
Events with low transformed BDT score are removed from the final analysis, and the transformed BDT binning is chosen to ensure that each bin contains an adequate number of events for the final fit.

\begin{table}[!htbp]
  \centering
  \topcaption{Regions in \mltau for the \zmutau and \zetau fits.}
  \begin{scotch}{ c  c  c }
    Region      & \multicolumn{2}{c}{\mltau bounds [\GeVns{}]}\\
                & \zltauh & \zltaul \\
    
    \hline
    \tautau     & [40,\,60] & [40,\,50] \\
    Signal-like & [60,\,85] & [50,\,100] \\
    \zll        & [85,\,100] & \NA \\
    misID (\jtotau and nonprompt $\Pe\PGm$) & [100,\,170] & [100,\,170] \\
  \end{scotch}
  \label{tab:Tau_Mass_Regions}
\end{table}

\section{Systematic uncertainties}
\label{sec:systematics}

The main sources of systematic uncertainty arise from the simulation (\eg, simulated object reconstruction efficiency or energy scale), from the 
background model uncertainties (\eg, validity of the model in the SR), and statistical uncertainties because of limited event counts in the simulated samples and CRs.  These uncertainties are discussed in the following paragraphs and summarized in Table~\ref{tab:systematics}.

In the \zemu search, the most significant sources are the purely statistical uncertainties from the parameters of the background functions in the fit.  These are followed by uncertainties arising from the discrete collection of background functions included in the envelope. The signal model uncertainties are negligible compared with those of the background model.

The largest sources of systematic uncertainty in the \zetau and \zmutau searches are the statistical uncertainties in the
background model templates and the light-lepton energy scale uncertainties.
The model-template statistical uncertainties arise largely from the \zee and \zmumu simulation samples used to estimate the $\Pell \to \tauh$ backgrounds,
along with those from the nonprompt $\Pe\PGm$ background estimate based on CRs in the data, where the subtraction of prompt leptons
from the data has a large statistical uncertainty. The statistical uncertainties in the \jtotau and embedded \tautau estimates
are much smaller because of their larger CR populations.
The impact of the electron and muon energy scale uncertainties is large because the \pt of the prompt \Pell is highly discriminating against leptonic \ztautau decays, as
well as being incorporated into several of the variables in the BDTs (dilepton mass and \pt, collinear mass, $\alpha_{i}$, $\mt(\Pell, \met)$, and the \pt ratio).
The uncertainties in the background estimate from CRs in data are also significant, mainly because of the corrections for the kinematic variation in the TF measurement regions and the
statistical uncertainties in these measurement regions.

The normalization of the simulated signal and minor background yields depends on the integrated luminosity, pileup corrections, and the theoretical modeling.  The integrated luminosities for the 2016, 2017, and 2018 data-taking years have individual uncertainties of 1.2--2.5\%~\cite{CMS-LUM-17-003,CMS-PAS-LUM-17-004,CMS-PAS-LUM-18-002}, while the overall uncertainty for the 2016--2018 period is 1.6\%.  To evaluate the uncertainty associated with the pileup reweighting, we vary the value of the total inelastic cross section by 5\%~\cite{CMS:2018mlc}.  Uncertainties arising from the theoretical model include the cross sections, the parton distribution functions, and the renormalization and factorization scales.  These uncertainties are propagated to the observed signal yields and are summarized in the row labeled ``Theory'' in Table~\ref{tab:systematics}.

\begin{table*}[!hbpt]
  \centering
  \topcaption{Sources of uncertainty and their impacts on the measured branching fraction, for each of the \PZ decay channels, and on the product of the production cross section and branching fraction for the \zprime resonance scan.  The uncertainty ranges for the \zprime resonance scan are ordered from the lowest to the highest \zprime mass point.  Entries to which the specified uncertainty does not apply are denoted with ``\NA''.}
  \cmsTable{
  \begin{scotch}{ p{0.28\linewidth} p{0.13\linewidth} p{0.13\linewidth} p{0.13\linewidth} p{0.205\linewidth} }
    Uncertainty source & $\mathcal{B}(\PZ\to\Pe\PGm)$\newline [$10^{-8}$] & $\mathcal{B}(\PZ\to\Pe\PGt)$\newline [$10^{-6}$] & $\mathcal{B}(\PZ\to\PGm\PGt)$\newline [$10^{-6}$] & $\sigma(\pp\to\PZpr + X)\times\newline\mathcal{B}(\PZpr\to\Pe\PGm)$ [fb]\\
    \hline
    Electron \& muon ID \& trigger & 0.1      &  0.3  & 0.1  & 0.1--0.01            \\
    Electron energy scale          & 0.2      &  3.5  & 0.2  & 0.6--0.01            \\
    Muon energy scale              & 0.1      &  0.1  & 0.9  & 0.3--0.01            \\
    Tau  ID                        & \NA      &  0.3  & 0.2  & \NA                   \\
    Tau energy scale               & \NA      &  0.4  & 0.4  & \NA                   \\
    $(\Pe,\,\PGm)\to \tauh$        & \NA      &  0.4  & 0.1  & \NA                   \\
    [\cmsTabSkip]                                                       
    Jet energy, \met               & 0.3      &  0.8  & 0.2  & 0.1--0.01            \\
    \PQb tagging                   & ${<}0.1$ &  0.2  & 0.2  & ${<}(0.1\text{--}0.01)$ \\
    Pileup                         & 0.1      &  0.2  & 0.1  & ${<}(0.1\text{--}0.01)$  \\
    Integrated luminosity          & 0.3      &  0.1  & 0.1  & 0.1--0.01            \\
    Theory                         & 0.1      &  0.5  & 0.2  & 0.1--$({<}0.01)$       \\
    [\cmsTabSkip]                                                       
    Parametric background          & 7.9      &  \NA  &  \NA & 0.3--0.06            \\
    Envelope                       & 2.5      &  \NA  &  \NA & 0.1--0.04            \\
    \zmumu yield                   & 0.5      &  \NA  &  \NA & \NA                   \\
    [\cmsTabSkip]                                                       
    Embedding energy resolution    & \NA      &  1.1  & 0.2  & \NA                   \\
    Embedding normalization        & \NA      &  1.5  & 0.2  & \NA                   \\
    \jtotau                        & \NA      &  2.8  & 0.9  & \NA                   \\
    Nonprompt $\Pe\PGm$            & \NA      &  2.7  & 0.4  & \NA                   \\
    [\cmsTabSkip]                                                       
    Template event counts          & ${<}0.1$ &  3.1  & 1.5  & 0.1--$({<}0.01)$       \\
    [\cmsTabSkip]                                                       
    Total systematic               & 8.3      &  5.7  & 2.2  & 0.8--0.08            \\
    Statistical                    & 5.6      &  1.9  & 1.6  & 1.9--0.18            \\
  \end{scotch}
  }
\label{tab:systematics}
\end{table*}

All uncertainties are included in the maximum likelihood fit to extract the branching fractions, which is implemented with the \textsc{Combine} statistical analysis package~\cite{CMS:2024onh}.
For an uncertainty that does not contain shape information, the uncertainty is implemented as a rate uncertainty with a log-normal
pdf. Uncertainties that impact the shape of the signal or background model are handled differently in the \zemu and \zltau searches.
In the \zemu search, the effect is split into a rate uncertainty and an impact on the pure shape effect.
A nuisance parameter is added as a shift in the corresponding model parameter drawn from a Gaussian pdf
centered on zero and with a width defined by the systematic source's effect on that model parameter.
In the \zetau and \zmutau searches, shape uncertainties are introduced via ``continuous morphing''~\cite{Conway:2011in} parametrized by upward and downward uncertainty templates. 
In the case of the \zemu background envelope, the discrete background function index parameter
is treated as a free parameter.
The fit selects the background function that maximizes the likelihood at any given point, profiling this discrete nuisance parameter~\cite{Dauncey:2014xga}.

\ifthenelse{\boolean{cms@external}}{\clearpage}{}
\section{Results}
\label{sec:results}

The branching fraction is extracted from a maximum likelihood fit of the background plus signal model to the data distribution in \memu, for the \zemu search, or in the BDT discriminant, for the \zltau searches.

For the \zemu search, the parameters of the background model developed from the \memu sidebands (Section~\ref{sec:emuFit}) are free in the fit to the full distributions.  The distributions for separate fits in each BDT region are shown along with the data in Fig.~\ref{fig:zemu_unblinded_fits}.
No excess of events beyond the SM expectation is observed.
We set upper limits on the branching fraction following the modified frequentist construction \CLs~\cite{Cowan:2010js,Junk:1999kv,CLs} implemented with the CMS statistical analysis tool \textsc{Combine}~\cite{CMS:2024onh}.
The observed and expected upper limits are shown in Fig.~\ref{fig:zemu_limits} and Table~\ref{tab:lfvz_limits} 
from the fits in each BDT score category and for the joint fit combining all categories,
along with the 68 and 95\% uncertainties.
The observed limit of $\mathcal{B}(\zemu) < 1.9 \times 10^{-7}$ represents the most stringent direct limit to date on this process.

The transformed BDT score distributions in the invariant mass and tau lepton final-state categories are shown
in Figs.~\ref{fig:zetauh_unblinded_fits} and~\ref{fig:zetaumu_unblinded_fits}
for the \zetau search and in Figs.~\ref{fig:zmutauh_unblinded_fits} and~\ref{fig:zmutaue_unblinded_fits} for the \zmutau search.
The pull appearing in the distributions in the lower panels for each BDT bin is the difference between the observed and fit yields divided by the uncertainty $\sqrt{\smash[b]{\sigma_{\text{obs}}^2 - \sigma_{\text{fit}}^2}}$.  The minus sign in the definition of this uncertainty accounts for the correlation between the observed and fit yields, which arises because the observed yield is included in the fit~\cite{James:2006edtwo}.  The large pulls that appear in some bins in the lower row of Fig.~\ref{fig:zetauh_unblinded_fits} reflect non-Gaussian uncertainties in one or more of the nuisance parameters.
The observed yield in the \etau (\mutau) channel is consistent with the SM background prediction at the $0.5\,(2.7)\,\sigma$ level.
The observed and expected 95\% \CL upper limits on the branching fractions for \zetau and \zmutau are shown in Fig.~\ref{fig:zxtau_limits} and Table~\ref{tab:lfvz_limits},
along with the 68 and 95\% uncertainties in the expected limit.

\begin{figure}[!htbp]
  \centering
  \includegraphics[width=0.49\textwidth]{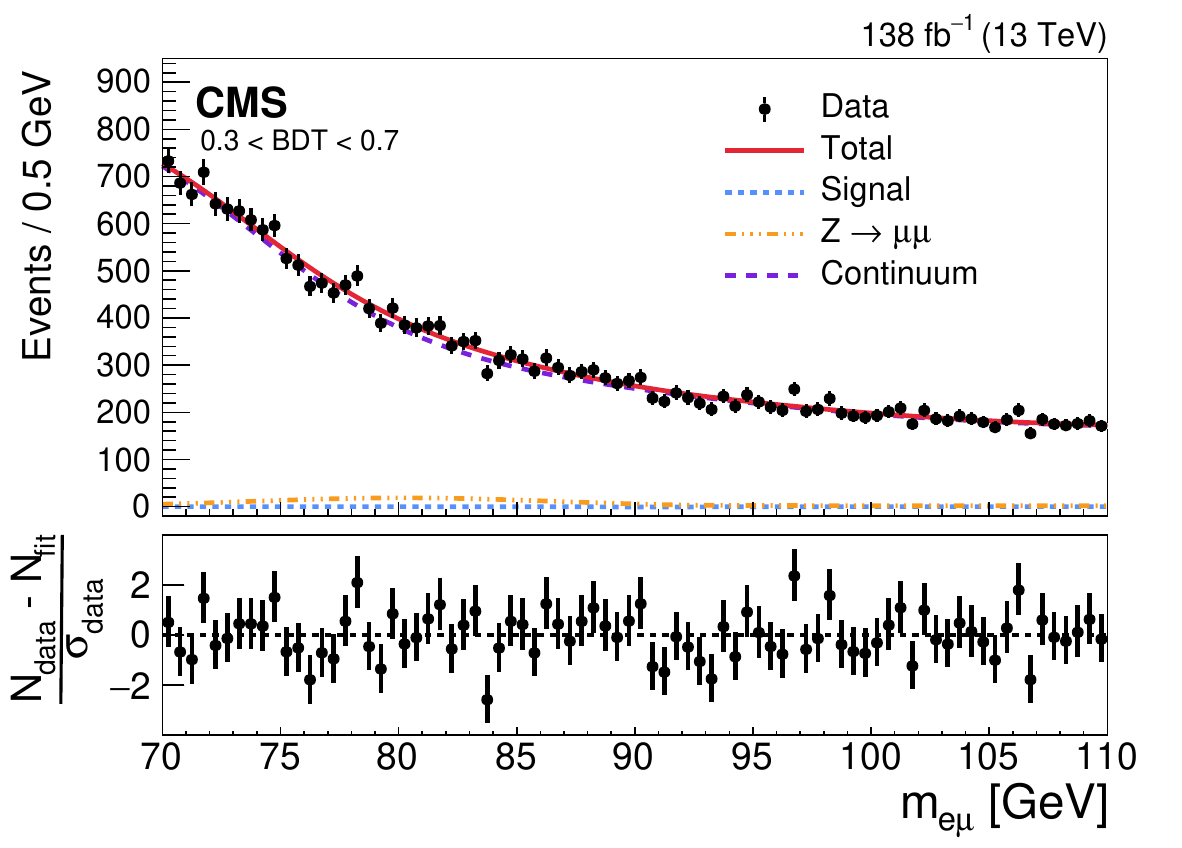}\\
  \includegraphics[width=0.49\textwidth]{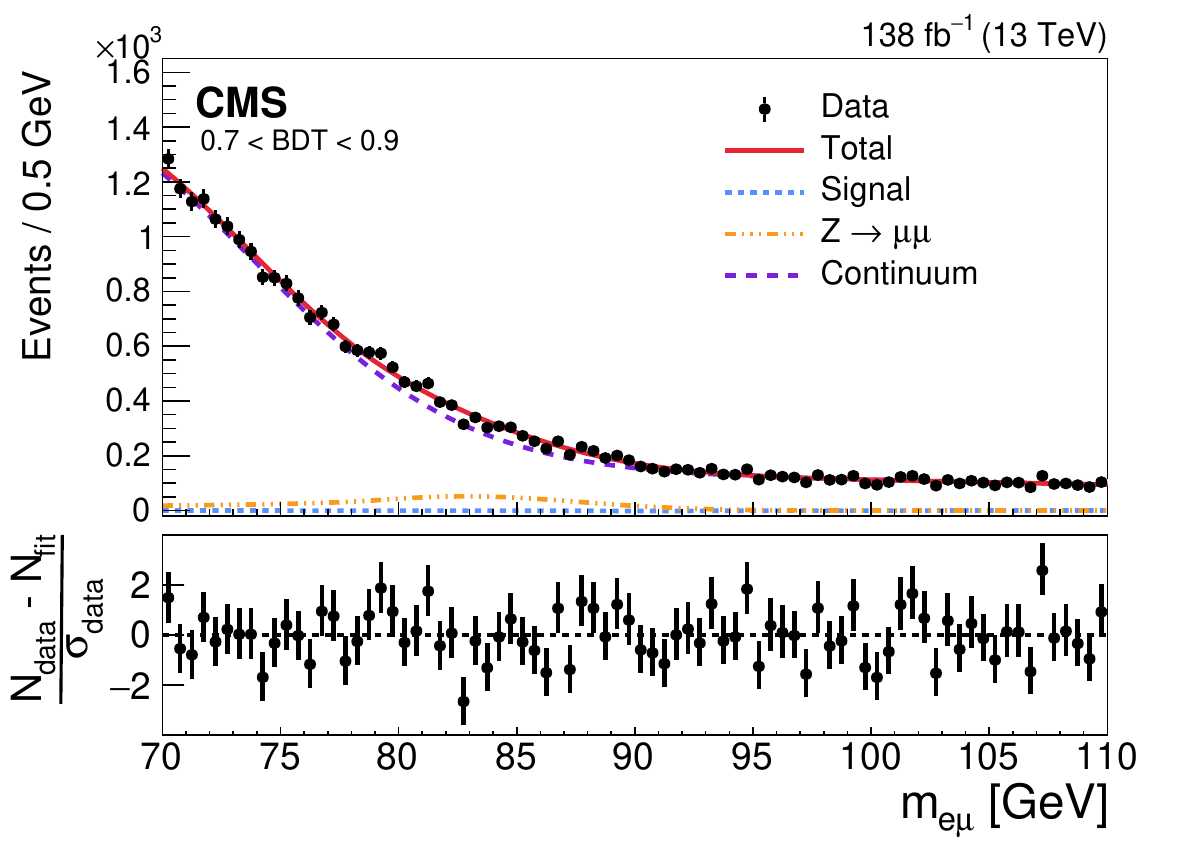}\\
  \includegraphics[width=0.49\textwidth]{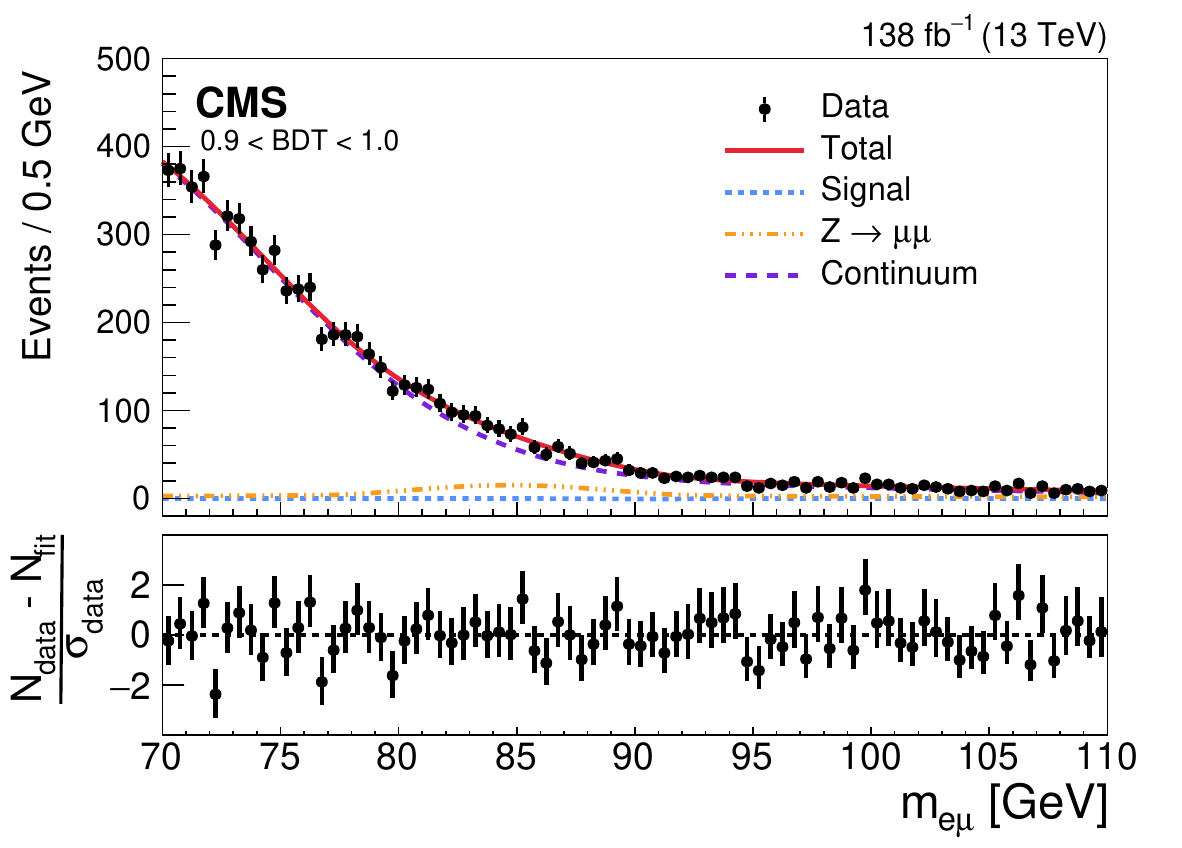}
  \caption{For the \zemu search, the invariant mass fit results for the BDT score ranges 0.3--0.7 (upper), 0.7--0.9 (middle), and 0.9--1.0 (lower).
    In each plot, the upper panel shows the data (points with bars showing the statistical uncertainties) together with the fit distribution curve (red) and its separate signal (blue dotted) \zmumu (yellow dash-dotted) and continuum background (purple dashed) components,
and the lower panel shows the deviations of the data from the fit function divided by the data uncertainty.
  }
  \label{fig:zemu_unblinded_fits}
\end{figure}

\begin{figure}[!htbp]
  \centering
  \includegraphics[width=0.49\textwidth]{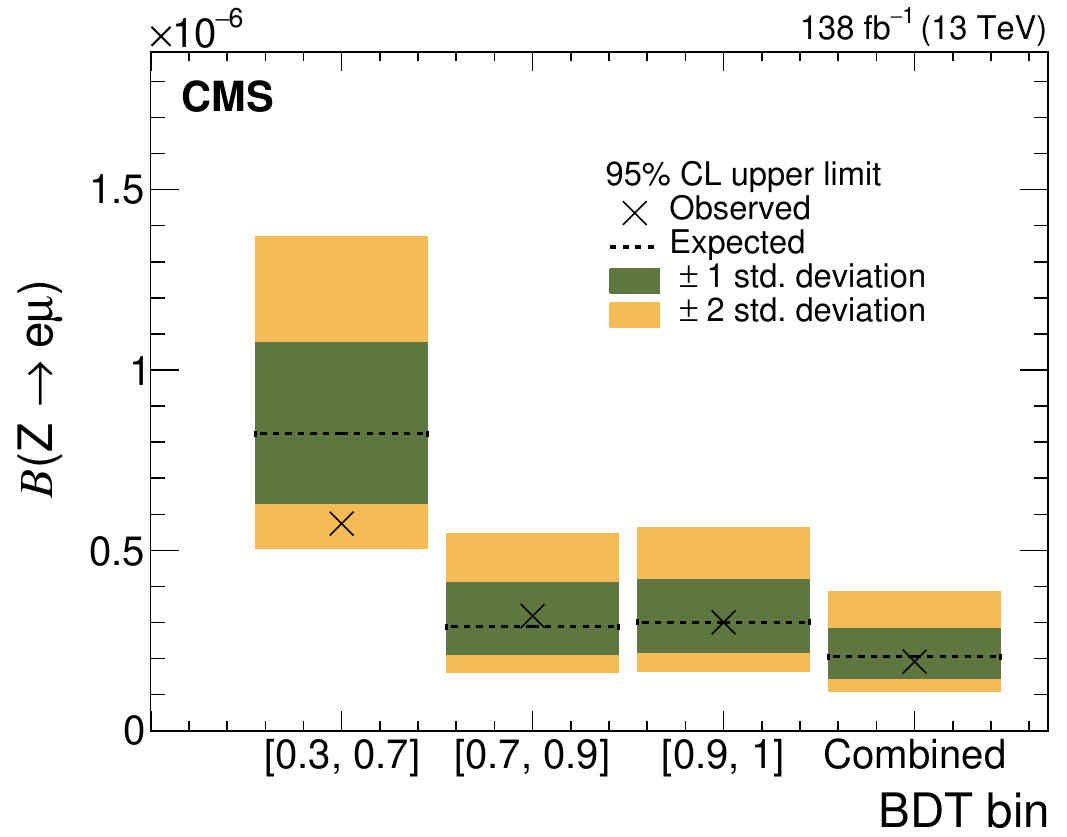}
  \caption{Upper limits at 95\% \CL on the branching fraction $\br(\zemu)$, for each BDT score range and for the final combined fit.  The observed limits are denoted by the markers, while the expected limits with their 68 and 95\% uncertainties are denoted by the horizontal dashed lines and green and yellow bands, respectively.
  }
  \label{fig:zemu_limits}
\end{figure}

\begin{figure*}[!htbp]
  \centering
  \includegraphics[width=0.48\textwidth]{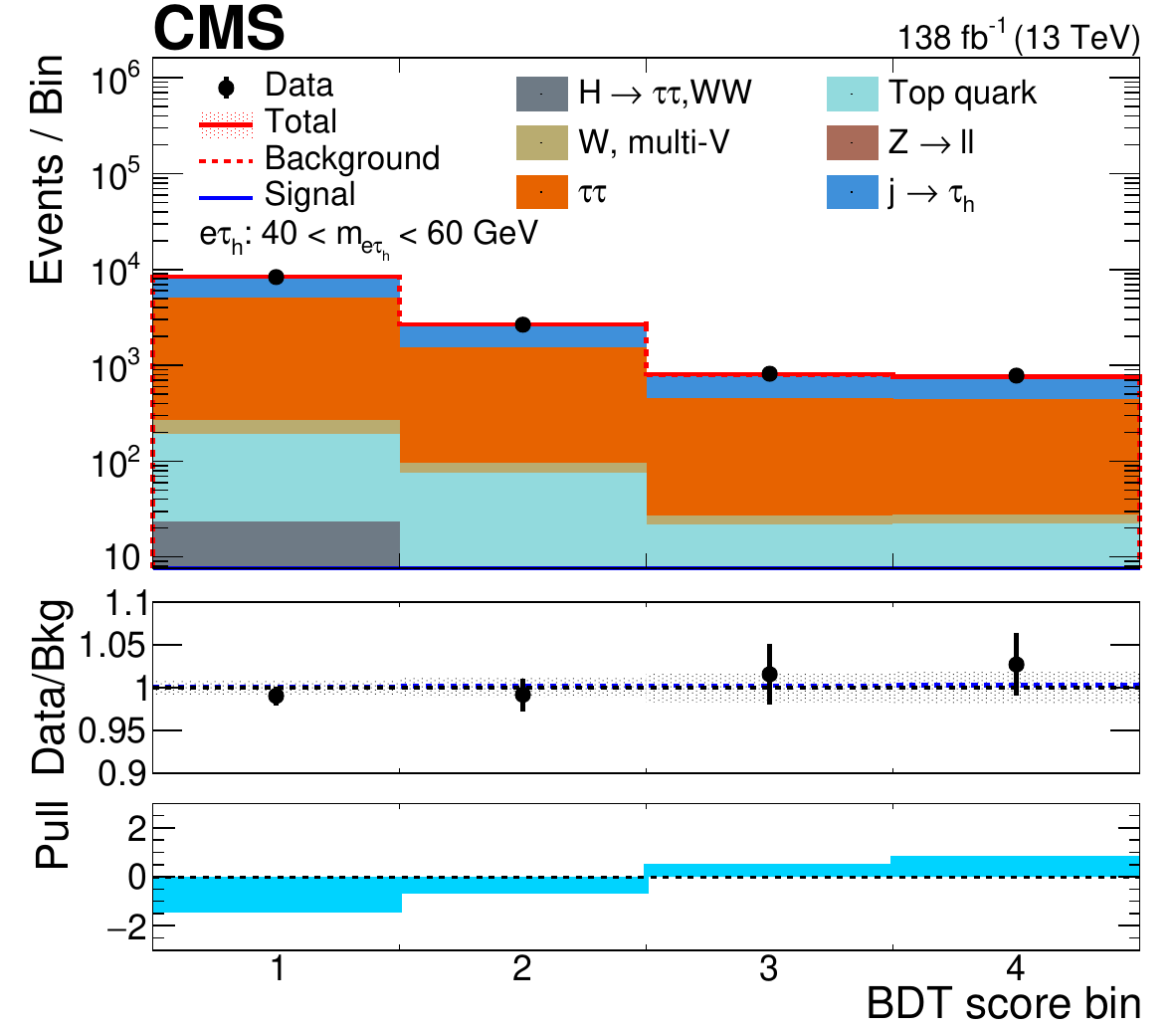}
  \includegraphics[width=0.48\textwidth]{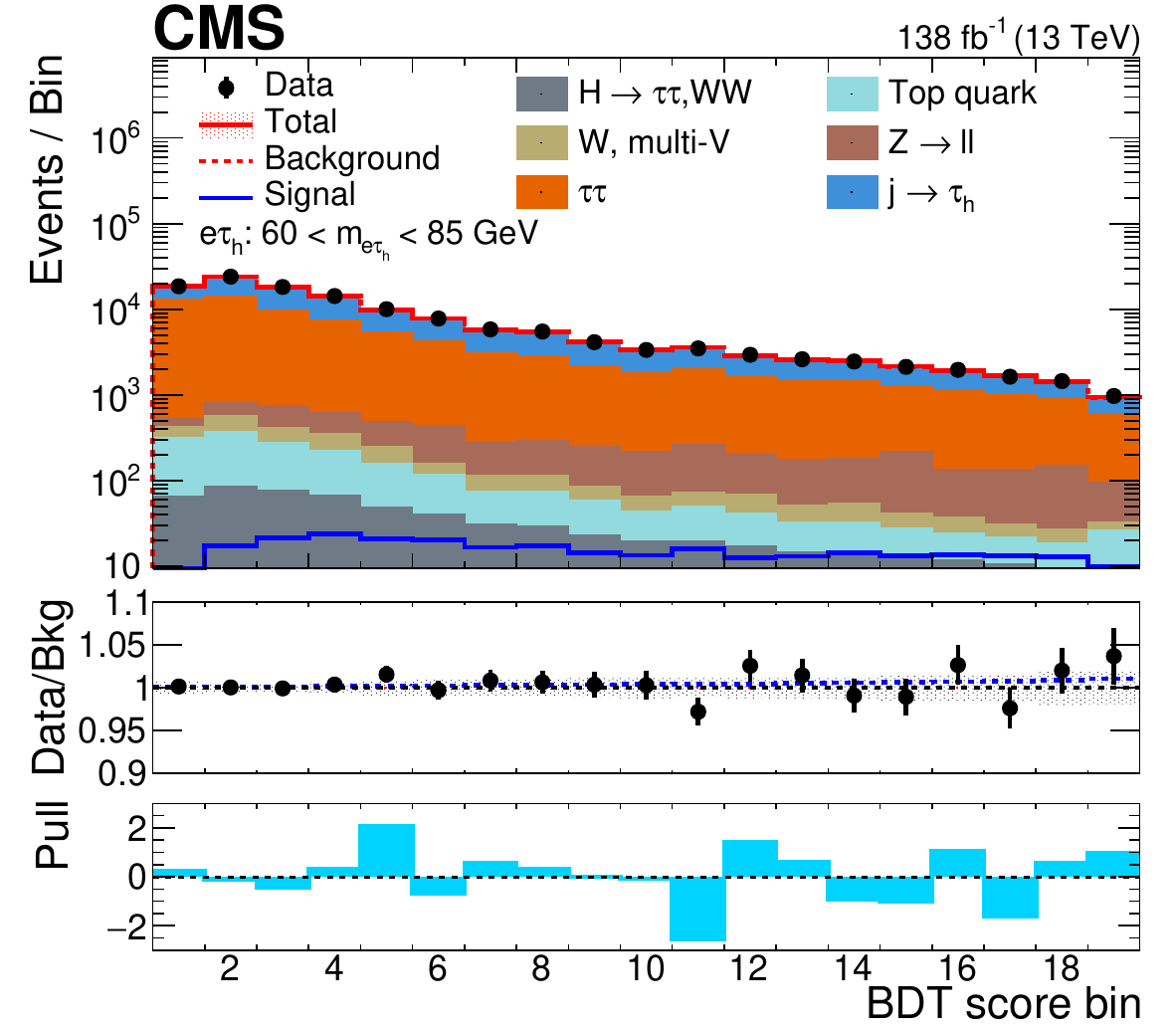} \\
  \includegraphics[width=0.48\textwidth]{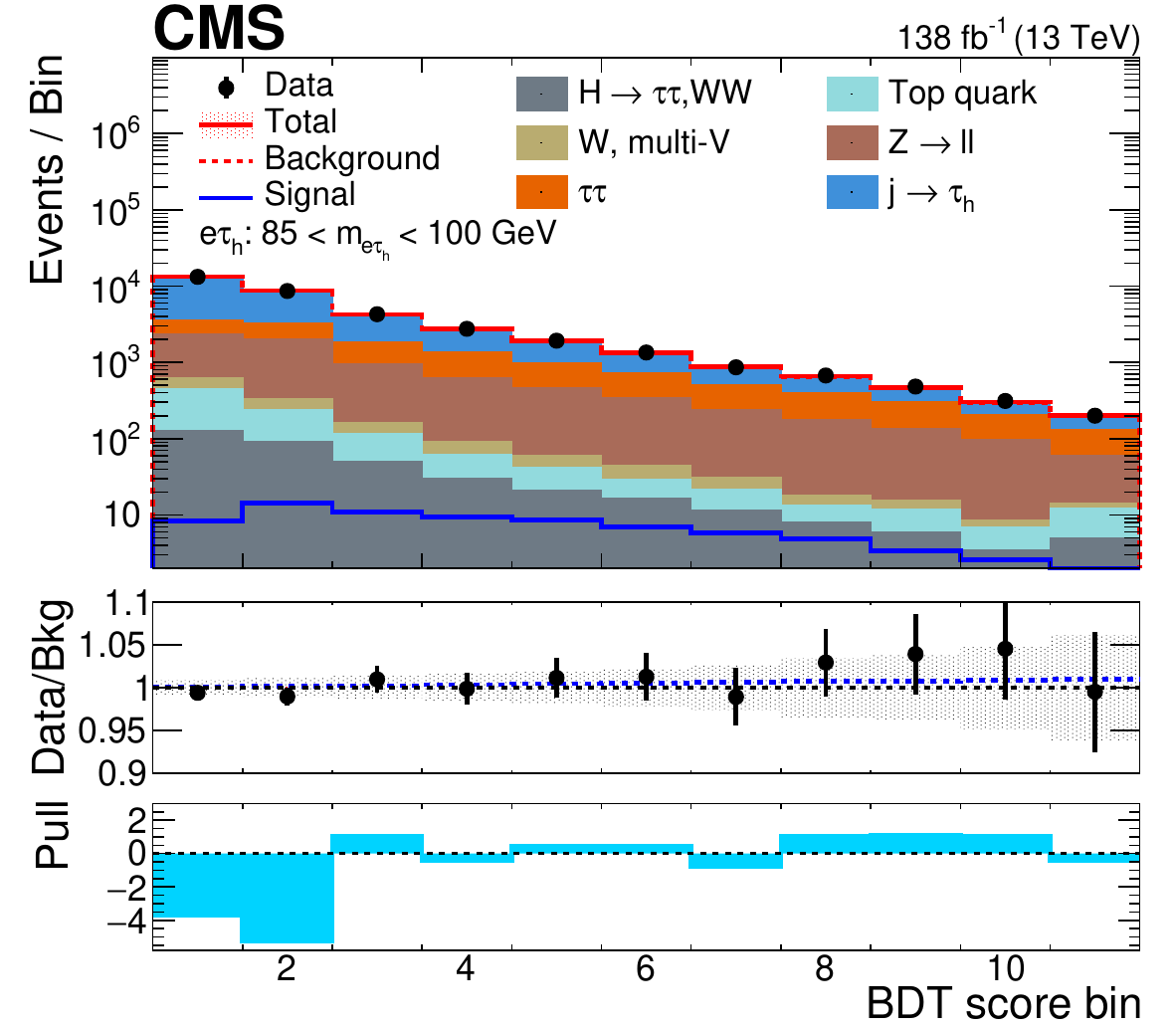}
  \includegraphics[width=0.48\textwidth]{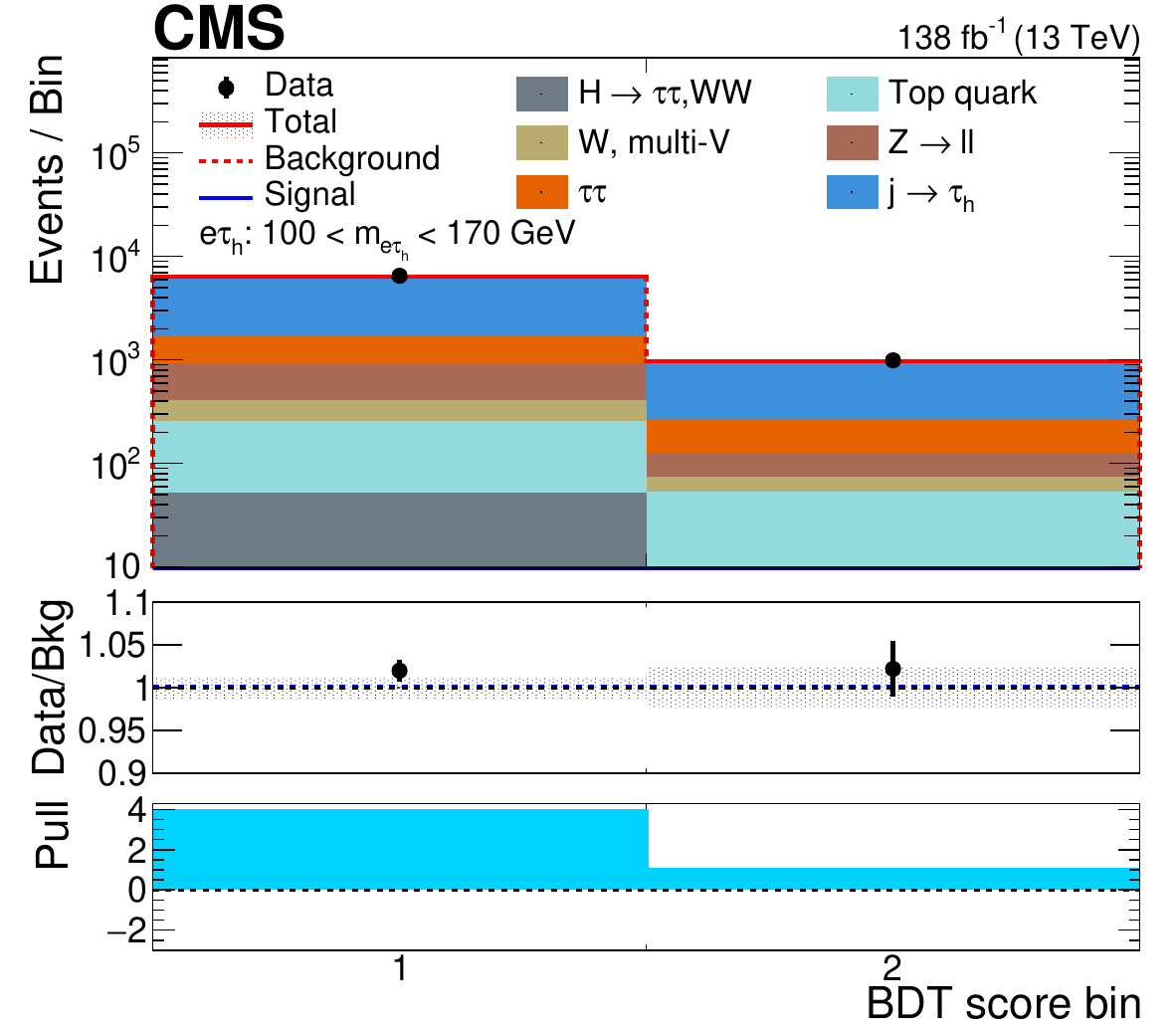}
  \caption{
    Transformed BDT score distributions for the \zetauh channels in the \metau ranges:  (upper left) 40--60\GeV, ``\tautau''; (upper right)  60--85\GeV, ``signal-like''; (lower left)  85--100\GeV, ``\zll''; (lower right)  100--170\GeV, ``misID''.
In each plot, the upper panel shows the data (points), the total yield from the signal + background fit (red open histogram), the signal component (blue open histogram), and the background components (stacked filled histograms).  The middle panel shows the ratio to the background of the data (points with bars showing the statistical uncertainty in the data) and the combined signal + background (blue dotted histogram).  The shaded band shows the systematic uncertainty in the background estimate.  The lower panel shows the pull defined in the text (light-blue filled histogram).
  }
  \label{fig:zetauh_unblinded_fits}
\end{figure*}

\begin{figure*}[!htbp]
  \centering
  \includegraphics[width=0.48\textwidth]{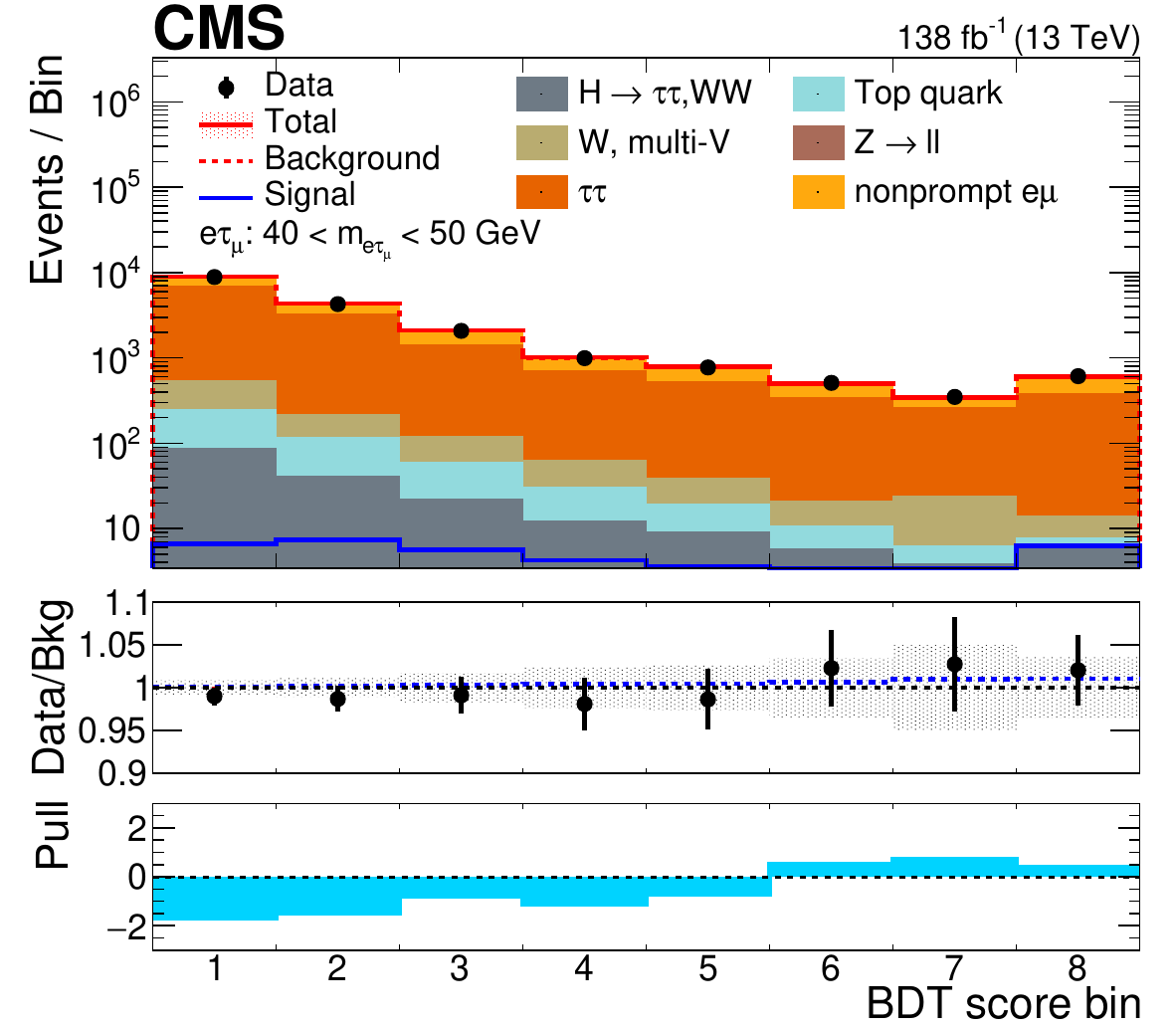}
  \includegraphics[width=0.48\textwidth]{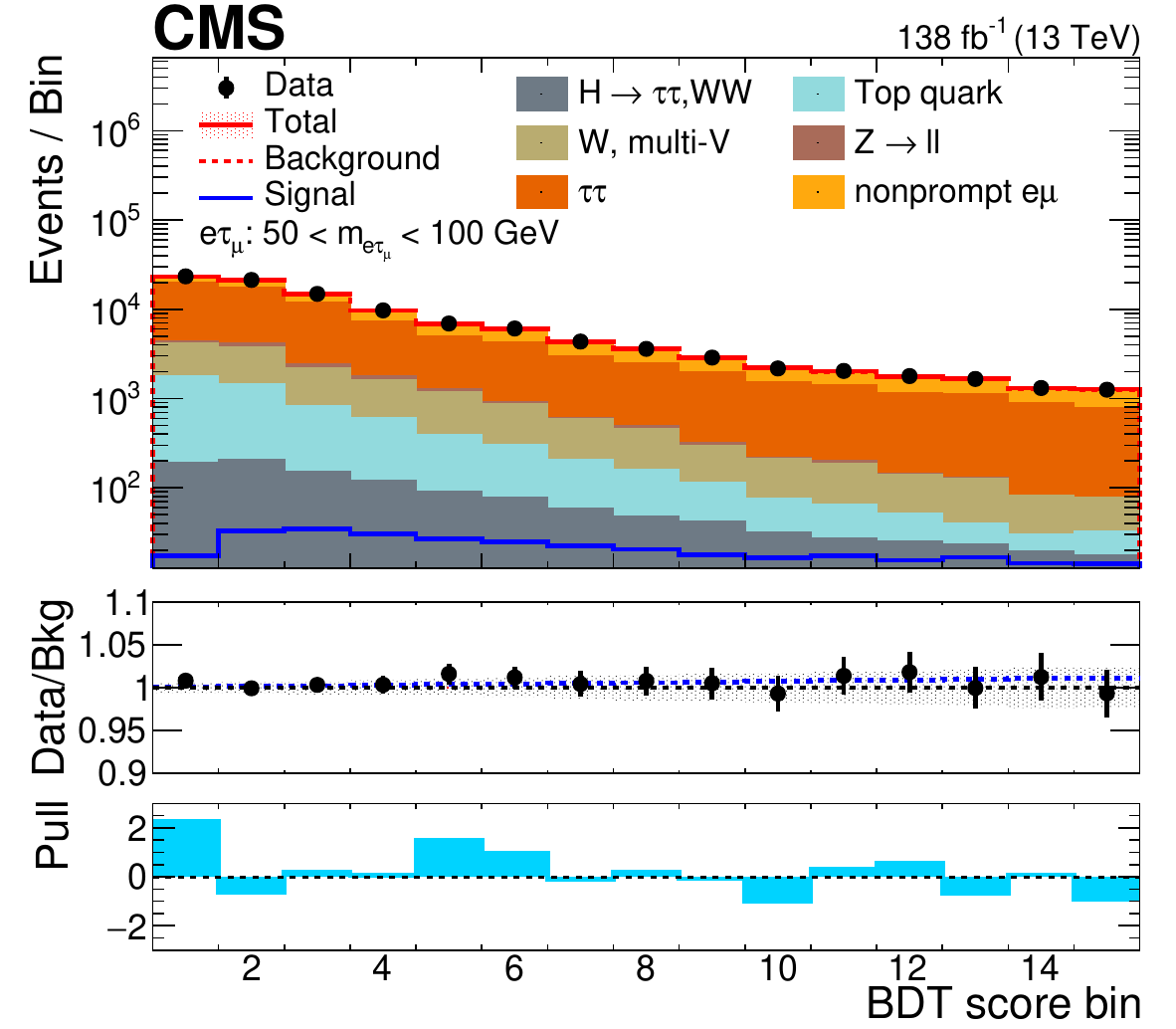}\\
  \includegraphics[width=0.48\textwidth]{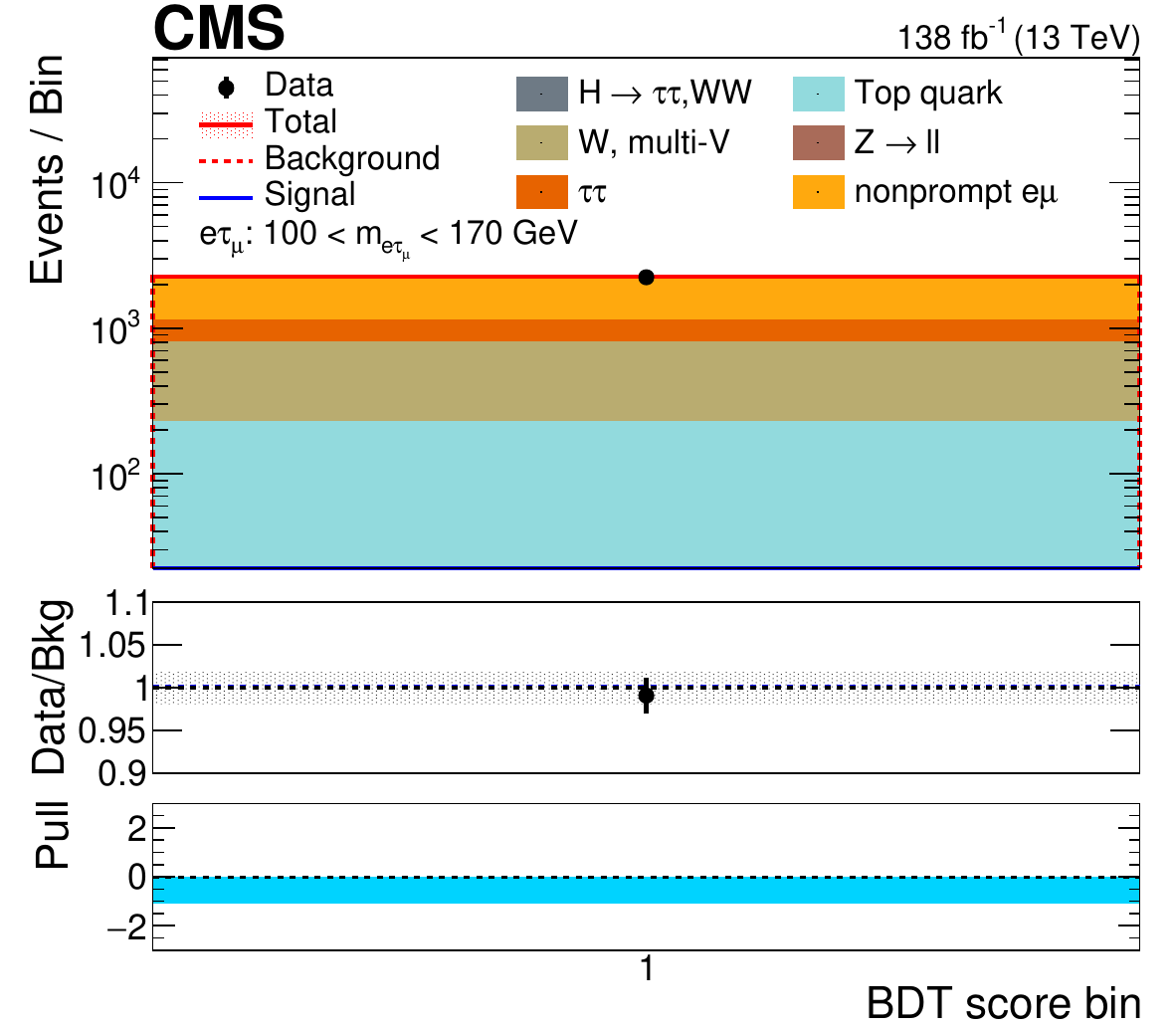}
  \caption{
    Transformed BDT score distributions for the \zetaumu channels in the \metau ranges:  (upper left) 40--50\GeV, ``\tautau''; (upper right)  50--100\GeV, ``signal-like''; (lower)  100--170\GeV, ``misID''.
In each plot, the upper panel shows the data (points), the total yield from the signal + background fit (red open histogram), the signal component (blue open histogram), and the background components (stacked filled histograms).  The middle panel shows the ratio to the background of the data (points with bars showing the statistical uncertainty in the data) and the combined signal + background (blue dotted histogram).  The shaded band shows the systematic uncertainty in the background estimate.  The lower panel shows the pull defined in the text (light-blue filled histogram).
  }
  \label{fig:zetaumu_unblinded_fits}
\end{figure*}

\begin{figure*}[!htbp]
  \centering
  \includegraphics[width=0.48\textwidth]{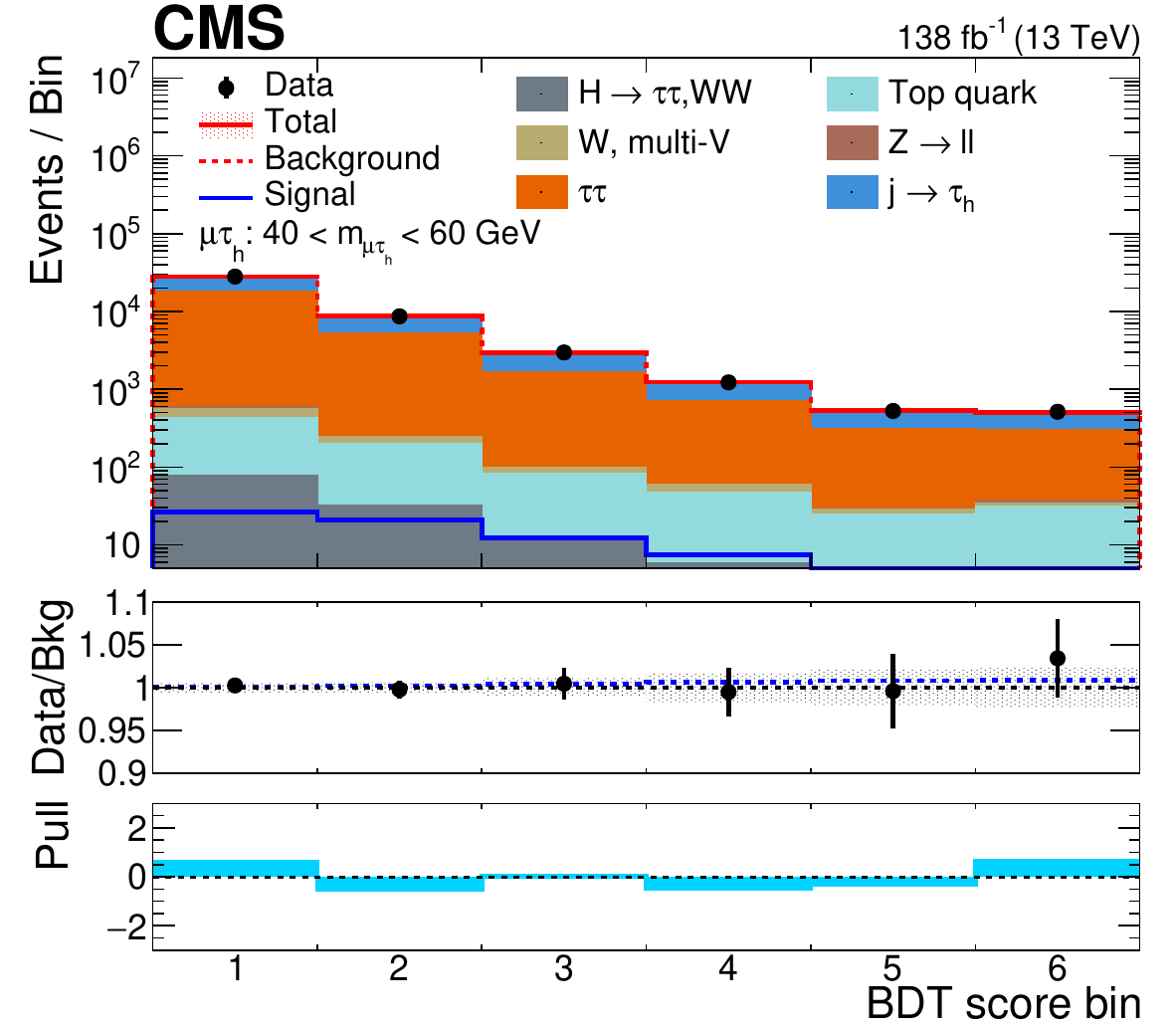}
  \includegraphics[width=0.48\textwidth]{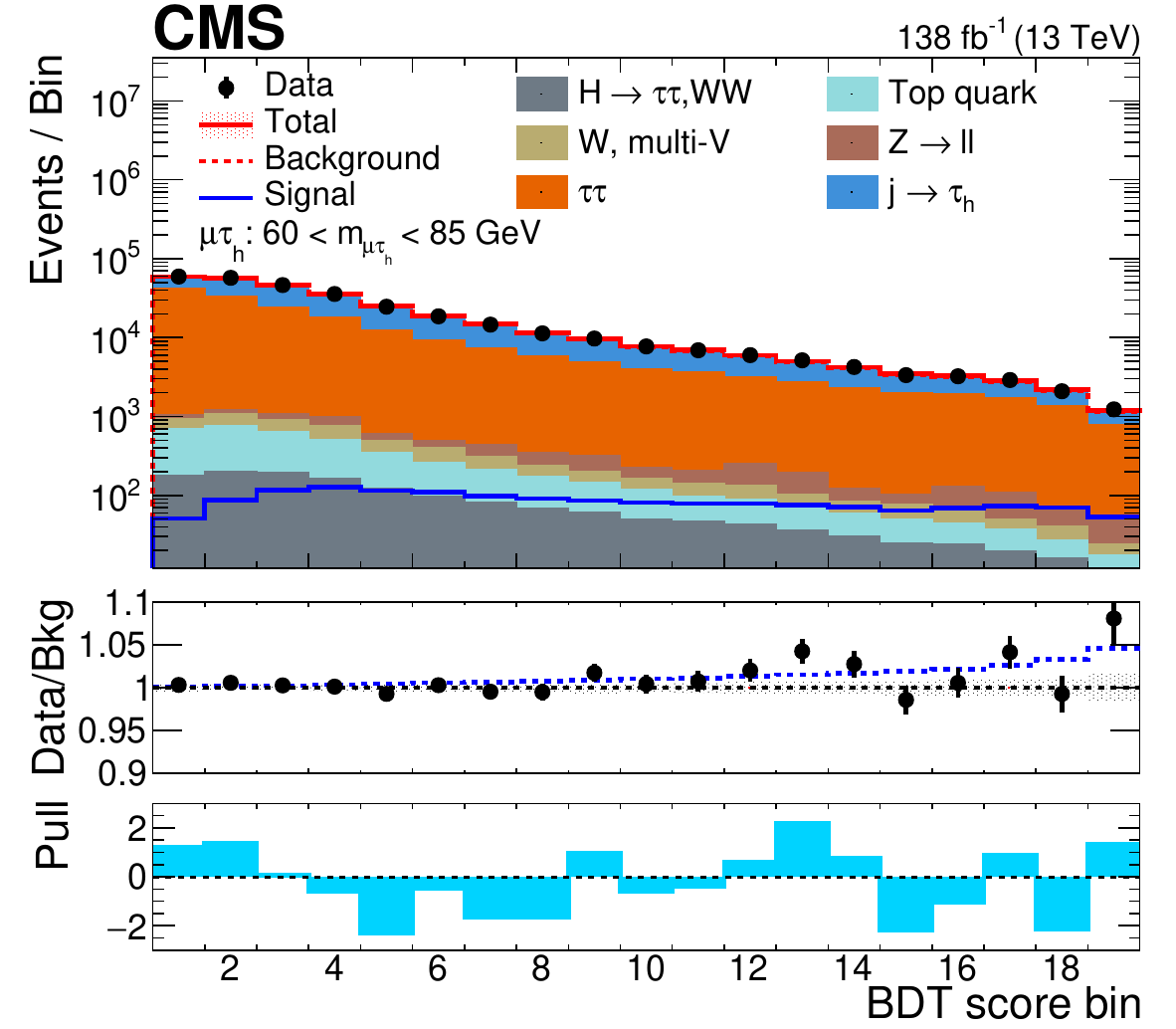} \\
  \includegraphics[width=0.48\textwidth]{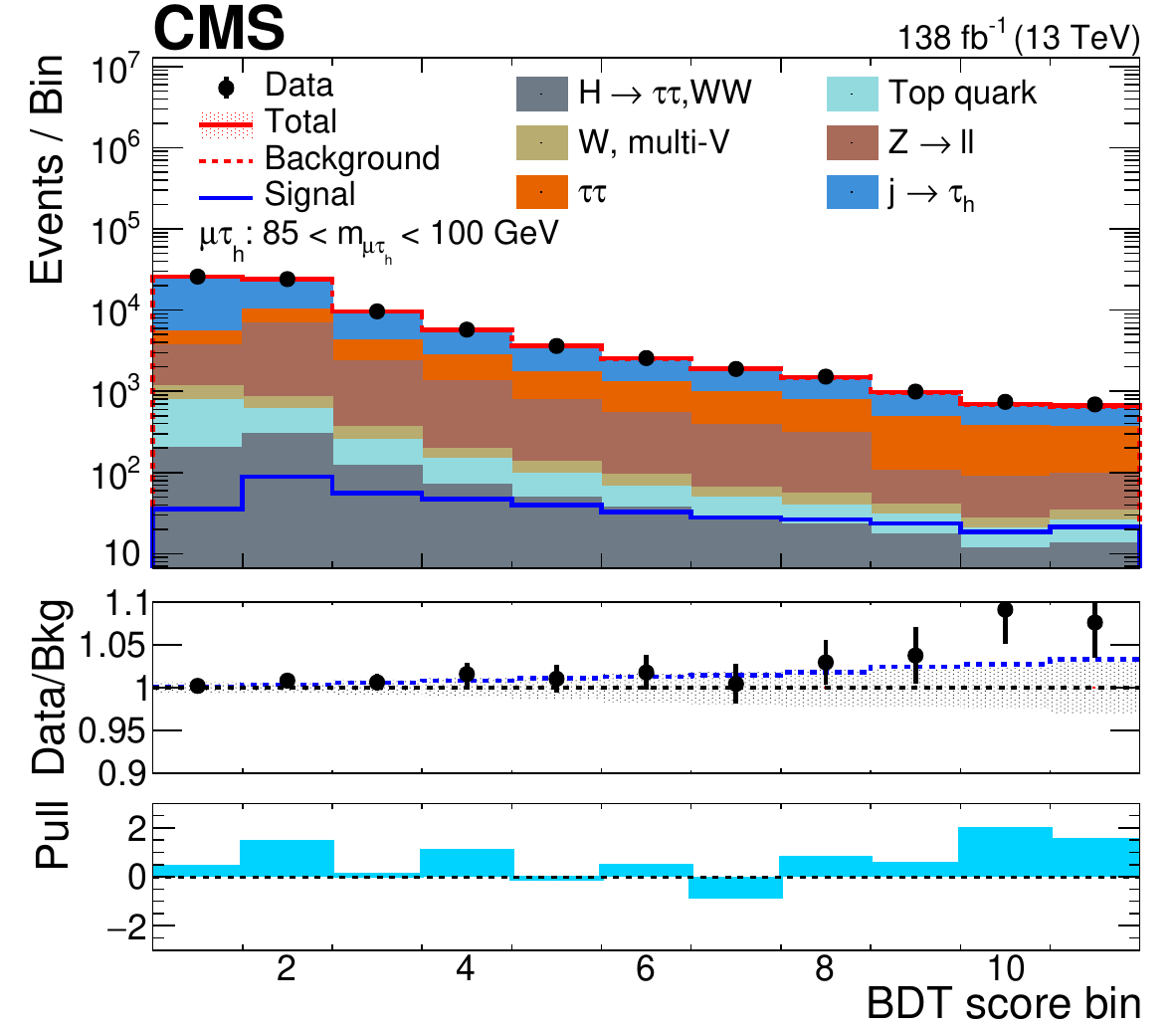}
  \includegraphics[width=0.48\textwidth]{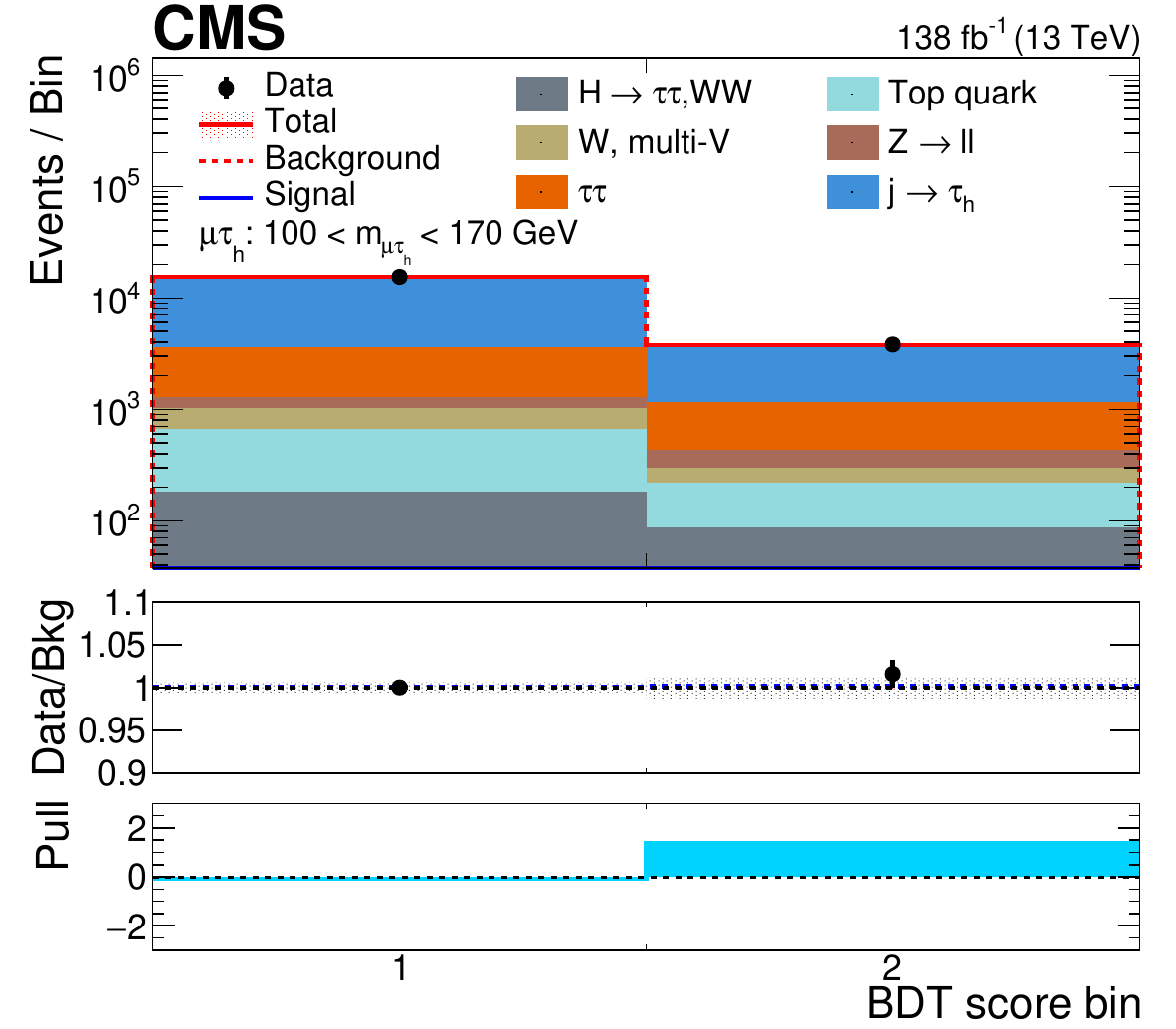}\\
  \caption{
    Transformed BDT score distributions for the \zmutauh channels in the \mmutau ranges:  (upper left) 40--60\GeV, ``\tautau''; (upper right)  60--85\GeV, ``signal-like''; (lower left)  85--100\GeV, ``\zll''; (lower right)  100--170\GeV, ``misID''.
In each plot, the upper panel shows the data (points), the total yield from the signal + background fit (red open histogram), the signal component (blue open histogram), and the background components (stacked filled histograms).  The middle panel shows the ratio to the background of the data (points with bars showing the statistical uncertainty in the data) and the combined signal + background (blue dotted histogram).  The shaded band shows the systematic uncertainty in the background estimate.  The lower panel shows the pull defined in the text (light-blue filled histogram).
  }
  \label{fig:zmutauh_unblinded_fits}
\end{figure*}

\begin{figure*}[!htbp]
  \includegraphics[width=0.48\textwidth]{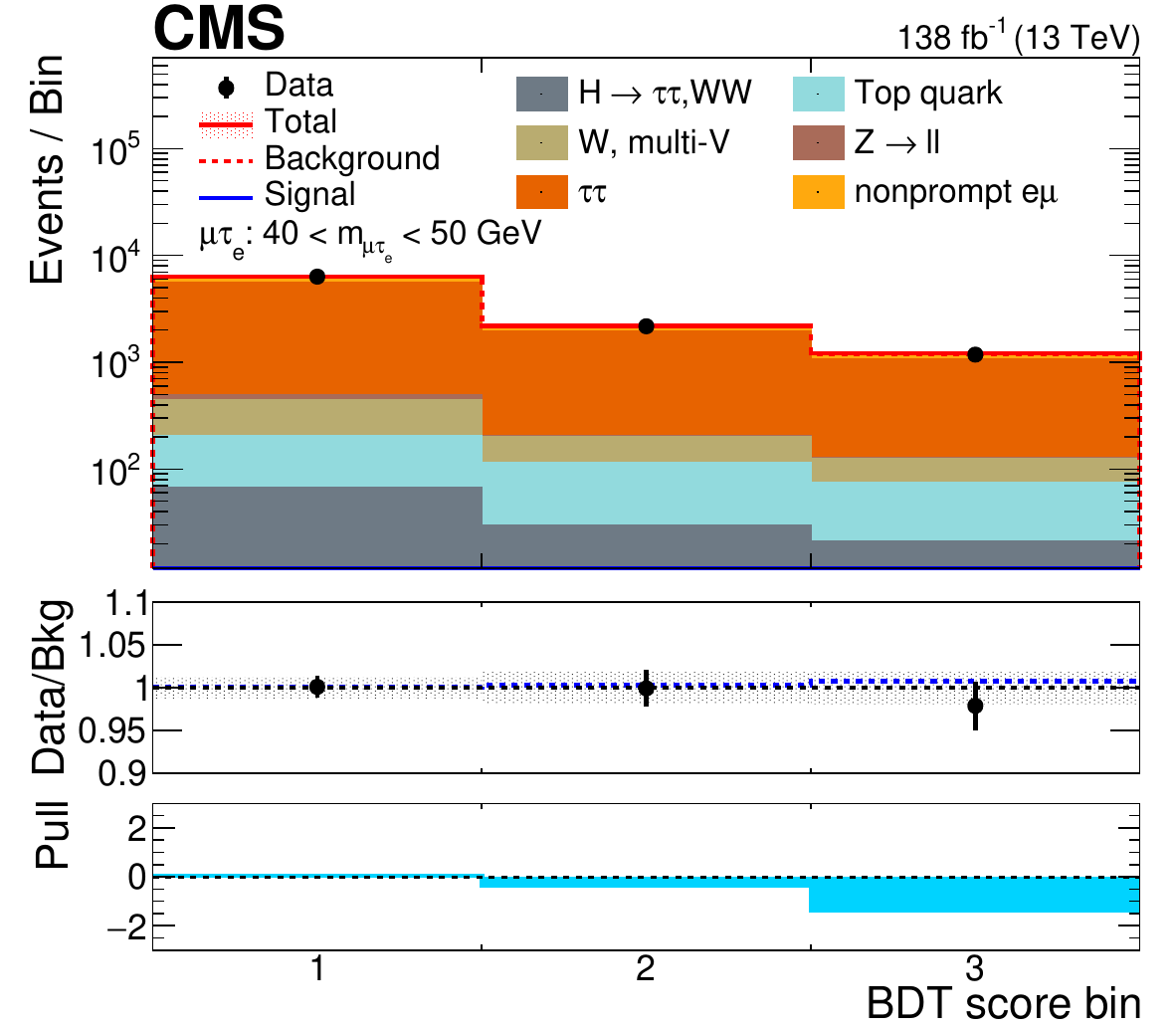}
  \includegraphics[width=0.48\textwidth]{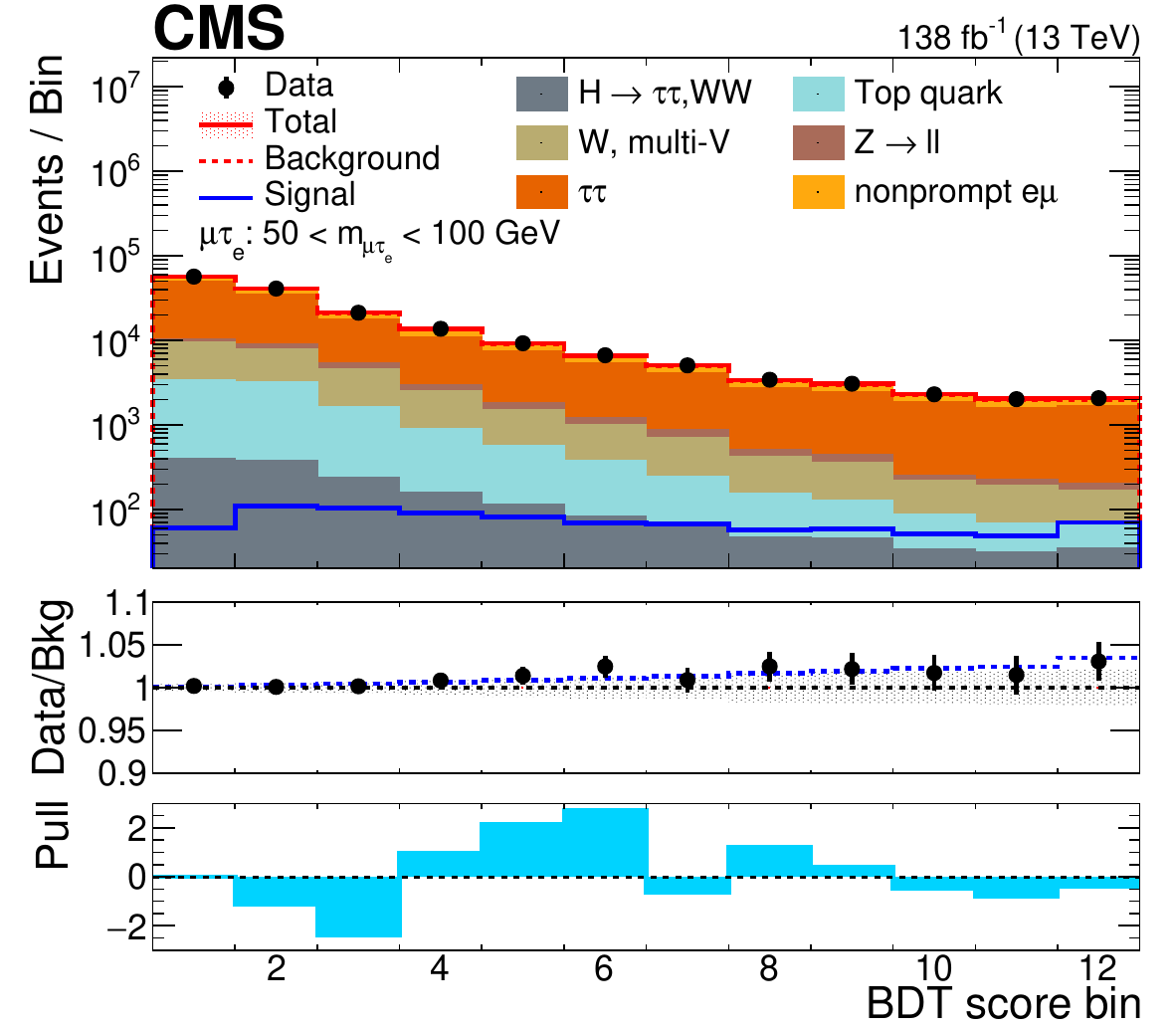}\\
  \includegraphics[width=0.48\textwidth]{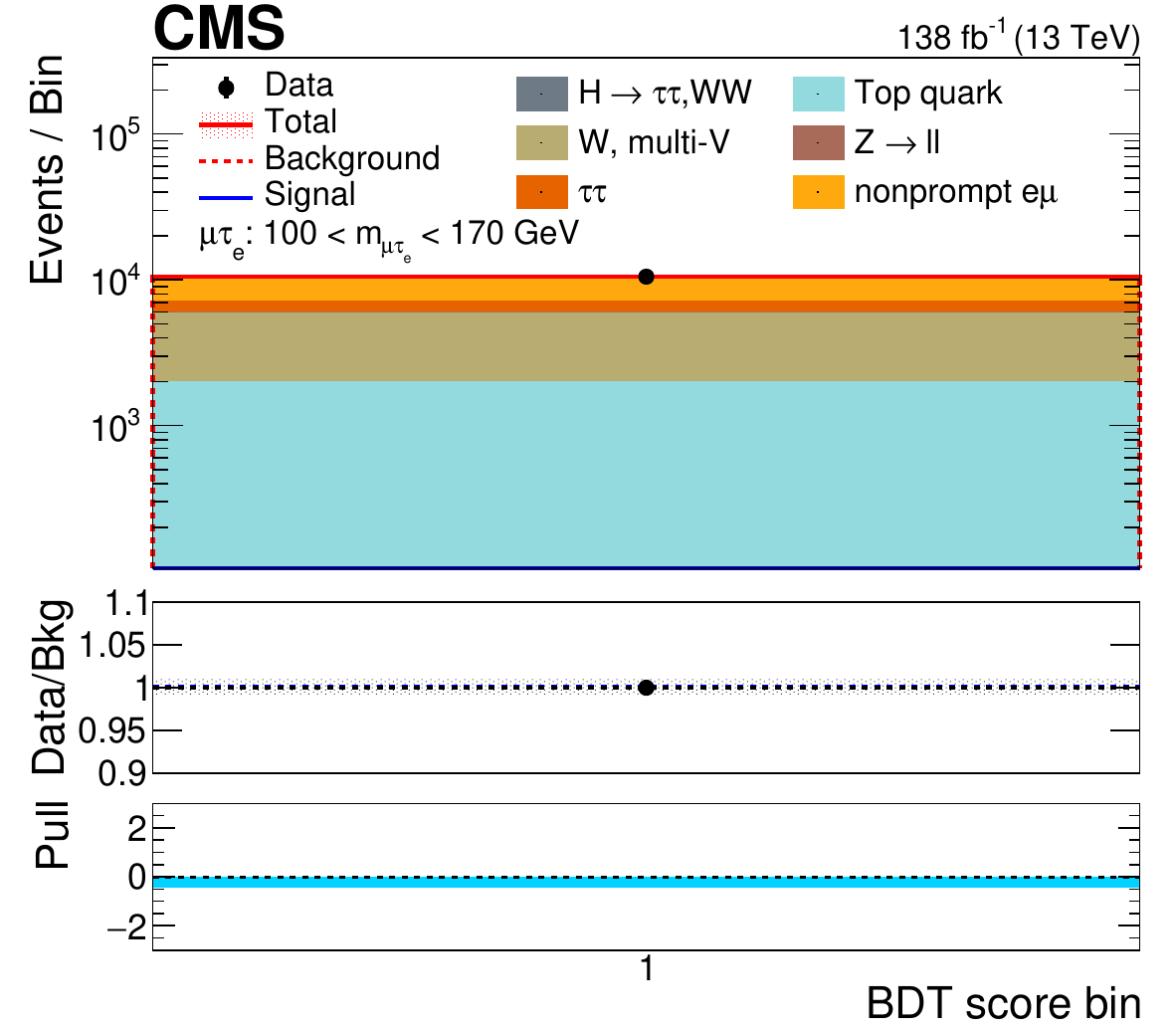}
  \caption{
    Transformed BDT score distributions for the \zmutaue channels in the \mmutau ranges:  (upper left) 40--50\GeV, ``\tautau''; (upper right) 50--100\GeV, ``signal-like''; (lower)  100--170\GeV, ``misID''.
In each plot, the upper panel shows the data (points), the total yield from the signal + background fit (red open histogram), the signal component (blue open histogram), and the background components (stacked filled histograms).  The middle panel shows the ratio to the background of the data (points with bars showing the statistical uncertainty in the data) and the combined signal + background (blue dotted histogram).  The shaded band shows the systematic uncertainty in the background estimate.  The lower panel shows the pull defined in the text (light-blue filled histogram).
  }
  \label{fig:zmutaue_unblinded_fits}
\end{figure*}

\begin{figure}[!htbp]
  \centering
  \includegraphics[width=0.48\textwidth]{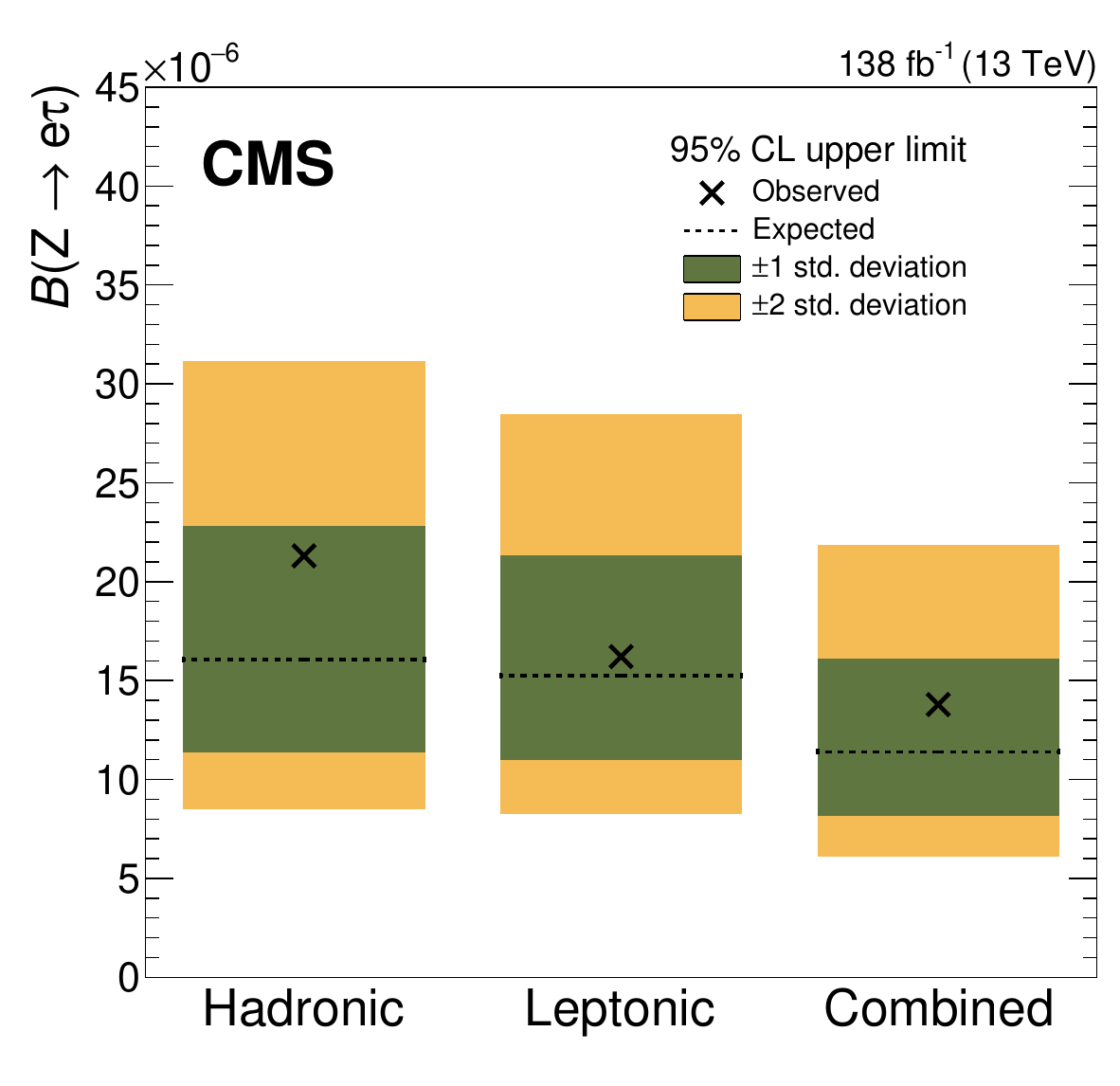}
  \includegraphics[width=0.48\textwidth]{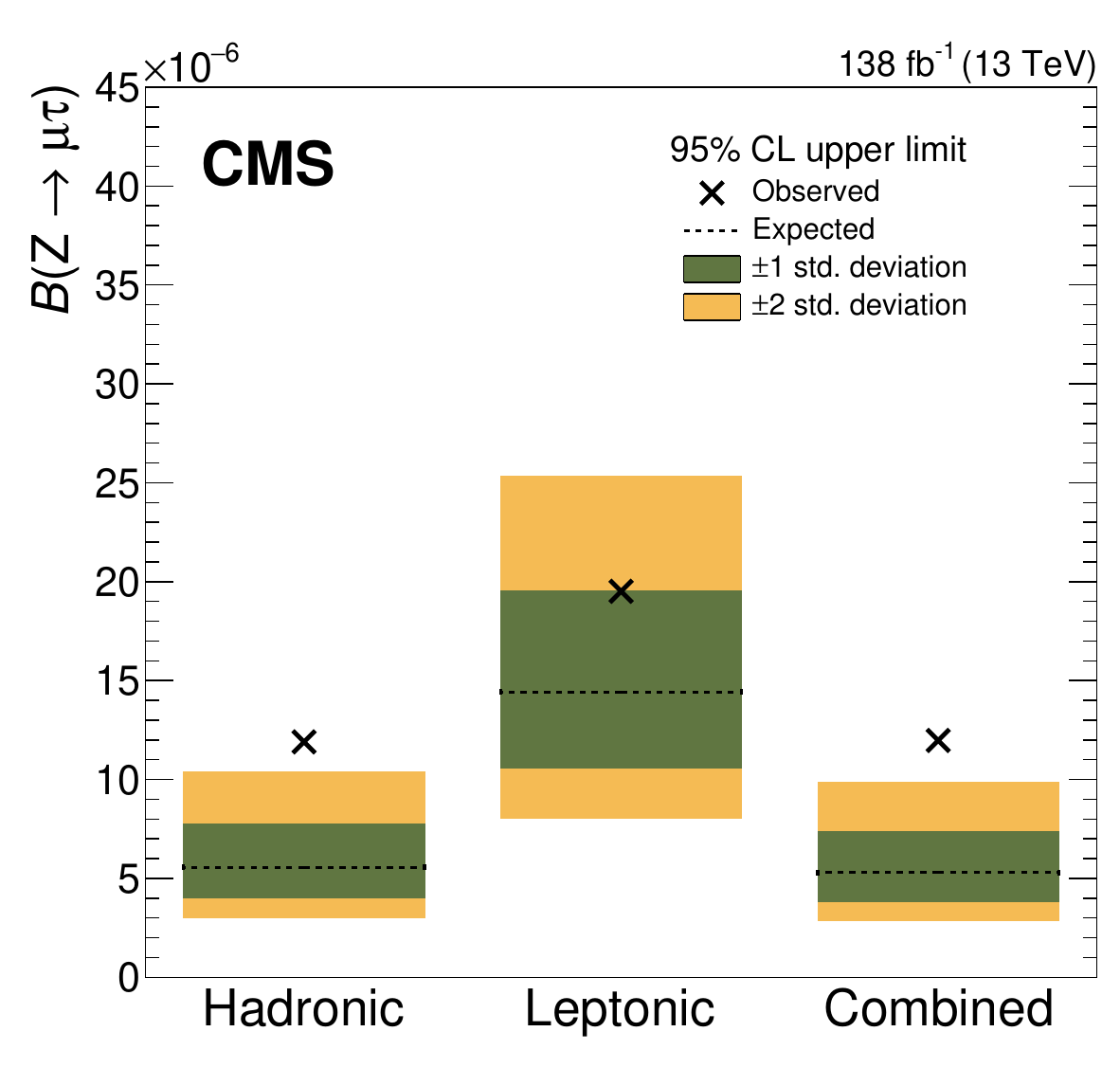}
  \caption{Observed and expected 95\% \CL upper limit by category, as well as for the final combined fit, for the \zetau (\cmsLeft) and \zmutau (\cmsRight) searches.  The observed limits are denoted by the markers, while the expected limits with their 68 and 95\% uncertainties are denoted by the horizontal dashed lines and green and yellow bands, respectively.}
  \label{fig:zxtau_limits}
\end{figure}

\begin{table*}[bhtp]
  \centering
  \topcaption{The measured branching fraction with its significance (signif.) and observed and expected 95\% \CL upper limits, for each of the \zemu, \zetau, and \zmutau decay channels.  The prior best published limits are also given for comparison.  Included are results for the separate BDT bins for \zemu and the separate \PGt decay subchannels for \zetau and \zmutau.}
  \renewcommand{\arraystretch}{1.2}
  \begin{scotch}{lcccc}
    Channel  &       Branching       & Signif. &     Observed     &     Prior       \\
             & fraction & [$\sigma$]    & (expected) limit &  (expected) limit \\
    \hline
    \zemu    & $-0.1^{+1.0}_{-1.0} \times 10^{-7}$ & $-0.1$ & $ 1.9 \,( 2.0^{+0.8}_{-0.6}) \times 10^{-7}$ & $2.6 \,( 2.4) \times 10^{-7}$~\cite{ATLAS:2022uhq} \\
    \quad $0.3 < {\rm BDT} < 0.7$ & $-3.4^{+2.8}_{-2.6}  \times 10^{-7}$ & $-1.2$ & $ 5.7 (8.2^{+2.5}_{-2.0}) \times 10^{-7}$ & \NA \\
    \quad $0.7 < {\rm BDT} < 0.9$ & $ 0.4^{+1.5}_{-1.4}  \times 10^{-7}$ & $+0.3$ & $ 3.2 (2.9^{+1.2}_{-0.8}) \times 10^{-7}$ & \NA \\
    \quad $0.9 < {\rm BDT} < 1.0$ & $ 0.0^{+1.5}_{-2.0}  \times 10^{-7}$ & $+0.0$ & $ 3.0 (3.0^{+1.2}_{-0.8}) \times 10^{-7}$ & \NA \\
    [\cmsTabSkip]
    \zetau   & $ 3.2^{+6.1}_{-6.0} \times 10^{-6}$ & $+0.5$ & $13.8 \,(11.4^{+4.7}_{-3.2}) \times 10^{-6}$ & $5.0 \,( 6.0) \times 10^{-6}$~\cite{ATLAS:2021bdj}\\
    \quad\zetauh  & $ 6.3^{+8.4}_{-8.2} \times 10^{-6}$ & $+0.8$ & $21.3 \,(16.1^{+6.7}_{-4.6}) \times 10^{-6}$ & $8.1 \,( 8.1) \times 10^{-6}$~\cite{ATLAS:2021bdj}\\
    \quad\zetaumu & $ 1.2^{+7.9}_{-8.1} \times 10^{-6}$ & $+0.2$ & $16.2 \,(15.3^{+6.1}_{-4.2}) \times 10^{-6}$ & $7.0 \,( 8.9) \times 10^{-6}$~\cite{ATLAS:2021bdj}\\
    [\cmsTabSkip]
    \zmutau  & $ 7.5^{+2.7}_{-2.7} \times 10^{-6}$ & $+2.7$ & $12.0 \,( 5.3^{+2.1}_{-1.5}) \times 10^{-6}$ & $6.5 \,( 5.3) \times 10^{-6}$~\cite{ATLAS:2021bdj}\\
    \quad\zmutauh & $ 7.2^{+2.8}_{-2.8} \times 10^{-6}$ & $+2.5$ & $11.9 \,( 5.6^{+2.2}_{-1.6}) \times 10^{-6}$ & $9.5 \,( 6.1) \times 10^{-6}$~\cite{ATLAS:2021bdj}\\
    \quad\zmutaue & $ 7.5^{+7.1}_{-7.9} \times 10^{-6}$ & $+1.0$ & $19.5 \,(14.4^{+5.1}_{-3.8}) \times 10^{-6}$ & $7.2 \,(10.0) \times 10^{-6}$~\cite{ATLAS:2021bdj}\\
  \end{scotch}
  \label{tab:lfvz_limits}
\end{table*}

Figure~\ref{fig:zprime_unblinded_fits} shows the \emu mass scan fit results for two example mass points.
No statistically significant excess is observed over the SM background.  We report 95\% \CL upper limits on the production cross section $\sigma(\pp\to\zprime\!+\!\PX)$ times
the branching fraction $\br(\zpemu)$ to the \emu final state.
The observed and expected limits for the entire mass scan
in the \emu final state are shown in Fig.~\ref{fig:scan_limits}.
All observed deviations from the background-only hypothesis have local $p$-values greater than 1\%.  The global $p$-value is greater than 39\%
after accounting for the look-elsewhere effect~\cite{cowan:statLHC}.

\begin{figure*}[!htbp]
  \centering
  \includegraphics[width=0.49\textwidth]{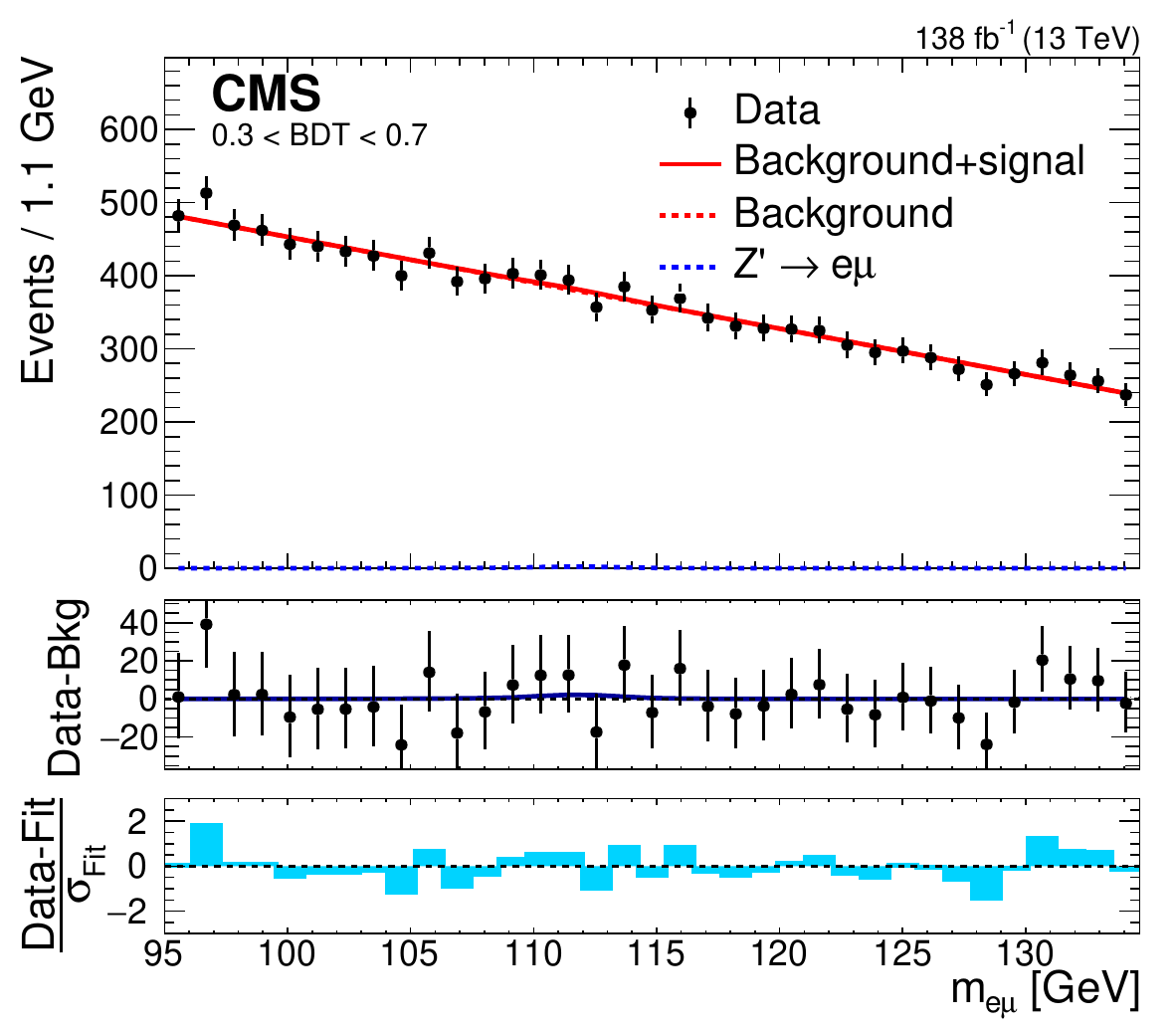}
  \includegraphics[width=0.49\textwidth]{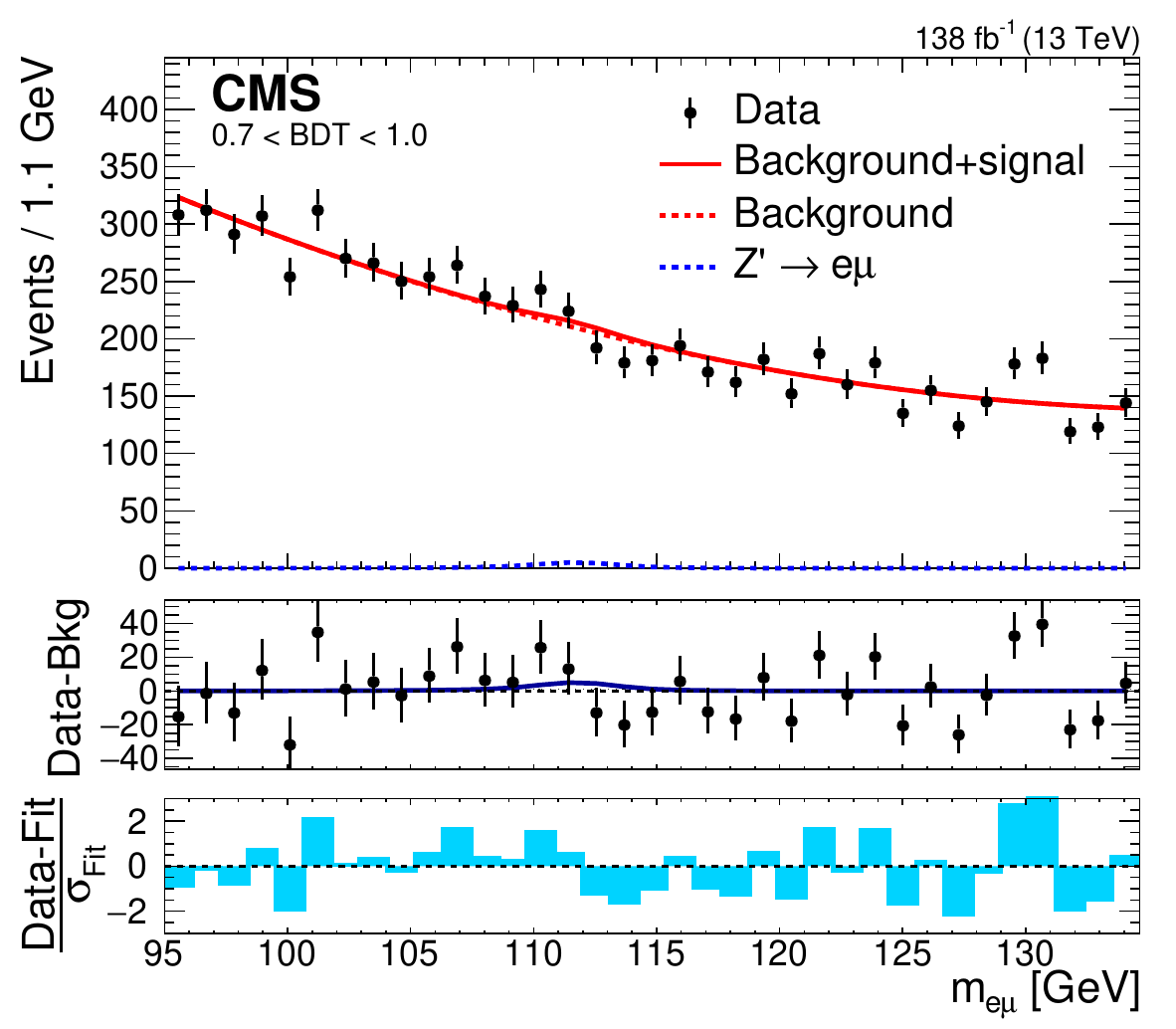}\\
  \includegraphics[width=0.49\textwidth]{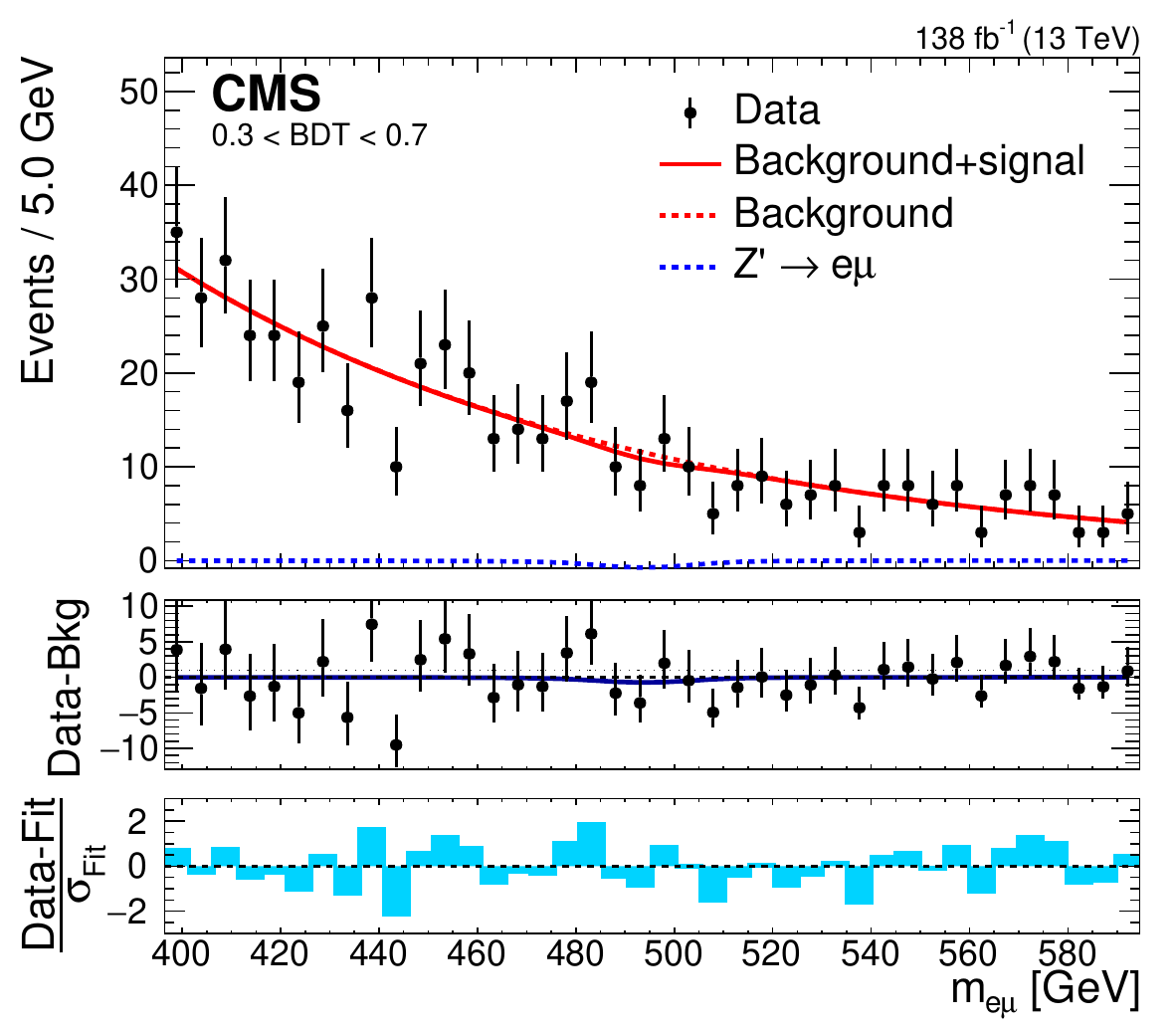}
  \includegraphics[width=0.49\textwidth]{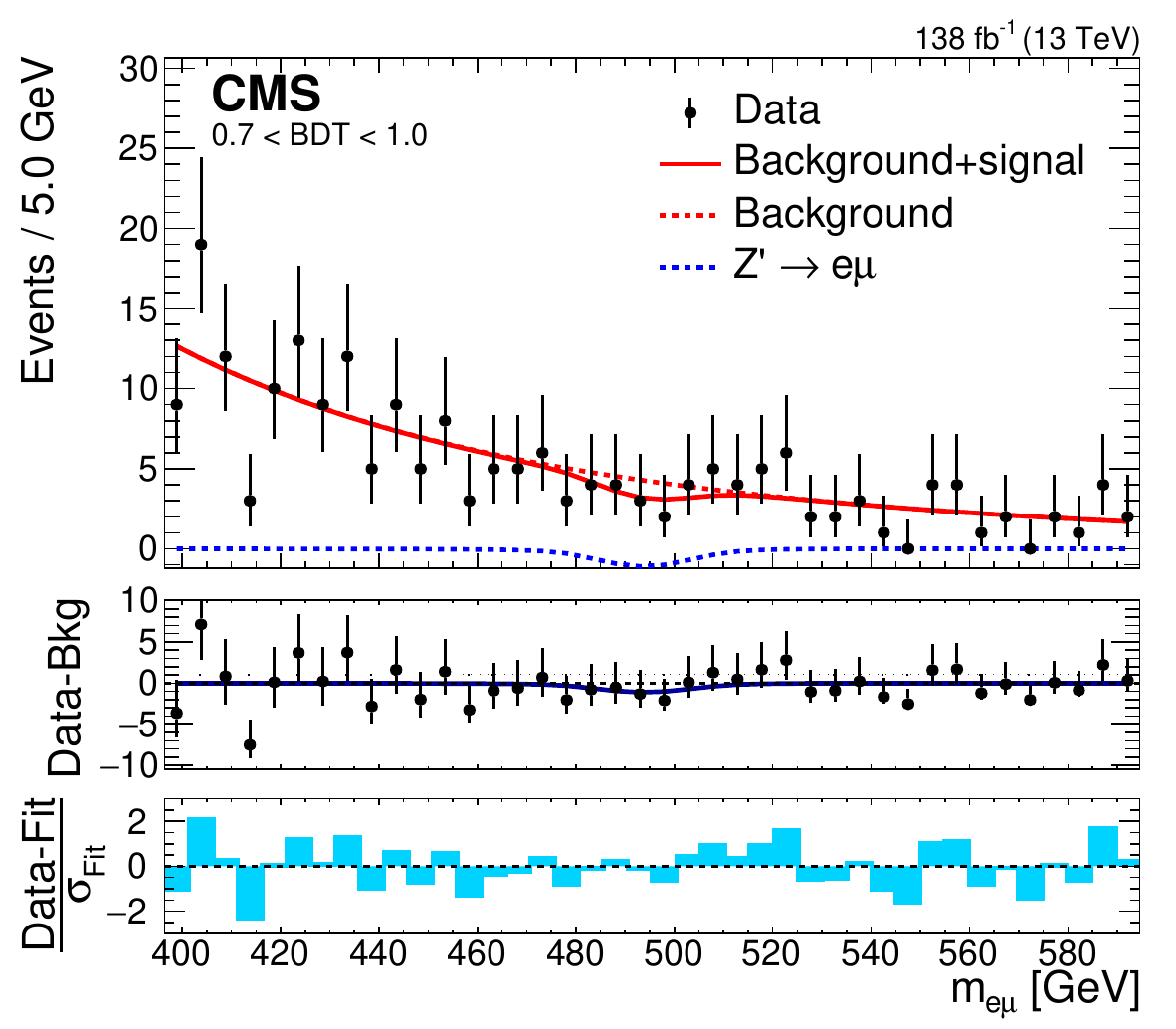}
  \caption{Distributions in \memu for the scan points 111\GeV (upper row) and 496\GeV (lower row) from the \zprime search.  In each row the BDT score range for the left (right) plot is 0.3--0.7 (0.7--1.0).  In each plot, the upper panel shows the data (points with bars showing the statistical uncertainty) together with the fit distribution curve (red solid) and its separate signal (blue dotted) and background (red dotted) components, the middle panel shows the background subtracted data with the fit signal distribution,
and the lower panel shows the deviations of the data from the fit function divided by the fit uncertainty.
  }
  \label{fig:zprime_unblinded_fits}
\end{figure*}

\begin{figure}[!htbp]
  \centering
  \includegraphics[width=0.49\textwidth]{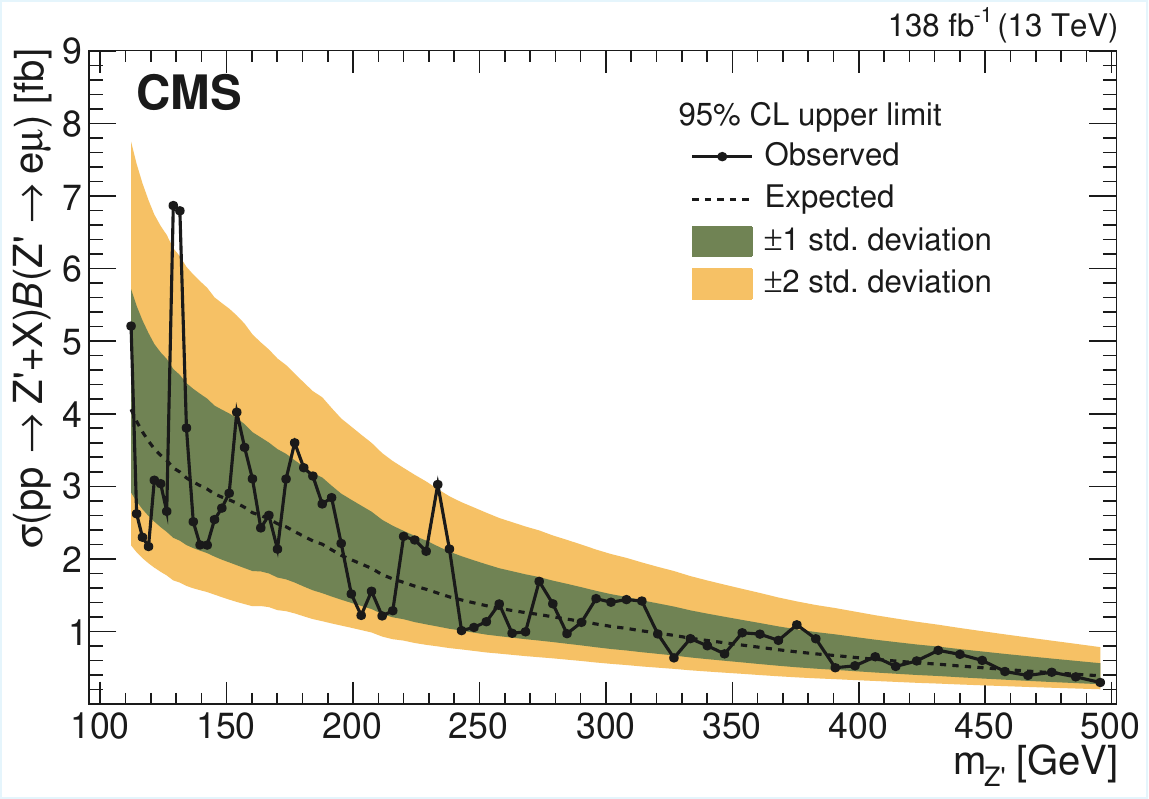}
  \caption{Expected and observed 95\% \CL upper limits on $\sigma(\pp\to\zprime\!+\!\PX)\br(\zpemu)$ for \zprime masses
    between 110 and 500 \GeV.  The solid black line connects filled circles representing the observed upper limits at the scan points, and the dashed line with filled error bands shows the expected limit.
  }
  \label{fig:scan_limits}
\end{figure}

The search presented here for CLFV \PZ boson decays represents an approximately 20\% improvement in sensitivity to the direct \zemu decay process, with respect to the best previous search~\cite{ATLAS:2022uhq}.
The \zmutau search has similar sensitivity as the prior limit, coming mainly from the \zmutauh channel.
The observed \zmutau limit is weaker than the expected one, mainly because of the $2.5\sigma$ excess we find in the \zmutauh channel.
The \zetau search presented here is limited by high electron trigger thresholds, lower electron identification efficiencies,
and large rates of \jtotau and $\Pe\to\tauh$ backgrounds, as compared with the \zmutau search, and the \zetau search in Ref.~\cite{ATLAS:2021bdj}.
Additionally, the electron energy scale has much larger uncertainties in the \tautau embedding simulation that significantly impacts the \zetau search
sensitivity.

As seen in Fig.~\ref{fig:scan_limits}, the \zpemu search yields an expected 95\% \CL upper limit on $\sigma(\pp\to\zprime\!+\!\PX)\br(\zpemu)$ 
of 4 fb at 110 \GeV and 0.4 fb at 500 \GeV, with an approximately exponential relation between the mass and the expected sensitivity.
The observed limits are consistent with these.
Searches for \zpemu over various mass ranges above 200\GeV have been reported by ATLAS~\cite{ATLAS:2015dva,ATLAS:2016loq} and CMS~\cite{CMS:2016awf,CMS:2022fsw}.  Where the ranges overlap, the present limits represent an improvement of at least a factor of five over these, and are either the first or the most stringent expected limits to date on the \zpemu process within the covered mass range.

\ifthenelse{\boolean{cms@external}}{\clearpage}{}
\section{Summary}
\label{sec:summary}

A search is presented for flavor violating decays of the \PZ boson to charged leptons, and for the presence of a heavier vector boson \zprime exhibiting such decays.  The data from proton-proton collisions at $\sqrt{s}=13\TeV$ were collected with the CMS detector at the LHC, and correspond to an integrated luminosity of $138\fbinv$.  The specific decay modes considered are $\zzprime\to\Pe\PGm$, \zetau, and \zmutau.  No significant excess of events over backgrounds from standard model processes is observed.  Observed (expected) upper limits of $1.9\times10^{-7}$, $1.38\times10^{-5}$, and $1.20\times10^{-5}$ ($2.0\times10^{-7}$, $1.14\times10^{-5}$, and $0.53\times10^{-5}$) at 95\% confidence level are set on the branching fractions for \zemu, \zetau, and \zmutau, respectively.  The limit for \zemu is the most restrictive to date, while for \zmutau the sensitivity in terms of the expected limit is the same as that of the previous best limit.  All of these limits are consistent with expectations from the standard model, and with constraints inferred from low-energy experimental limits.  For \zprime boson masses in the range 110--500\GeV, upper limits are set on the cross section times the branching fraction to $\Pe\PGm$ that range from 0.3 to 7\unit{fb}, and are the most restrictive to date for this mass range.  Future studies can benefit from additional data, since even the systematic uncertainties arise mainly from statistical ones in control samples.

\begin{acknowledgments}
We congratulate our colleagues in the CERN accelerator departments for the excellent performance of the LHC and thank the technical and administrative staffs at CERN and at other CMS institutes for their contributions to the success of the CMS effort. In addition, we gratefully acknowledge the computing centers and personnel of the Worldwide LHC Computing Grid and other centers for delivering so effectively the computing infrastructure essential to our analyses. Finally, we acknowledge the enduring support for the construction and operation of the LHC, the CMS detector, and the supporting computing infrastructure provided by the following funding agencies: SC (Armenia), BMBWF and FWF (Austria); FNRS and FWO (Belgium); CNPq, CAPES, FAPERJ, FAPERGS, and FAPESP (Brazil); MES and BNSF (Bulgaria); CERN; CAS, MoST, and NSFC (China); MINCIENCIAS (Colombia); MSES and CSF (Croatia); RIF (Cyprus); SENESCYT (Ecuador); ERC PRG, TARISTU24-TK10 and MoER TK202 (Estonia); Academy of Finland, MEC, and HIP (Finland); CEA and CNRS/IN2P3 (France); SRNSF (Georgia); BMBF, DFG, and HGF (Germany); GSRI (Greece); NKFIH (Hungary); DAE and DST (India); IPM (Iran); SFI (Ireland); INFN (Italy); MSIT and NRF (Republic of Korea); MES (Latvia); LMTLT (Lithuania); MOE and UM (Malaysia); BUAP, CINVESTAV, CONACYT, LNS, SEP, and UASLP-FAI (Mexico); MOS (Montenegro); MBIE (New Zealand); PAEC (Pakistan); MES and NSC (Poland); FCT (Portugal); MESTD (Serbia); MICIU/AEI and PCTI (Spain); MOSTR (Sri Lanka); Swiss Funding Agencies (Switzerland); MST (Taipei); MHESI and NSTDA (Thailand); TUBITAK and TENMAK (T\"{u}rkiye); NASU (Ukraine); STFC (United Kingdom); 

\hyphenation{Rachada-pisek} Individuals have received support from the Marie-Curie program and the European Research Council and Horizon 2020 Grant, contract Nos.\ 675440, 724704, 752730, 758316, 765710, 824093, 101115353, 101002207, 101001205, and COST Action CA16108 (European Union); the Leventis Foundation; the Alfred P.\ Sloan Foundation; the Alexander von Humboldt Foundation; the Science Committee, project no. 22rl-037 (Armenia); the Fonds pour la Formation \`a la Recherche dans l'Industrie et dans l'Agriculture (FRIA-Belgium); the Beijing Municipal Science \& Technology Commission, No. Z191100007219010, the Fundamental Research Funds for the Central Universities, the Ministry of Science and Technology of China under Grant No. 2023YFA1605804, and the Natural Science Foundation of China under Grant No. 12061141002 (China); the Ministry of Education, Youth and Sports (MEYS) of the Czech Republic; the Shota Rustaveli National Science Foundation, grant FR-22-985 (Georgia); the Deutsche Forschungsgemeinschaft (DFG), among others, under Germany's Excellence Strategy -- EXC 2121 ``Quantum Universe" -- 390833306, and under project number 400140256 - GRK2497; the Hellenic Foundation for Research and Innovation (HFRI), Project Number 2288 (Greece); the Hungarian Academy of Sciences, the New National Excellence Program - \'UNKP, the NKFIH research grants K 131991, K 133046, K 138136, K 143460, K 143477, K 146913, K 146914, K 147048, 2020-2.2.1-ED-2021-00181, TKP2021-NKTA-64, and 2021-4.1.2-NEMZ\_KI-2024-00036 (Hungary); the Council of Science and Industrial Research, India; ICSC -- National Research Center for High Performance Computing, Big Data and Quantum Computing, FAIR -- Future Artificial Intelligence Research, and CUP I53D23001070006 (Mission 4 Component 1), funded by the NextGenerationEU program (Italy); the Latvian Council of Science; the Ministry of Education and Science, project no. 2022/WK/14, and the National Science Center, contracts Opus 2021/41/B/ST2/01369 and 2021/43/B/ST2/01552 (Poland); the Funda\c{c}\~ao para a Ci\^encia e a Tecnologia, grant CEECIND/01334/2018 (Portugal); MICIU/AEI/10.13039/501100011033, ERDF/EU, "European Union NextGenerationEU/PRTR", and Programa Severo Ochoa del Principado de Asturias (Spain); the Chulalongkorn Academic into Its 2nd Century Project Advancement Project, and the National Science, Research and Innovation Fund via the Program Management Unit for Human Resources \& Institutional Development, Research and Innovation, grant B39G680009 (Thailand); the Kavli Foundation; the Nvidia Corporation; the SuperMicro Corporation; the Welch Foundation, contract C-1845; and the Weston Havens Foundation (USA).
\end{acknowledgments}

\bibliography{auto_generated}
\cleardoublepage \appendix\section{The CMS Collaboration \label{app:collab}}\begin{sloppypar}\hyphenpenalty=5000\widowpenalty=500\clubpenalty=5000\cmsinstitute{Yerevan Physics Institute, Yerevan, Armenia}
{\tolerance=6000
A.~Hayrapetyan, V.~Makarenko\cmsorcid{0000-0002-8406-8605}, A.~Tumasyan\cmsAuthorMark{1}\cmsorcid{0009-0000-0684-6742}
\par}
\cmsinstitute{Institut f\"{u}r Hochenergiephysik, Vienna, Austria}
{\tolerance=6000
W.~Adam\cmsorcid{0000-0001-9099-4341}, J.W.~Andrejkovic, L.~Benato\cmsorcid{0000-0001-5135-7489}, T.~Bergauer\cmsorcid{0000-0002-5786-0293}, M.~Dragicevic\cmsorcid{0000-0003-1967-6783}, C.~Giordano, P.S.~Hussain\cmsorcid{0000-0002-4825-5278}, M.~Jeitler\cmsAuthorMark{2}\cmsorcid{0000-0002-5141-9560}, N.~Krammer\cmsorcid{0000-0002-0548-0985}, A.~Li\cmsorcid{0000-0002-4547-116X}, D.~Liko\cmsorcid{0000-0002-3380-473X}, I.~Mikulec\cmsorcid{0000-0003-0385-2746}, J.~Schieck\cmsAuthorMark{2}\cmsorcid{0000-0002-1058-8093}, R.~Sch\"{o}fbeck\cmsAuthorMark{2}\cmsorcid{0000-0002-2332-8784}, D.~Schwarz\cmsorcid{0000-0002-3821-7331}, M.~Shooshtari, M.~Sonawane\cmsorcid{0000-0003-0510-7010}, W.~Waltenberger\cmsorcid{0000-0002-6215-7228}, C.-E.~Wulz\cmsAuthorMark{2}\cmsorcid{0000-0001-9226-5812}
% ...
\par}
\vskip\cmsinstskip
\dag:~Deceased\\
$^{1}$Also at Yerevan State University, Yerevan, Armenia\\
$^{2}$Also at TU Wien, Vienna, Austria\\
% ...
\end{sloppypar}
\end{document}